

%
%
%
%
%
%
%

\catcode`\@=11
\font\tensmc=cmcsc10
\def\smc{\tensmc}

\def\hcorrection#1{\advance\hoffset by #1 }
\def\vcorrection#1{\advance\voffset by #1 }
\def\wlog#1{}
\newif\iftitle@

\outer\def\heading{\bigbreak\bgroup\let\\=\cr\tabskip\centering
    \halign to \hsize\bgroup\smc\hfill\ignorespaces##\unskip\hfill\cr}
\def\endheading{\cr\egroup\egroup\nobreak\medskip}

\outer\def\endproclaim{\par\ifdim\lastskip<\medskipamount\removelastskip
  \penalty 55 \fi\medskip\rm}
\outer\def\demo#1{\par\ifdim\lastskip<\smallskipamount\removelastskip
    \smallskip\fi\noindent{\smc\ignorespaces#1\unskip:\enspace}\rm
      \ignorespaces}

\hyphenation{man-u-script man-u-scripts ap-pen-dix ap-pen-di-ces}
\hyphenation{data-base data-bases}
\ifx\amstexloaded@\relax\catcode`\@=13
   \else\let\amstexloaded@=\relax\fi
\newlinechar=`\^^J
\def\eat@#1{}
\def\Space@.{\futurelet\Space@\relax}
\Space@. %
\newhelp\athelp@
{Only certain combinations beginning with @ make sense to me.^^J
Perhaps you wanted \string\@\space for a printed @?^^J
I've ignored the character or group after @.}
\def\futureletnextat@{\futurelet\next\at@}
{\catcode`\@=\active
\lccode`\Z=`\@ \lowercase
{\gdef@{\expandafter\csname futureletnextatZ\endcsname}
\expandafter\gdef\csname atZ\endcsname
   {\ifcat\noexpand\next a\def\next{\csname atZZ\endcsname}\else
   \ifcat\noexpand\next0\def\next{\csname atZZ\endcsname}\else
    \def\next{\csname atZZZ\endcsname}\fi\fi\next}
\expandafter\gdef\csname atZZ\endcsname#1{\expandafter
   \ifx\csname #1Zat\endcsname\relax\def\next
     {\errhelp\expandafter=\csname athelpZ\endcsname
      \errmessage{Invalid use of \string@}}\else
       \def\next{\csname #1Zat\endcsname}\fi\next}
\expandafter\gdef\csname atZZZ\endcsname#1{\errhelp
    \expandafter=\csname athelpZ\endcsname
      \errmessage{Invalid use of \string@}}}}
\def\atdef@#1{\expandafter\def\csname #1@at\endcsname}
\newhelp\defahelp@{If you typed \string\define\space cs instead of
\string\define\string\cs\space^^J
I've substituted an inaccessible control sequence so that your^^J
definition will be completed without mixing me up too badly.^^J
If you typed \string\define{\string\cs} the inaccessible control sequence^^J
was defined to be \string\cs, and the rest of your^^J
definition appears as input.}
\newhelp\defbhelp@{I've ignored your definition, because it might^^J
conflict with other uses that are important to me.}
\def\define{\futurelet\next\define@}
\def\define@{\ifcat\noexpand\next\relax
  \def\next{\define@@}%
  \else\errhelp=\defahelp@
  \errmessage{\string\define\space must be followed by a control
     sequence}\def\next{\def\garbage@}\fi\next}
\def\undefined@{}
\def\preloaded@{}
\def\define@@#1{\ifx#1\relax\errhelp=\defbhelp@
   \errmessage{\string#1\space is already defined}\def\next{\def\garbage@}%
   \else\expandafter\ifx\csname\expandafter\eat@\string
	 #1@\endcsname\undefined@\errhelp=\defbhelp@
   \errmessage{\string#1\space can't be defined}\def\next{\def\garbage@}%
   \else\expandafter\ifx\csname\expandafter\eat@\string#1\endcsname\relax
     \def\next{\def#1}\else\errhelp=\defbhelp@
     \errmessage{\string#1\space is already defined}\def\next{\def\garbage@}%
      \fi\fi\fi\next}
\def\famzero{\fam\z@}

\def\cos{\mathop{\famzero cos}\nolimits}

\def\exp{\mathop{\famzero exp}\nolimits}

\def\lim{\mathop{\famzero lim}}

\def\log{\mathop{\famzero log}\nolimits}

\def\min{\mathop{\famzero min}}

\def\sec{\mathop{\famzero sec}\nolimits}
\def\sin{\mathop{\famzero sin}\nolimits}

\def\tan{\mathop{\famzero tan}\nolimits}
\def\tanh{\mathop{\famzero tanh}\nolimits}
\def\textfont@#1#2{\def#1{\relax\ifmmode
    \errmessage{Use \string#1\space only in text}\else#2\fi}}
\let\ic@=\/
\def\/{\unskip\ic@}
\def\textfonti{\the\textfont1 }
\def\t#1#2{{\edef\next{\the\font}\textfonti\accent"7F \next#1#2}}
\let\B=\=
\let\D=\.
\def~{\unskip\nobreak\ \ignorespaces}
{\catcode`\@=\active
\gdef\@{\char'100 }}
\atdef@-{\leavevmode\futurelet\next\athyph@}
\def\athyph@{\ifx\next-\let\next=\athyph@@
  \else\let\next=\athyph@@@\fi\next}
\def\athyph@@@{\hbox{-}}
\def\athyph@@#1{\futurelet\next\athyph@@@@}
\def\athyph@@@@{\if\next-\def\next##1{\hbox{---}}\else
    \def\next{\hbox{--}}\fi\next}
\def\.{.\spacefactor=\@m}
\atdef@.{\null.}
\atdef@,{\null,}
\atdef@;{\null;}
\atdef@:{\null:}
\atdef@?{\null?}
\atdef@!{\null!}
\def\srdr@{\thinspace}
\def\drsr@{\kern.02778em}
\def\sldl@{\kern.02778em}
\def\dlsl@{\thinspace}
\atdef@"{\unskip\futurelet\next\atqq@}
\def\atqq@{\ifx\next\Space@\def\next. {\atqq@@}\else
	 \def\next.{\atqq@@}\fi\next.}
\def\atqq@@{\futurelet\next\atqq@@@}
\def\atqq@@@{\ifx\next`\def\next`{\atqql@}\else\def\next'{\atqqr@}\fi\next}
\def\atqql@{\futurelet\next\atqql@@}
\def\atqql@@{\ifx\next`\def\next`{\sldl@``}\else\def\next{\dlsl@`}\fi\next}
\def\atqqr@{\futurelet\next\atqqr@@}
\def\atqqr@@{\ifx\next'\def\next'{\srdr@''}\else\def\next{\drsr@'}\fi\next}

\def\textfontii{\the\textfont2 }
\def\{{\relax\ifmmode\lbrace\else
    {\textfontii f}\spacefactor=\@m\fi}
\def\}{\relax\ifmmode\rbrace\else
    \let\@sf=\empty\ifhmode\edef\@sf{\spacefactor=\the\spacefactor}\fi
      {\textfontii g}\@sf\relax\fi}
\def\nonhmodeerr@#1{\errmessage
     {\string#1\space allowed only within text}}
\def\linebreak{\relax\ifhmode\unskip\break\else
    \nonhmodeerr@\linebreak\fi}
\def\allowlinebreak{\relax
   \ifhmode\allowbreak\else\nonhmodeerr@\allowlinebreak\fi}
\newskip\saveskip@
\def\nolinebreak{\relax\ifhmode\saveskip@=\lastskip\unskip
  \nobreak\ifdim\saveskip@>\z@\hskip\saveskip@\fi
   \else\nonhmodeerr@\nolinebreak\fi}
\def\newline{\relax\ifhmode\null\hfil\break
    \else\nonhmodeerr@\newline\fi}
\def\nonmathaerr@#1{\errmessage
     {\string#1\space is not allowed in display math mode}}
\def\nonmathberr@#1{\errmessage{\string#1\space is allowed only in math mode}}
\def\mathbreak{\relax\ifmmode\ifinner\break\else
   \nonmathaerr@\mathbreak\fi\else\nonmathberr@\mathbreak\fi}
\def\nomathbreak{\relax\ifmmode\ifinner\nobreak\else
    \nonmathaerr@\nomathbreak\fi\else\nonmathberr@\nomathbreak\fi}
\def\allowmathbreak{\relax\ifmmode\ifinner\allowbreak\else
     \nonmathaerr@\allowmathbreak\fi\else\nonmathberr@\allowmathbreak\fi}
\def\pagebreak{\relax\ifmmode
   \ifinner\errmessage{\string\pagebreak\space
     not allowed in non-display math mode}\else\postdisplaypenalty-\@M\fi
   \else\ifvmode\penalty-\@M\else\edef\spacefactor@
       {\spacefactor=\the\spacefactor}\vadjust{\penalty-\@M}\spacefactor@
	\relax\fi\fi}
\def\nopagebreak{\relax\ifmmode
     \ifinner\errmessage{\string\nopagebreak\space
    not allowed in non-display math mode}\else\postdisplaypenalty\@M\fi
    \else\ifvmode\nobreak\else\edef\spacefactor@
	{\spacefactor=\the\spacefactor}\vadjust{\penalty\@M}\spacefactor@
	 \relax\fi\fi}
\def\newpage{\relax\ifvmode\vfill\penalty-\@M\else\nonvmodeerr@\newpage\fi}
\def\nonvmodeerr@#1{\errmessage
    {\string#1\space is allowed only between paragraphs}}
\def\smallpagebreak{\relax\ifvmode\smallbreak
      \else\nonvmodeerr@\smallpagebreak\fi}
\def\medpagebreak{\relax\ifvmode\medbreak
       \else\nonvmodeerr@\medpagebreak\fi}
\def\bigpagebreak{\relax\ifvmode\bigbreak
      \else\nonvmodeerr@\bigpagebreak\fi}
\newdimen\captionwidth@
\captionwidth@=\hsize
\advance\captionwidth@ by -1.5in
\def\caption#1{}
\def\topspace#1{\gdef\thespace@{#1}\ifvmode\def\next
    {\futurelet\next\topspace@}\else\def\next{\nonvmodeerr@\topspace}\fi\next}
\def\topspace@{\ifx\next\Space@\def\next. {\futurelet\next\topspace@@}\else
     \def\next.{\futurelet\next\topspace@@}\fi\next.}
\def\topspace@@{\ifx\next\caption\let\next\topspace@@@\else
    \let\next\topspace@@@@\fi\next}
 \def\topspace@@@@{\topinsert\vbox to
       \thespace@{}\endinsert}
\def\topspace@@@\caption#1{\topinsert\vbox to
    \thespace@{}\nobreak
      \smallskip
    \setbox\z@=\hbox{\noindent\ignorespaces#1\unskip}%
   \ifdim\wd\z@>\captionwidth@
   \centerline{\vbox{\hsize=\captionwidth@\noindent\ignorespaces#1\unskip}}%
   \else\centerline{\box\z@}\fi\endinsert}
\def\midspace#1{\gdef\thespace@{#1}\ifvmode\def\next
    {\futurelet\next\midspace@}\else\def\next{\nonvmodeerr@\midspace}\fi\next}
\def\midspace@{\ifx\next\Space@\def\next. {\futurelet\next\midspace@@}\else
     \def\next.{\futurelet\next\midspace@@}\fi\next.}
\def\midspace@@{\ifx\next\caption\let\next\midspace@@@\else
    \let\next\midspace@@@@\fi\next}
 \def\midspace@@@@{\midinsert\vbox to
       \thespace@{}\endinsert}
\def\midspace@@@\caption#1{\midinsert\vbox to
    \thespace@{}\nobreak
      \smallskip
      \setbox\z@=\hbox{\noindent\ignorespaces#1\unskip}%
      \ifdim\wd\z@>\captionwidth@
    \centerline{\vbox{\hsize=\captionwidth@\noindent\ignorespaces#1\unskip}}%
    \else\centerline{\box\z@}\fi\endinsert}
\mathchardef\prime@="0230
\def\prime{{{}\prime@{}}}
\def\prim@s{\prime@\futurelet\next\pr@m@s}

\def\,{\relax\ifmmode\mskip\thinmuskip\else\thinspace\fi}
\def\!{\relax\ifmmode\mskip-\thinmuskip\else\negthinspace\fi}
\def\frac#1#2{{#1\over#2}}

\def\:{\nobreak\hskip.1111em{:}\hskip.3333em plus .0555em\relax}
\def\intic@{\mathchoice{\hskip5\p@}{\hskip4\p@}{\hskip4\p@}{\hskip4\p@}}
\def\negintic@{\mathchoice{\hskip-5\p@}{\hskip-4\p@}{\hskip-4\p@}{\hskip-4\p@}}
\def\intkern@{\mathchoice{\!\!\!}{\!\!}{\!\!}{\!\!}}
\def\intdots@{\mathchoice{\cdots}{{\cdotp}\mkern1.5mu
    {\cdotp}\mkern1.5mu{\cdotp}}{{\cdotp}\mkern1mu{\cdotp}\mkern1mu
      {\cdotp}}{{\cdotp}\mkern1mu{\cdotp}\mkern1mu{\cdotp}}}
\newcount\intno@
\def\iint{\intno@=\tw@\futurelet\next\ints@}
\def\iiint{\intno@=\thr@@\futurelet\next\ints@}
\def\iiiint{\intno@=4 \futurelet\next\ints@}
\def\idotsint{\intno@=\z@\futurelet\next\ints@}
\def\ints@{\findlimits@\ints@@}
\newif\iflimtoken@
\newif\iflimits@
\def\findlimits@{\limtoken@false\limits@false\ifx\next\limits
 \limtoken@true\limits@true\else\ifx\next\nolimits\limtoken@true\limits@false
    \fi\fi}
\def\multintlimits@{\intop\ifnum\intno@=\z@\intdots@
  \else\intkern@\fi
    \ifnum\intno@>\tw@\intop\intkern@\fi
     \ifnum\intno@>\thr@@\intop\intkern@\fi\intop}
\def\multint@{\int\ifnum\intno@=\z@\intdots@\else\intkern@\fi
   \ifnum\intno@>\tw@\int\intkern@\fi
    \ifnum\intno@>\thr@@\int\intkern@\fi\int}
\def\ints@@{\iflimtoken@\def\ints@@@{\iflimits@
   \negintic@\mathop{\intic@\multintlimits@}\limits\else
    \multint@\nolimits\fi\eat@}\else
     \def\ints@@@{\multint@\nolimits}\fi\ints@@@}
\def\Sb{_\bgroup\vspace@
	\baselineskip=\fontdimen10 \scriptfont\tw@
	\advance\baselineskip by \fontdimen12 \scriptfont\tw@
	\lineskip=\thr@@\fontdimen8 \scriptfont\thr@@
	\lineskiplimit=\thr@@\fontdimen8 \scriptfont\thr@@
	\Let@\vbox\bgroup\halign\bgroup \hfil$\scriptstyle
	    {##}$\hfil\cr}
\def\endSb{\crcr\egroup\egroup\egroup}
\def\Sp{^\bgroup\vspace@
	\baselineskip=\fontdimen10 \scriptfont\tw@
	\advance\baselineskip by \fontdimen12 \scriptfont\tw@
	\lineskip=\thr@@\fontdimen8 \scriptfont\thr@@
	\lineskiplimit=\thr@@\fontdimen8 \scriptfont\thr@@
	\Let@\vbox\bgroup\halign\bgroup \hfil$\scriptstyle
	    {##}$\hfil\cr}
\def\endSp{\crcr\egroup\egroup\egroup}
\def\Let@{\relax\iffalse{\fi\let\\=\cr\iffalse}\fi}
\def\vspace@{\def\vspace##1{\noalign{\vskip##1 }}}
\def\aligned{\,\vcenter\bgroup\vspace@\Let@\openup\jot\m@th\ialign
  \bgroup \strut\hfil$\displaystyle{##}$&$\displaystyle{{}##}$\hfil\crcr}
\def\endaligned{\crcr\egroup\egroup}
\def\matrix{\,\vcenter\bgroup\Let@\vspace@
    \normalbaselines
  \m@th\ialign\bgroup\hfil$##$\hfil&&\quad\hfil$##$\hfil\crcr
    \mathstrut\crcr\noalign{\kern-\baselineskip}}
\def\endmatrix{\crcr\mathstrut\crcr\noalign{\kern-\baselineskip}\egroup
		\egroup\,}
\newtoks\hashtoks@
\hashtoks@={#}
\def\format{\crcr\egroup\iffalse{\fi\ifnum`}=0 \fi\format@}
\def\format@#1\\{\def\preamble@{#1}%
  \def\c{\hfil$\the\hashtoks@$\hfil}%
  \def\r{\hfil$\the\hashtoks@$}%
  \def\l{$\the\hashtoks@$\hfil}%
  \setbox\z@=\hbox{\xdef\Preamble@{\preamble@}}\ifnum`{=0 \fi\iffalse}\fi
   \ialign\bgroup\span\Preamble@\crcr}

\newif\iftagsleft@
\tagsleft@true
\def\TagsOnRight{\global\tagsleft@false}
\def\tag#1$${\iftagsleft@\leqno\else\eqno\fi
 \hbox{\def\pagebreak{\global\postdisplaypenalty-\@M}%
 \def\nopagebreak{\global\postdisplaypenalty\@M}\rm(#1\unskip)}%
  $$\postdisplaypenalty\z@\ignorespaces}
\interdisplaylinepenalty=\@M
\def\allowdisplaybreak@{\def\allowdisplaybreak{\noalign{\allowbreak}}}
\def\displaybreak@{\def\displaybreak{\noalign{\break}}}
\def\align#1\endalign{\def\tag{&}\vspace@\allowdisplaybreak@\displaybreak@
  \iftagsleft@\lalign@#1\endalign\else
   \ralign@#1\endalign\fi}
\def\ralign@#1\endalign{\displ@y\Let@\tabskip\centering\halign to\displaywidth
     {\hfil$\displaystyle{##}$\tabskip=\z@&$\displaystyle{{}##}$\hfil
       \tabskip=\centering&\llap{\hbox{(\rm##\unskip)}}\tabskip\z@\crcr
	     #1\crcr}}
\def\lalign@#1\endalign{\displ@y\Let@\tabskip\centering\halign to \displaywidth
   {\hfil$\displaystyle{##}$\tabskip=\z@&$\displaystyle{{}##}$\hfil
   \tabskip=\centering&\kern-\displaywidth
	\rlap{\hbox{(\rm##\unskip)}}\tabskip=\displaywidth\crcr
	       #1\crcr}}
\def\overrightarrow{\mathpalette\overrightarrow@}
\def\overrightarrow@#1#2{\vbox{\ialign{$##$\cr
    #1{-}\mkern-6mu\cleaders\hbox{$#1\mkern-2mu{-}\mkern-2mu$}\hfill
     \mkern-6mu{\to}\cr
     \noalign{\kern -1\p@\nointerlineskip}
     \hfil#1#2\hfil\cr}}}
\def\overleftarrow{\mathpalette\overleftarrow@}
\def\overleftarrow@#1#2{\vbox{\ialign{$##$\cr
     #1{\leftarrow}\mkern-6mu\cleaders\hbox{$#1\mkern-2mu{-}\mkern-2mu$}\hfill
      \mkern-6mu{-}\cr
     \noalign{\kern -1\p@\nointerlineskip}
     \hfil#1#2\hfil\cr}}}
\def\overleftrightarrow{\mathpalette\overleftrightarrow@}
\def\overleftrightarrow@#1#2{\vbox{\ialign{$##$\cr
     #1{\leftarrow}\mkern-6mu\cleaders\hbox{$#1\mkern-2mu{-}\mkern-2mu$}\hfill
       \mkern-6mu{\to}\cr
    \noalign{\kern -1\p@\nointerlineskip}
      \hfil#1#2\hfil\cr}}}
\def\underrightarrow{\mathpalette\underrightarrow@}
\def\underrightarrow@#1#2{\vtop{\ialign{$##$\cr
    \hfil#1#2\hfil\cr
     \noalign{\kern -1\p@\nointerlineskip}
    #1{-}\mkern-6mu\cleaders\hbox{$#1\mkern-2mu{-}\mkern-2mu$}\hfill
     \mkern-6mu{\to}\cr}}}
\def\underleftarrow{\mathpalette\underleftarrow@}
\def\underleftarrow@#1#2{\vtop{\ialign{$##$\cr
     \hfil#1#2\hfil\cr
     \noalign{\kern -1\p@\nointerlineskip}
     #1{\leftarrow}\mkern-6mu\cleaders\hbox{$#1\mkern-2mu{-}\mkern-2mu$}\hfill
      \mkern-6mu{-}\cr}}}
\def\underleftrightarrow{\mathpalette\underleftrightarrow@}
\def\underleftrightarrow@#1#2{\vtop{\ialign{$##$\cr
      \hfil#1#2\hfil\cr
    \noalign{\kern -1\p@\nointerlineskip}
     #1{\leftarrow}\mkern-6mu\cleaders\hbox{$#1\mkern-2mu{-}\mkern-2mu$}\hfill
       \mkern-6mu{\to}\cr}}}
\def\sqrt#1{\radical"270370 {#1}}
\def\dots{\relax\ifmmode\let\next=\ldots\else\let\next=\tdots@\fi\next}
\def\tdots@{\unskip\ \tdots@@}
\def\tdots@@{\futurelet\next\tdots@@@}
\def\tdots@@@{$\mathinner{\ldotp\ldotp\ldotp}\,
   \ifx\next,$\else
   \ifx\next.\,$\else
   \ifx\next;\,$\else
   \ifx\next:\,$\else
   \ifx\next?\,$\else
   \ifx\next!\,$\else
   $ \fi\fi\fi\fi\fi\fi}
\def\text{\relax\ifmmode\let\next=\text@\else\let\next=\text@@\fi\next}
\def\text@@#1{\hbox{#1}}
\def\text@#1{\mathchoice
 {\hbox{\everymath{\displaystyle}\def\textfonti{\the\textfont1 }%
    \def\textfontii{\the\textfont2 }\textdef@@ T#1}}
 {\hbox{\everymath{\textstyle}\def\textfonti{\the\textfont1 }%
    \def\textfontii{\the\textfont2 }\textdef@@ T#1}}
 {\hbox{\everymath{\scriptstyle}\def\textfonti{\the\scriptfont1 }%
   \def\textfontii{\the\scriptfont2 }\textdef@@ S\rm#1}}
 {\hbox{\everymath{\scriptscriptstyle}\def\textfonti{\the\scriptscriptfont1 }%
   \def\textfontii{\the\scriptscriptfont2 }\textdef@@ s\rm#1}}}
\def\textdef@@#1{\textdef@#1\rm \textdef@#1\bf
   \textdef@#1\sl \textdef@#1\it}

\def\textdef@#1#2{\def\next{\csname\expandafter\eat@\string#2fam\endcsname}%
\if S#1\edef#2{\the\scriptfont\next\relax}%
 \else\if s#1\edef#2{\the\scriptscriptfont\next\relax}%
 \else\edef#2{\the\textfont\next\relax}\fi\fi}
\scriptfont\itfam=\tenit \scriptscriptfont\itfam=\tenit
\scriptfont\slfam=\tensl \scriptscriptfont\slfam=\tensl
\mathcode`\0="0030
\mathcode`\1="0031
\mathcode`\2="0032
\mathcode`\3="0033
\mathcode`\4="0034
\mathcode`\5="0035
\mathcode`\6="0036
\mathcode`\7="0037
\mathcode`\8="0038
\mathcode`\9="0039
\def\Cal{\relax\ifmmode\let\next=\Cal@\else
     \def\next{\errmessage{Use \string\Cal\space only in math mode}}\fi\next}
\def\Cal@#1{{\fam2 #1}}
\def\bold{\relax\ifmmode\let\next=\bold@\else
   \def\next{\errmessage{Use \string\bold\space only in math
      mode}}\fi\next}\def\bold@#1{{\fam\bffam #1}}
\mathchardef\Gamma="0000
\mathchardef\Delta="0001
\mathchardef\Theta="0002
\mathchardef\Lambda="0003
\mathchardef\Xi="0004
\mathchardef\Pi="0005
\mathchardef\Sigma="0006
\mathchardef\Upsilon="0007
\mathchardef\Phi="0008
\mathchardef\Psi="0009
\mathchardef\Omega="000A
\mathchardef\varGamma="0100
\mathchardef\varDelta="0101
\mathchardef\varTheta="0102
\mathchardef\varLambda="0103
\mathchardef\varXi="0104
\mathchardef\varPi="0105
\mathchardef\varSigma="0106
\mathchardef\varUpsilon="0107
\mathchardef\varPhi="0108
\mathchardef\varPsi="0109
\mathchardef\varOmega="010A
\font\dummyft@=dummy
\fontdimen1 \dummyft@=\z@
\fontdimen2 \dummyft@=\z@
\fontdimen3 \dummyft@=\z@
\fontdimen4 \dummyft@=\z@
\fontdimen5 \dummyft@=\z@
\fontdimen6 \dummyft@=\z@
\fontdimen7 \dummyft@=\z@
\fontdimen8 \dummyft@=\z@
\fontdimen9 \dummyft@=\z@
\fontdimen10 \dummyft@=\z@
\fontdimen11 \dummyft@=\z@
\fontdimen12 \dummyft@=\z@
\fontdimen13 \dummyft@=\z@
\fontdimen14 \dummyft@=\z@
\fontdimen15 \dummyft@=\z@
\fontdimen16 \dummyft@=\z@
\fontdimen17 \dummyft@=\z@
\fontdimen18 \dummyft@=\z@
\fontdimen19 \dummyft@=\z@
\fontdimen20 \dummyft@=\z@
\fontdimen21 \dummyft@=\z@
\fontdimen22 \dummyft@=\z@
\def\fontlist@{\\{\tenrm}\\{\sevenrm}\\{\fiverm}\\{\teni}\\{\seveni}%
 \\{\fivei}\\{\tensy}\\{\sevensy}\\{\fivesy}\\{\tenex}\\{\tenbf}\\{\sevenbf}%
 \\{\fivebf}\\{\tensl}\\{\tenit}\\{\tensmc}}
\def\dodummy@{{\def\\##1{\global\let##1=\dummyft@}\fontlist@}}
\newif\ifsyntax@
\newcount\countxviii@
\def\newtoks@{\alloc@5\toks\toksdef\@cclvi}
\def\nopages@{\output={\setbox\z@=\box\@cclv \deadcycles=\z@}\newtoks@\output}
\def\syntax{\syntax@true\dodummy@\countxviii@=\count18
\loop \ifnum\countxviii@ > \z@ \textfont\countxviii@=\dummyft@
   \scriptfont\countxviii@=\dummyft@ \scriptscriptfont\countxviii@=\dummyft@
     \advance\countxviii@ by-\@ne\repeat
\dummyft@\tracinglostchars=\z@
  \nopages@\frenchspacing\hbadness=\@M}
\def\wlog#1{\immediate\write-1{#1}}
\catcode`\@=\active
%
%
%
%
%


\def\ltwid{\raise.3ex\hbox{$<$\kern-.75em\lower1ex\hbox{$\sim$}}}
\def\gl{\raise.5ex\hbox{$>$}\kern-.8em\lower.5ex\hbox{$<$}}
\def\gtwid{\raise.3ex\hbox{$>$\kern-.75em\lower1ex\hbox{$\sim$}}}

\def\3he{$^3\text{He}$}

\def\eg{{\it e.g.}}

\def\etal{{\it et al}}


\def\pd{\partial}

\def\Tc{$T_c$}

\def\a0{\text{ \AA}}
\def\Yba{YBa$_2$\-Cu$_3$O$_{7-\delta}$}

%
%
%
%
%
%
%
%
%
%
%
%
%
\catcode`\@=11 
\let\rel@x=\relax
\let\n@expand=\relax
\def\pr@tect{\let\n@expand=\noexpand}
\let\protect=\pr@tect
\let\gl@bal=\global
%
%
%
\newfam\cpfam
\newdimen\b@gheight             \b@gheight=12pt
\newcount\f@ntkey               \f@ntkey=0
\def\f@m{\afterassignment\samef@nt\f@ntkey=}
\def\samef@nt{\fam=\f@ntkey \the\textfont\f@ntkey\rel@x}
\def\setstr@t{\setbox\strutbox=\hbox{\vrule height 0.85\b@gheight
				depth 0.35\b@gheight width\z@ }}
%

%
%
%
%
%

\font\fourteenrm  =cmr10 scaled\magstep2
\font\twelverm    =cmr10 scaled\magstep1
\font\ninerm      =cmr9
\font\sixrm       =cmr6

\font\fourteenbf  =cmbx10 scaled\magstep3
\font\twelvebf    =cmbx10 scaled\magstep1
\font\ninebf      =cmbx9
\font\sixbf       =cmbx6
\font\seventeeni  =cmmi10 scaled\magstep3    \skewchar\seventeeni='177
\font\fourteeni   =cmmi10 scaled\magstep2     \skewchar\fourteeni='177
\font\twelvei     =cmmi10 scaled\magstep1       \skewchar\twelvei='177
\font\ninei       =cmmi9                          \skewchar\ninei='177
\font\sixi        =cmmi6                           \skewchar\sixi='177
\font\seventeensy =cmsy10 scaled\magstep3    \skewchar\seventeensy='60
\font\fourteensy  =cmsy10 scaled\magstep2     \skewchar\fourteensy='60
\font\twelvesy    =cmsy10 scaled\magstep1       \skewchar\twelvesy='60
\font\ninesy      =cmsy9                          \skewchar\ninesy='60
\font\sixsy       =cmsy6                           \skewchar\sixsy='60

\font\fourteenex  =cmex10 scaled\magstep2
\font\twelveex    =cmex10 scaled\magstep1

\font\fourteensl  =cmsl10 scaled\magstep2
\font\twelvesl    =cmsl10 scaled\magstep1
\font\ninesl      =cmsl9

\font\fourteenit  =cmti10 scaled\magstep2
\font\twelveit    =cmti10 scaled\magstep1
\font\nineit      =cmti9
\font\fourteentt  =cmtt10 scaled\magstep2
\font\twelvett    =cmtt10 scaled\magstep1
\font\fourteencp  =cmcsc10 scaled\magstep2
\font\twelvecp    =cmcsc10 scaled\magstep1
\font\tencp       =cmcsc10
%
%
\def\fourteenf@nts{\relax
    \textfont0=\fourteenrm          \scriptfont0=\tenrm
      \scriptscriptfont0=\sevenrm
    \textfont1=\fourteeni           \scriptfont1=\teni
      \scriptscriptfont1=\seveni
    \textfont2=\fourteensy          \scriptfont2=\tensy
      \scriptscriptfont2=\sevensy
    \textfont3=\fourteenex          \scriptfont3=\twelveex
      \scriptscriptfont3=\tenex
    \textfont\itfam=\fourteenit     \scriptfont\itfam=\tenit
    \textfont\slfam=\fourteensl     \scriptfont\slfam=\tensl
    \textfont\bffam=\fourteenbf     \scriptfont\bffam=\tenbf
      \scriptscriptfont\bffam=\sevenbf
    \textfont\ttfam=\fourteentt
    \textfont\cpfam=\fourteencp }
\def\twelvef@nts{\relax
    \textfont0=\twelverm          \scriptfont0=\ninerm
      \scriptscriptfont0=\sixrm
    \textfont1=\twelvei           \scriptfont1=\ninei
      \scriptscriptfont1=\sixi
    \textfont2=\twelvesy           \scriptfont2=\ninesy
      \scriptscriptfont2=\sixsy
    \textfont3=\twelveex          \scriptfont3=\tenex
      \scriptscriptfont3=\tenex
    \textfont\itfam=\twelveit     \scriptfont\itfam=\nineit
    \textfont\slfam=\twelvesl     \scriptfont\slfam=\ninesl
    \textfont\bffam=\twelvebf     \scriptfont\bffam=\ninebf
      \scriptscriptfont\bffam=\sixbf
    \textfont\ttfam=\twelvett
    \textfont\cpfam=\twelvecp }
\def\tenf@nts{\relax
    \textfont0=\tenrm          \scriptfont0=\sevenrm
      \scriptscriptfont0=\fiverm
    \textfont1=\teni           \scriptfont1=\seveni
      \scriptscriptfont1=\fivei
    \textfont2=\tensy          \scriptfont2=\sevensy
      \scriptscriptfont2=\fivesy
    \textfont3=\tenex          \scriptfont3=\tenex
      \scriptscriptfont3=\tenex
    \textfont\itfam=\tenit     \scriptfont\itfam=\seveni  
    \textfont\slfam=\tensl     \scriptfont\slfam=\sevenrm 
    \textfont\bffam=\tenbf     \scriptfont\bffam=\sevenbf
      \scriptscriptfont\bffam=\fivebf
    \textfont\ttfam=\tentt
    \textfont\cpfam=\tencp }
%
%
%
\def\rm{\n@expand\f@m0 }
\def\mit{\n@expand\f@m1 }         
\def\cal{\n@expand\f@m2 }
\def\it{\n@expand\f@m\itfam}
\def\sl{\n@expand\f@m\slfam}
\def\bf{\n@expand\f@m\bffam}
\def\tt{\n@expand\f@m\ttfam}
\def\caps{\n@expand\f@m\cpfam}    
\def\em@{\rel@x\ifnum\f@ntkey=0 \it \else
	\ifnum\f@ntkey=\bffam \it \else \rm \fi \fi }
\def\em{\n@expand\em@}
\def\fourteenpoint{\fourteenf@nts \samef@nt \b@gheight=14pt \setstr@t }
\def\twelvepoint{\twelvef@nts \samef@nt \b@gheight=12pt \setstr@t }
\def\tenpoint{\tenf@nts \samef@nt \b@gheight=10pt \setstr@t }
\normalbaselineskip = 20pt plus 0.2pt minus 0.1pt
\normallineskip = 1.5pt plus 0.1pt minus 0.1pt
\normallineskiplimit = 1.5pt
\newskip\normaldisplayskip
\normaldisplayskip = 20pt plus 5pt minus 10pt
\newskip\normaldispshortskip
\normaldispshortskip = 6pt plus 5pt
\newskip\normalparskip
\normalparskip = 6pt plus 2pt minus 1pt
\newskip\skipregister
\skipregister = 5pt plus 2pt minus 1.5pt
\newif\ifsingl@
\newif\ifdoubl@
\newif\iftwelv@  \twelv@true
\def\singlespace{\singl@true\doubl@false\spaces@t}
\def\doublespace{\singl@false\doubl@true\spaces@t}
\def\normalspace{\singl@false\doubl@false\spaces@t}
\def\Tenpoint{\tenpoint\twelv@false\spaces@t}
\def\Twelvepoint{\twelvepoint\twelv@true\spaces@t}
\def\spaces@t{\rel@x
      \iftwelv@ \ifsingl@\subspaces@t3:4;\else\subspaces@t1:1;\fi
       \else \ifsingl@\subspaces@t3:5;\else\subspaces@t4:5;\fi \fi
      \ifdoubl@ \multiply\baselineskip by 5
	 \divide\baselineskip by 4 \fi }
\def\subspaces@t#1:#2;{
      \baselineskip = \normalbaselineskip
      \multiply\baselineskip by #1 \divide\baselineskip by #2
      \lineskip = \normallineskip
      \multiply\lineskip by #1 \divide\lineskip by #2
      \lineskiplimit = \normallineskiplimit
      \multiply\lineskiplimit by #1 \divide\lineskiplimit by #2
      \parskip = \normalparskip
      \multiply\parskip by #1 \divide\parskip by #2
      \abovedisplayskip = \normaldisplayskip
      \multiply\abovedisplayskip by #1 \divide\abovedisplayskip by #2
      \belowdisplayskip = \abovedisplayskip
      \abovedisplayshortskip = \normaldispshortskip
      \multiply\abovedisplayshortskip by #1
	\divide\abovedisplayshortskip by #2
      \belowdisplayshortskip = \abovedisplayshortskip
      \advance\belowdisplayshortskip by \belowdisplayskip
      \divide\belowdisplayshortskip by 2
      \smallskipamount = \skipregister
      \multiply\smallskipamount by #1 \divide\smallskipamount by #2
      \medskipamount = \smallskipamount \multiply\medskipamount by 2
      \bigskipamount = \smallskipamount \multiply\bigskipamount by 4 }
\def\normalbaselines{ \baselineskip=\normalbaselineskip
   \lineskip=\normallineskip \lineskiplimit=\normallineskip
   \iftwelv@\else \multiply\baselineskip by 4 \divide\baselineskip by 5
     \multiply\lineskiplimit by 4 \divide\lineskiplimit by 5
     \multiply\lineskip by 4 \divide\lineskip by 5 \fi }
\Twelvepoint  
\interlinepenalty=50
\interfootnotelinepenalty=5000
\predisplaypenalty=9000
\postdisplaypenalty=500
\hfuzz=1pt
\vfuzz=0.2pt
\newdimen\HOFFSET  \HOFFSET=0pt
\newdimen\VOFFSET  \VOFFSET=0pt
\newdimen\HSWING   \HSWING=0pt
\dimen\footins=8in
%
%
%
\newskip\pagebottomfiller
\pagebottomfiller=\z@ plus \z@ minus \z@
\def\pagecontents{
   \ifvoid\topins\else\unvbox\topins\vskip\skip\topins\fi
   \dimen@ = \dp255 \unvbox255
   \vskip\pagebottomfiller
   \ifvoid\footins\else\vskip\skip\footins\footrule\unvbox\footins\fi
   \ifr@ggedbottom \kern-\dimen@ \vfil \fi }
\def\makeheadline{\vbox to 0pt{ \skip@=\topskip
      \advance\skip@ by -12pt \advance\skip@ by -2\normalbaselineskip
      \vskip\skip@ \line{\vbox to 12pt{}\the\headline} \vss
      }\nointerlineskip}
\def\makefootline{\baselineskip = 1.5\normalbaselineskip
		 \line{\the\footline}}
\newif\iffrontpage
\newif\ifp@genum
\def\nopagenumbers{\p@genumfalse}
\def\pagenumbers{\p@genumtrue}
\pagenumbers
\newtoks\paperheadline
\newtoks\paperfootline
\newtoks\letterheadline
\newtoks\letterfootline
\newtoks\letterinfo
\newtoks\date
\paperheadline={\hfil}
\paperfootline={\hss\iffrontpage\else\ifp@genum\tenrm\folio\hss\fi\fi}
\letterheadline{\iffrontpage \hfil \else
    \rm \ifp@genum page~~\folio\fi \hfil\the\date \fi}
\letterfootline={\iffrontpage\the\letterinfo\else\hfil\fi}
\letterinfo={\hfil}
\def\monthname{\rel@x\ifcase\month 0/\or January\or February\or
   March\or April\or May\or June\or July\or August\or September\or
   October\or November\or December\else\number\month/\fi}
\def\today{\monthname~\number\day, \number\year}
\date={\today}
\headline=\paperheadline 
\footline=\paperfootline 
\countdef\pageno=1      \countdef\pagen@=0
\countdef\pagenumber=1  \pagenumber=1
\def\advancepageno{\gl@bal\advance\pagen@ by 1
   \ifnum\pagenumber<0 \gl@bal\advance\pagenumber by -1
    \else\gl@bal\advance\pagenumber by 1 \fi
    \gl@bal\frontpagefalse  \swing@ }
\def\folio{\ifnum\pagenumber<0 \romannumeral-\pagenumber
	   \else \number\pagenumber \fi }
\def\swing@{\ifodd\pagenumber \gl@bal\advance\hoffset by -\HSWING
	     \else \gl@bal\advance\hoffset by \HSWING \fi }
\def\footrule{\dimen@=\prevdepth\nointerlineskip
   \vbox to 0pt{\vskip -0.25\baselineskip \hrule width 0.35\hsize \vss}
   \prevdepth=\dimen@ }
\let\footnotespecial=\rel@x
\newdimen\footindent
\footindent=24pt
\def\Textindent#1{\noindent\llap{#1\enspace}\ignorespaces}
\def\Vfootnote#1{\insert\footins\bgroup
   \interlinepenalty=\interfootnotelinepenalty \floatingpenalty=20000
   \singl@true\doubl@false\Tenpoint
   \splittopskip=\ht\strutbox \boxmaxdepth=\dp\strutbox
   \leftskip=\footindent \rightskip=\z@skip
   \parindent=0.5\footindent \parfillskip=0pt plus 1fil
   \spaceskip=\z@skip \xspaceskip=\z@skip \footnotespecial
   \Textindent{#1}\footstrut\futurelet\next\fo@t}

\def\vfootnote#1{\Vfootnote{${#1}$}}
\def\footnote#1{\attach{#1}\vfootnote{#1}}

\let\footsymbol=\star
\newcount\lastf@@t           \lastf@@t=-1
\newcount\footsymbolcount    \footsymbolcount=0
\newif\ifPhysRev
\def\bumpfootsymbolcount{\rel@x
   \iffrontpage \bumpfootsymbolpos \else \advance\lastf@@t by 1
     \ifPhysRev \bumpfootsymbolneg \else \bumpfootsymbolpos \fi \fi
   \gl@bal\lastf@@t=\pagen@ }
\def\bumpfootsymbolpos{\ifnum\footsymbolcount <0
			    \gl@bal\footsymbolcount =0 \fi
    \ifnum\lastf@@t<\pagen@ \gl@bal\footsymbolcount=0
     \else \gl@bal\advance\footsymbolcount by 1 \fi }
\def\bumpfootsymbolneg{\ifnum\footsymbolcount >0
	     \gl@bal\footsymbolcount =0 \fi
	 \gl@bal\advance\footsymbolcount by -1 }
\def\fd@f#1 {\xdef\footsymbol{\mathchar"#1 }}
\def\generatefootsymbol{\ifcase\footsymbolcount \fd@f 13F \or \fd@f 279
	\or \fd@f 27A \or \fd@f 278 \or \fd@f 27B \else
	\ifnum\footsymbolcount <0 \fd@f{023 \number-\footsymbolcount }
	 \else \fd@f 203 {\loop \ifnum\footsymbolcount >5
		\fd@f{203 \footsymbol } \advance\footsymbolcount by -1
		\repeat }\fi \fi }

\def\nonfrenchspacing{\sfcode`\.=3001 \sfcode`\!=3000 \sfcode`\?=3000
	\sfcode`\:=2000 \sfcode`\;=1500 \sfcode`\,=1251 }
\nonfrenchspacing
\newdimen\d@twidth
{\setbox0=\hbox{s.} \gl@bal\d@twidth=\wd0 \setbox0=\hbox{s}
	\gl@bal\advance\d@twidth by -\wd0 }
\def\removehglue{\loop \unskip \ifdim\lastskip >\z@ \repeat }
\def\roll@ver#1{\removehglue \nobreak \count255 =\spacefactor \dimen@=\z@
	\ifnum\count255 =3001 \dimen@=\d@twidth \fi
	\ifnum\count255 =1251 \dimen@=\d@twidth \fi
    \iftwelv@ \kern-\dimen@ \else \kern-0.83\dimen@ \fi
   #1\spacefactor=\count255 }
\def\step@ver#1{\rel@x \ifmmode #1\else \ifhmode
	\roll@ver{${}#1$}\else {\setbox0=\hbox{${}#1$}}\fi\fi }
\def\attach#1{\step@ver{^{\mkern 2mu #1} }}

%
%
%
\newcount\chapternumber      \chapternumber=0
\newcount\sectionnumber      \sectionnumber=0
\newcount\equanumber         \equanumber=0
\let\chapterlabel=\rel@x
\let\sectionlabel=\rel@x
\newtoks\chapterstyle        \chapterstyle={\Number}
\newtoks\sectionstyle        \sectionstyle={\Number}
\newskip\chapterskip         \chapterskip=\bigskipamount
\newskip\sectionskip         \sectionskip=\medskipamount
\newskip\headskip            \headskip=8pt plus 3pt minus 3pt
\newdimen\chapterminspace    \chapterminspace=15pc
\newdimen\sectionminspace    \sectionminspace=10pc
\newdimen\referenceminspace  \referenceminspace=20pc
\newif\ifcn@                 \cn@true
\newif\ifcn@@                \cn@@false
\def\numberedchapters{\cn@true}
\def\unnumberedchapters{\cn@false\sequentialequations}
\def\chapterreset{\gl@bal\advance\chapternumber by 1
   \ifnum\equanumber<0 \else\gl@bal\equanumber=0\fi
   \gl@bal\sectionnumber=0 \let\sectionlabel=\rel@x
   \ifcn@ \gl@bal\cn@@true {\pr@tect
       \xdef\chapterlabel{\the\chapterstyle{\the\chapternumber}}}%
    \else \gl@bal\cn@@false \gdef\chapterlabel{\rel@x}\fi }
\def\@alpha#1{\count255='140 \advance\count255 by #1\char\count255}
 \def\alphabetic{\n@expand\@alpha}
\def\@Alpha#1{\count255='100 \advance\count255 by #1\char\count255}
 \def\Alphabetic{\n@expand\@Alpha}
\def\@Roman#1{\uppercase\expandafter{\romannumeral #1}}
 \def\Roman{\n@expand\@Roman}
\def\@roman#1{\romannumeral #1}    \def\roman{\n@expand\@roman}
\def\@number#1{\number #1}         \def\Number{\n@expand\@number}
\def\BLANK#1{\rel@x}               
\def\titleparagraphs{\interlinepenalty=9999
     \leftskip=0.03\hsize plus 0.22\hsize minus 0.03\hsize
     \rightskip=\leftskip \parfillskip=0pt
     \hyphenpenalty=9000 \exhyphenpenalty=9000
     \tolerance=9999 \pretolerance=9000
     \spaceskip=0.333em \xspaceskip=0.5em }
\def\titlestyle#1{\par\begingroup \titleparagraphs
     \iftwelv@\fourteenpoint\else\twelvepoint\fi
   \noindent #1\par\endgroup }
\def\spacecheck#1{\dimen@=\pagegoal\advance\dimen@ by -\pagetotal
   \ifdim\dimen@<#1 \ifdim\dimen@>0pt \vfil\break \fi\fi}
\def\chapter#1{\par \penalty-300 \vskip\chapterskip
   \chapterreset \titlestyle{\ifcn@@\chapterlabel.~\fi #1}
   \nobreak\vskip\headskip \penalty 30000
   {\pr@tect\wlog{\string\chapter\space \chapterlabel}} }

\def\section#1{\par \ifnum\lastpenalty=30000\else
   \penalty-200\vskip\sectionskip \spacecheck\sectionminspace\fi
   \gl@bal\advance\sectionnumber by 1
   {\pr@tect
   \ifcn@@\expandafter\toks@\expandafter{\chapterlabel.}\else\toks@={}\fi
   \xdef\sectionlabel{\the\toks@\the\sectionstyle{\the\sectionnumber}}%
   \wlog{\string\section\space \sectionlabel}}%
   \noindent {\caps\enspace\sectionlabel.~~#1}\par
   \nobreak\vskip\headskip \penalty 30000 }
\def\subsection#1{\par
   \ifnum\the\lastpenalty=30000\else \penalty-100\smallskip \fi
   \noindent\undertext{#1}\enspace \vadjust{\penalty5000}}

\def\undertext#1{\vtop{\hbox{#1}\kern 1pt \hrule}}
\def\ACK{\par\penalty-100\medskip \spacecheck\sectionminspace
   \line{\fourteenrm\hfil ACKNOWLEDGEMENTS\hfil}\nobreak\vskip\headskip }

\def\APPENDIX#1#2{\par\penalty-300\vskip\chapterskip
   \spacecheck\chapterminspace \chapterreset \xdef\chapterlabel{#1}
   \titlestyle{APPENDIX #2} \nobreak\vskip\headskip \penalty 30000
   \wlog{\string\Appendix~\chapterlabel} }
\def\Appendix#1{\APPENDIX{#1}{#1}}
\def\appendix{\APPENDIX{A}{}}
%
%
%
%
%
\def\eqname#1{\rel@x {\pr@tect
  \ifnum\equanumber<0 \xdef#1{{(\number-\equanumber)}}%
     \gl@bal\advance\equanumber by -1
  \else \gl@bal\advance\equanumber by 1
   \ifcn@@ \toks@=\expandafter{\chapterlabel.}\else\toks@={}\fi
   \xdef#1{{(\the\toks@\number\equanumber)}}\fi #1}}

\def\eqn{\eqno\eqname}

\def\eqinsert#1{\noalign{\dimen@=\prevdepth \nointerlineskip
   \setbox0=\hbox to\displaywidth{\hfil #1}
   \vbox to 0pt{\kern 0.5\baselineskip\hbox{$\!\box0\!$}\vss}
   \prevdepth=\dimen@}}
%

%
%
\def\GENITEM#1;#2{\par \hangafter=0 \hangindent=#1
    \Textindent{$ #2 $}\ignorespaces}
\outer\def\newitem#1=#2;{\gdef#1{\GENITEM #2;}}

\newdimen\itemsize                \itemsize=30pt
\newitem\item=1\itemsize;
\newitem\sitem=1.75\itemsize;     
\newitem\ssitem=2.5\itemsize;     
\outer\def\newlist#1=#2&#3&#4;{\toks0={#2}\toks1={#3}%
   \count255=\escapechar \escapechar=-1
   \alloc@0\list\countdef\insc@unt\listcount     \listcount=0
   \edef#1{\par
      \countdef\listcount=\the\allocationnumber
      \advance\listcount by 1
      \hangafter=0 \hangindent=#4
      \Textindent{\the\toks0{\listcount}\the\toks1}}
   \expandafter\expandafter\expandafter
    \edef\c@t#1{begin}{\par
      \countdef\listcount=\the\allocationnumber \listcount=1
      \hangafter=0 \hangindent=#4
      \Textindent{\the\toks0{\listcount}\the\toks1}}
   \expandafter\expandafter\expandafter
    \edef\c@t#1{con}{\par \hangafter=0 \hangindent=#4 \noindent}
   \escapechar=\count255}
\def\c@t#1#2{\csname\string#1#2\endcsname}
\newlist\point=\Number&.&1.0\itemsize;
\newlist\subpoint=(\alphabetic&)&1.75\itemsize;
\newlist\subsubpoint=(\roman&)&2.5\itemsize;
%

%
%
%
%
\newcount\referencecount     \referencecount=0
\newcount\lastrefsbegincount \lastrefsbegincount=0
\newif\ifreferenceopen       \newwrite\referencewrite
\newdimen\refindent          \refindent=30pt
%
\def\normalrefmark#1{[#1]}
\let\PRrefmark=\attach
\def\NPrefmark#1{\step@ver{{\;[#1]}}}
\def\refmark#1{\rel@x\ifPhysRev\PRrefmark{#1}\else\normalrefmark{#1}\fi}
\def\refend@{\refmark{\number\referencecount}}
\def\refend{\refend@{}\space }
\def\refsend{\refmark{\count255=\referencecount
   \advance\count255 by-\lastrefsbegincount
   \ifcase\count255 \number\referencecount
   \or \number\lastrefsbegincount,\number\referencecount
   \else \number\lastrefsbegincount-\number\referencecount \fi}\space }
\def\REFNUM#1{\rel@x \gl@bal\advance\referencecount by 1
    \xdef#1{\the\referencecount }}
\def\Refnum#1{\REFNUM #1\refend@ } 
\def\REF#1{\REFNUM #1\R@FWRITE\ignorespaces}
\def\Ref#1{\Refnum #1\REFWRITE }
\def\ref{\Ref\?}
\def\REFS#1{\REFNUM #1\gl@bal\lastrefsbegincount=\referencecount
    \REFWRITE }

       \let\REFSCON=\REF
\def\r@fitem#1{\par \hangafter=0 \hangindent=\refindent \Textindent{#1}}
\def\refitem#1{\r@fitem{#1.}}
\def\NPrefitem#1{\r@fitem{[#1]}}
\def\NPrefs{\let\refmark=\NPrefmark \let\refitem=NPrefitem}
\def\REFWRITE{\R@FWRITE\rel@x }
\def\R@FWRITE#1{\ifreferenceopen \else \gl@bal\referenceopentrue
     \immediate\openout\referencewrite=\jobname.refs
     \toks@={\begingroup \refoutspecials \catcode`\^^M=10 }%
     \immediate\write\referencewrite{\the\toks@}\fi
    \immediate\write\referencewrite{\noexpand\refitem %
				    {\the\referencecount}}%
    \p@rse@ndwrite \referencewrite #1}
\begingroup
 \catcode`\^^M=\active \let^^M=\relax %
 \gdef\p@rse@ndwrite#1#2{\begingroup \catcode`\^^M=12 \newlinechar=`\^^M%
	 \chardef\rw@write=#1\sc@nlines#2}%
 \gdef\sc@nlines#1#2{\sc@n@line \g@rbage #2^^M\endsc@n \endgroup #1}%
 \gdef\sc@n@line#1^^M{\expandafter\toks@\expandafter{\deg@rbage #1}%
	 \immediate\write\rw@write{\the\toks@}%
	 \futurelet\n@xt \sc@ntest }%
\endgroup
\def\sc@ntest{\ifx\n@xt\endsc@n \let\n@xt=\rel@x
       \else \let\n@xt=\sc@n@notherline \fi \n@xt }
\def\sc@n@notherline{\sc@n@line \g@rbage }
\def\deg@rbage#1{}
\let\g@rbage=\relax    \let\endsc@n=\relax
\def\refout{\par\penalty-400\vskip\chapterskip
   \spacecheck\referenceminspace
   \ifreferenceopen \Closeout\referencewrite \referenceopenfalse \fi
   \line{\fourteenrm\hfil REFERENCES\hfil}\vskip\headskip
   \input \jobname.refs
   }
\def\refoutspecials{\sfcode`\.=1000 \interlinepenalty=1000
	 \rightskip=\z@ plus 1em minus \z@ }
\def\Closeout#1{\toks0={\par\endgroup}\immediate\write#1{\the\toks0}%
   \immediate\closeout#1}
%
%
\newcount\figurecount     \figurecount=0
\newcount\tablecount      \tablecount=0
\newif\iffigureopen       \newwrite\figurewrite
\newif\iftableopen        \newwrite\tablewrite
\def\FIGNUM#1{\rel@x \gl@bal\advance\figurecount by 1
    \xdef#1{\the\figurecount}}
\def\FIGURE#1{\FIGNUM #1\F@GWRITE\ignorespaces }
\let\FIG=\FIGURE

\def\figitem#1{\r@fitem{Fig. #1.}}  
\def\FIGWRITE{\F@GWRITE\rel@x }
\def\TABNUM#1{\rel@x \gl@bal\advance\tablecount by 1
    \xdef#1{\the\tablecount}}
\def\TABLE#1{\TABNUM #1\T@BWRITE\ignorespaces }

\def\tabitem#1{\r@fitem{#1:}}
\def\TABWRITE{\T@BWRITE\rel@x }
\def\F@GWRITE#1{\iffigureopen \else \gl@bal\figureopentrue
     \immediate\openout\figurewrite=\jobname.figs
     \toks@={\begingroup \catcode`\^^M=10 }%
     \immediate\write\figurewrite{\the\toks@}\fi
    \immediate\write\figurewrite{\noexpand\figitem %
				 {\the\figurecount}}%
    \p@rse@ndwrite \figurewrite #1}
\def\T@BWRITE#1{\iftableopen \else \gl@bal\tableopentrue
     \immediate\openout\tablewrite=\jobname.tabs
     \toks@={\begingroup \catcode`\^^M=10 }%
     \immediate\write\tablewrite{\the\toks@}\fi
    \immediate\write\tablewrite{\noexpand\tabitem %
				 {\the\tablecount}}%
    \p@rse@ndwrite \tablewrite #1}
\def\figout{\par\penalty-400
   \vskip\chapterskip\spacecheck\referenceminspace
   \iffigureopen \Closeout\figurewrite \figureopenfalse \fi
   \line{\fourteenrm\hfil FIGURE CAPTIONS\hfil}\vskip\headskip
   \input \jobname.figs
   }
\def\tabout{\par\penalty-400
   \vskip\chapterskip\spacecheck\referenceminspace
   \iftableopen \Closeout\tablewrite \tableopenfalse \fi
   \line{\fourteenrm\hfil TABLE CAPTIONS\hfil}\vskip\headskip
   \input \jobname.tabs
   }
%
%
%
\newbox\picturebox
\def\p@cht{\ht\picturebox }
\def\p@cwd{\wd\picturebox }
\def\p@cdp{\dp\picturebox }
\newdimen\xshift
\newdimen\yshift
\newdimen\captionwidth
\newskip\captionskip
\captionskip=15pt plus 5pt minus 3pt
\def\fullwidth{\captionwidth=\hsize }
\newtoks\Caption
\newif\ifcaptioned
\newif\ifselfcaptioned
\def\caption{\captionedtrue \Caption }
\newcount\linesabove
\newif\iffileexists
\newtoks\picfilename
\def\fil@#1 {\fileexiststrue \picfilename={#1}}
\def\file#1{\if=#1\let\n@xt=\fil@ \else \def\n@xt{\fil@ #1}\fi \n@xt }
\def\pl@t{\begingroup \pr@tect
    \setbox\picturebox=\hbox{}\fileexistsfalse
    \let\height=\p@cht \let\width=\p@cwd \let\depth=\p@cdp
    \xshift=\z@ \yshift=\z@ \captionwidth=\z@
    \Caption={}\captionedfalse
    \linesabove =0 \picturedefault }
\def\plot{\pl@t \selfcaptionedfalse }
\def\Picture#1{\gl@bal\advance\figurecount by 1
    \xdef#1{\the\figurecount}\pl@t \selfcaptionedtrue }

\def\s@vepicture{\iffileexists \parsefilename \redopicturebox \fi
   \ifdim\captionwidth>\z@ \else \captionwidth=\p@cwd \fi
   \xdef\lastpicture{\iffileexists
	\setbox0=\hbox{\raise\the\yshift \vbox{%
	      \moveright\the\xshift\hbox{\picturedefinition}}}%
	\else \setbox0=\hbox{}\fi
	 \ht0=\the\p@cht \wd0=\the\p@cwd \dp0=\the\p@cdp
	 \vbox{\hsize=\the\captionwidth \line{\hss\box0 \hss }%
	      \ifcaptioned \vskip\the\captionskip \noexpand\Tenpoint
		\ifselfcaptioned Figure~\the\figurecount.\enspace \fi
		\the\Caption \fi }}%
    \endgroup }
\let\endpicture=\s@vepicture
\def\savepicture#1{\s@vepicture \global\let#1=\lastpicture }
\def\displaypicture{\fullwidth \s@vepicture $$\lastpicture $${}}
\def\toppicture{\fullwidth \s@vepicture \topinsert
    \lastpicture \medskip \endinsert }
\def\midpicture{\fullwidth \s@vepicture \midinsert
    \lastpicture \endinsert }
%
%
\def\leftpicture{\pres@tpicture
    \dimen@i=\hsize \advance\dimen@i by -\dimen@ii
    \setbox\picturebox=\hbox to \hsize {\box0 \hss }%
    \wr@paround }
\def\rightpicture{\pres@tpicture
    \dimen@i=\z@
    \setbox\picturebox=\hbox to \hsize {\hss \box0 }%
    \wr@paround }
\def\pres@tpicture{\gl@bal\linesabove=\linesabove
    \s@vepicture \setbox\picturebox=\vbox{
	 \kern \linesabove\baselineskip \kern 0.3\baselineskip
	 \lastpicture \kern 0.3\baselineskip }%
    \dimen@=\p@cht \dimen@i=\dimen@
    \advance\dimen@i by \pagetotal
    \par \ifdim\dimen@i>\pagegoal \vfil\break \fi
    \dimen@ii=\hsize
    \advance\dimen@ii by -\parindent \advance\dimen@ii by -\p@cwd
    \setbox0=\vbox to\z@{\kern-\baselineskip \unvbox\picturebox \vss }}
\def\wr@paround{\Caption={}\count255=1
    \loop \ifnum \linesabove >0
	 \advance\linesabove by -1 \advance\count255 by 1
	 \advance\dimen@ by -\baselineskip
	 \expandafter\Caption \expandafter{\the\Caption \z@ \hsize }%
      \repeat
    \loop \ifdim \dimen@ >\z@
	 \advance\count255 by 1 \advance\dimen@ by -\baselineskip
	 \expandafter\Caption \expandafter{%
	     \the\Caption \dimen@i \dimen@ii }%
      \repeat
    \edef\n@xt{\parshape=\the\count255 \the\Caption \z@ \hsize }%
    \par\noindent \n@xt \strut \vadjust{\box\picturebox }}
\let\picturedefault=\relax
\let\parsefilename=\relax
\def\redopicturebox{\let\picturedefinition=\rel@x
   \errhelp=\disabledpictures
   \errmessage{This version of TeX cannot handle pictures.  Sorry.}}
\newhelp\disabledpictures
     {You will get a blank box in place of your picture.}
%
%
%
%
%
%
%
%
%
%
\def\FRONTPAGE{\ifvoid255\else\vfill\penalty-20000\fi
   \gl@bal\pagenumber=1     \gl@bal\chapternumber=0
   \gl@bal\equanumber=0     \gl@bal\sectionnumber=0
   \gl@bal\referencecount=0 \gl@bal\figurecount=0
   \gl@bal\tablecount=0     \gl@bal\frontpagetrue
   \gl@bal\lastf@@t=0       \gl@bal\footsymbolcount=0
   \gl@bal\cn@@false }

\def\papers{\papersize\headline=\paperheadline\footline=\paperfootline}
\def\papersize{\hsize=35pc \vsize=50pc \hoffset=0pc \voffset=1pc
   \advance\hoffset by\HOFFSET \advance\voffset by\VOFFSET
   \pagebottomfiller=0pc
   \skip\footins=\bigskipamount \normalspace }
\papers  
%
%
\newskip\lettertopskip       \lettertopskip=20pt plus 50pt
\newskip\letterbottomskip    \letterbottomskip=\z@ plus 100pt
\newskip\signatureskip       \signatureskip=40pt plus 3pt
\def\lettersize{\hsize=6.5in \vsize=8.5in \hoffset=0in \voffset=0.5in
   \advance\hoffset by\HOFFSET \advance\voffset by\VOFFSET
   \pagebottomfiller=\letterbottomskip
   \skip\footins=\smallskipamount \multiply\skip\footins by 3
   \singlespace }
\def\MEMO{\lettersize \headline=\letterheadline \footline={\hfil }%
   \let\rule=\memorule \FRONTPAGE \memohead }

\def\memodate{\afterassignment\MEMO \date }
\def\memit@m#1{\smallskip \hangafter=0 \hangindent=1in
    \Textindent{\caps #1}}
\def\subject{\memit@m{Subject:}}
\def\topic{\memit@m{Topic:}}
\def\from{\memit@m{From:}}
\def\to{\rel@x \ifmmode \rightarrow \else \memit@m{To:}\fi }
\def\memorule{\medskip\hrule height 1pt\bigskip}  
\def\memohead{\centerline{\fourteenrm MEMORANDUM}}
\newwrite\labelswrite
\newtoks\rw@toks
\def\letters{\lettersize
   \headline=\letterheadline \footline=\letterfootline
   \immediate\openout\labelswrite=\jobname.lab}

\let\letterhead=\rel@x
\def\addressee#1{\medskip\line{\hskip 0.75\hsize plus\z@ minus 0.25\hsize
			       \the\date \hfil }%
   \vskip \lettertopskip
   \ialign to\hsize{\strut ##\hfil\tabskip 0pt plus \hsize \crcr #1\crcr}
   \writelabel{#1}\medskip \noindent\hskip -\spaceskip \ignorespaces }
\def\rwl@begin#1\cr{\rw@toks={#1\crcr}\rel@x
   \immediate\write\labelswrite{\the\rw@toks}\futurelet\n@xt\rwl@next}
\def\rwl@next{\ifx\n@xt\rwl@end \let\n@xt=\rel@x
      \else \let\n@xt=\rwl@begin \fi \n@xt}
\let\rwl@end=\rel@x
\def\writelabel#1{\immediate\write\labelswrite{\noexpand\labelbegin}
     \rwl@begin #1\cr\rwl@end
     \immediate\write\labelswrite{\noexpand\labelend}}
\newtoks\FromAddress         \FromAddress={}
\newtoks\sendername          \sendername={}
\newbox\FromLabelBox
\newdimen\labelwidth          \labelwidth=6in
\def\makelabels{\afterassignment\Makelabels \sendersname=}
\def\Makelabels{\FRONTPAGE \letterinfo={\hfil } \MakeFromBox
     \immediate\closeout\labelswrite  \input \jobname.lab\vfil\eject}
\let\labelend=\rel@x
\def\labelbegin#1\labelend{\setbox0=\vbox{\ialign{##\hfil\cr #1\crcr}}
     \MakeALabel }
\def\MakeFromBox{\gl@bal\setbox\FromLabelBox=\vbox{\Tenpoint
     \ialign{##\hfil\cr \the\sendername \the\FromAddress \crcr }}}
\def\MakeALabel{\vskip 1pt \hbox{\vrule \vbox{
	\hsize=\labelwidth \hrule\bigskip
	\leftline{\hskip 1\parindent \copy\FromLabelBox}\bigskip
	\centerline{\hfil \box0 } \bigskip \hrule
	}\vrule } \vskip 1pt plus 1fil }
\def\signed#1{\par \nobreak \bigskip \dt@pfalse \begingroup
  \everycr={\noalign{\nobreak
	    \ifdt@p\vskip\signatureskip\gl@bal\dt@pfalse\fi }}%
  \tabskip=0.5\hsize plus \z@ minus 0.5\hsize
  \halign to\hsize {\strut ##\hfil\tabskip=\z@ plus 1fil minus \z@\crcr
	  \noalign{\gl@bal\dt@ptrue}#1\crcr }%
  \endgroup \bigskip }
\newbox\letterb@x
\def\lettertext{\par \vskip\parskip \unvcopy\letterb@x \par }
\def\multiletter{\setbox\letterb@x=\vbox\bgroup
      \everypar{\vrule height 1\baselineskip depth 0pt width 0pt }
      \singlespace \topskip=\baselineskip }
\def\letterend{\par\egroup}
%
%
%
\newskip\frontpageskip
\newtoks\Pubnum   
\newtoks\Pubtype  \let\pubtype=\Pubtype
\newif\ifp@bblock  \p@bblocktrue
\def\PH@SR@V{\doubl@true \baselineskip=24.1pt plus 0.2pt minus 0.1pt
	     \parskip= 3pt plus 2pt minus 1pt }
\def\PHYSREV{\papers\PhysRevtrue\PH@SR@V}

\def\titlepage{\FRONTPAGE\papers\ifPhysRev\PH@SR@V\fi
   \ifp@bblock\p@bblock \else\hrule height\z@ \rel@x \fi }
\def\nopubblock{\p@bblockfalse}
\def\endpage{\vfil\break}
\frontpageskip=12pt plus .5fil minus 2pt
\Pubtype={}
\Pubnum={}
\def\p@bblock{\begingroup \tabskip=\hsize minus \hsize
   \baselineskip=1.5\ht\strutbox \topspace-2\baselineskip
   \halign to\hsize{\strut ##\hfil\tabskip=0pt\crcr
       \the\Pubnum\crcr\the\date\crcr\the\pubtype\crcr}\endgroup}
\def\title#1{\vskip\frontpageskip \titlestyle{#1} \vskip\headskip }
\def\author#1{\vskip\frontpageskip\titlestyle{\twelvecp #1}\nobreak}

\def\address#1{\par\kern 5pt\titlestyle{\twelvepoint\it #1}}
\def\andaddress{\par\kern 5pt \centerline{\sl and} \address}

\def\abstract{\par\dimen@=\prevdepth \hrule height\z@ \prevdepth=\dimen@
   \vskip\frontpageskip\centerline{\fourteenrm ABSTRACT}\vskip\headskip }

%
%
%
       
\def\eg{\hbox{\it e.g.}}       
\def\etal{\hbox{\it et al.}}   
\def\\{\rel@x \ifmmode \backslash \else {\tt\char`\\}\fi }
\def\sequentialequations{\rel@x \if\equanumber<0 \else
  \gl@bal\equanumber=-\equanumber \gl@bal\advance\equanumber by -1 \fi }
%

%
\def\journal#1&#2&#3(#4){\begingroup \let\journal=\dummyj@urnal
    \unskip~\sl #1\unskip~\bf\ignorespaces #2\rm
    \unskip,~\ignorespaces #3
    (\afterassignment\j@ur \count255=#4)\endgroup\ignorespaces }
\def\j@ur{\ifnum\count255<100 \advance\count255 by 1900 \fi
	  \number\count255 }
\def\dummyj@urnal{%
    \toks@={Reference foul up: nested \journal macros}%
    \errhelp={Your forgot & or ( ) after the last \journal}%
    \errmessage{\the\toks@ }}

\def\topspace{\hrule height 0pt depth 0pt \vskip}

\def\Buildrel#1\under#2{\mathrel{\mathop{#2}\limits_{#1}}}
\def\becomes#1{\mathchoice{\becomes@\scriptstyle{#1}}
   {\becomes@\scriptstyle{#1}} {\becomes@\scriptscriptstyle{#1}}
   {\becomes@\scriptscriptstyle{#1}}}
\def\becomes@#1#2{\mathrel{\setbox0=\hbox{$\m@th #1{\,#2\,}$}%
	\mathop{\hbox to \wd0 {\rightarrowfill}}\limits_{#2}}}

\let\int=\intop         
\def\lsim{\mathrel{\mathpalette\@versim<}}
\def\gsim{\mathrel{\mathpalette\@versim>}}
\def\@versim#1#2{\vcenter{\offinterlineskip
	\ialign{$\m@th#1\hfil##\hfil$\crcr#2\crcr\sim\crcr } }}
\def\big#1{{\hbox{$\left#1\vbox to 0.85\b@gheight{}\right.\n@space$}}}
\def\Big#1{{\hbox{$\left#1\vbox to 1.15\b@gheight{}\right.\n@space$}}}
\def\bigg#1{{\hbox{$\left#1\vbox to 1.45\b@gheight{}\right.\n@space$}}}
\def\Bigg#1{{\hbox{$\left#1\vbox to 1.75\b@gheight{}\right.\n@space$}}}
\def\){\mskip 2mu\nobreak }
%
%
%
\let\sec@nt=\sec
\def\sec{\rel@x\ifmmode\let\n@xt=\sec@nt\else\let\n@xt\section\fi\n@xt}
\def\obsolete#1{\message{Macro \string #1 is obsolete.}}
\def\firstsec#1{\obsolete\firstsec \section{#1}}
\def\firstsubsec#1{\obsolete\firstsubsec \subsection{#1}}
\def\thispage#1{\obsolete\thispage \gl@bal\pagenumber=#1\frontpagefalse}
\def\thischapter#1{\obsolete\thischapter \gl@bal\chapternumber=#1}
\def\splitout{\obsolete\splitout\rel@x}
\def\prop{\obsolete\prop \propto }
\def\nextequation#1{\obsolete\nextequation \gl@bal\equanumber=#1
   \ifnum\the\equanumber>0 \gl@bal\advance\equanumber by 1 \fi}
\def\BOXITEM{\afterassigment\B@XITEM\setbox0=}
\def\B@XITEM{\par\hangindent\wd0 \noindent\box0 }
%
%
%
\def\phyzzx{PHY\setbox0=\hbox{Z}\copy0 \kern-0.5\wd0 \box0 X}
        
\everyjob{\xdef\today{\monthname~\number\day, \number\year}
}
%
%
%
\catcode`\@=12 
%
%
\HOFFSET=0.375truein

\input epsf.tex
\PHYSREV 
\nopubblock 
\titlepage 
%
%
\singlespace

%

\def\upt3{UPt$_3$} 
\def\ube13{UBe$_{13}$}

\def\cecu2si2{CeCu$_2$\-Si$_2$} 
\def\he3{$^3$He}
 
\def\dx2y2{$d_{x^2-y^2}$}

\title{\bf Experimental Constraints on the Pairing State of the Cuprate
Superconductors: an Emerging Consensus}\footnote{}{To appear in 
{\sl Physical Properties of High Temperature Superconductors}, Vol. 5,
D.M. Ginsberg (ed.), (World Scientific, Singapore, 1996).}

\author{James Annett$^1$, Nigel Goldenfeld$^2$ and Anthony J.
Leggett$^{2}$}

\address{$^1$H.H. Wills Physics Laboratory, University of Bristol,
Royal Fort, Tyndall Avenue, Bristol BS8 1TL, UK.}

\address{$^2$Department of Physics, University of Illinois at
Urbana-Champaign, 1110 West Green Street, Urbana, Il. 61801-3080, USA.
}


\vfil 
\abstract\noindent 
We present a critical discussion of recent experimental probes of the
pairing state of the high temperature superconductors, focusing
primarily, but not exclusively, on \Yba, where the best data currently
exist.  Penetration depth measurements near \Tc\ give no indication of
an extra transition, indicating that the pairing state is a
one-dimensional representation of the crystal symmetry.  Penetration
depth measurements at low temperatures provide strong evidence for a
change in sign of the gap function over the Fermi surface.  Quantum
mechanical phase interference experiments generally confirm this and in
addition show that the nodal positions are consistent with a
\dx2y2\ pairing state.  This pairing state is consistent with photoemission
measurements of the gap function, Raman scattering, the effect on
\Tc\ of impurities, and many other data (reviewed by two of us
previously) which indicate the presence of low lying excitations in the
superconducting state.  We also discuss evidence that apparently 
does not fit in with a \dx2y2\ pairing state, and we describe possible
alternative scenarios.

\bigskip\noindent 
Pacs Numbers: 74.72.-h

\endpage

{\bf \chapter{Introduction} }

The pairing state symmetry of the high temperature superconductors is a
topic that over the last 5 or 6 years has moved from the wings to
centre stage.  Although the star billing in the tragicomical history of
high temperature superconductivity must surely go to the mechanism for
pair formation, the identification of the pairing state may well be a
bit part that has stolen the show.  Indeed no other aspect of the high
temperature superconductors has ignited as much controversy; and some
of the most elegant and sophisticated experiments in condensed matter
physics have been performed to illuminate this issue.

In 1990, writing in Volume II of this series, Annett, Goldenfeld and
Renn (hereafter referred to as AGR)\REFS\agr{J.F. Annett, N. Goldenfeld
and S.R. Renn, in {\sl Physical Properties of High Temperature
Superconductors II}, D.M. Ginsberg (ed.) (World Scientific, New Jersey,
1990), Chapter 9, p. 571.}\REFSCON\ag{For a brief follow-up to AGR, see
J.F. Annett and N. Goldenfeld, \journal J. Low Temp. Phys. &89&197(92).}
\refsend\ presented a lengthy review of the experimentally determined
properties of the superconducting state of \Yba, in which they
described a variety of anomalous features which would be naturally
explained if the pairing state were an unconventional singlet state
with line nodes.  These features included the presence of low energy
excitations in NMR, Raman scattering, infra-red absorption and the
absence of a Hebel-Slichter coherence peak.  On the other hand, AGR
were prevented from concluding that a d-wave state was indeed
indicated, because of the apparent close agreement between the best
available measurements\Ref\fiory{A.T. Fiory, A.F. Hebard, P.M.
Mankiewich, and R.E. Howard, \journal Phys. Rev. Lett. &61&1419(88).}
of the temperature dependent penetration depth $\lambda (T)$ below 20 K
and the exponential form for this quantity
expected for a pairing state without nodes.  In contrast, a clean
unconventional singlet state with line nodes would have yielded a
linear dependence on temperature for the quantity measured.  Thus,
AGR's original article ended with the paragraph

{\narrower\smallskip\noindent\it In summary, given the presently
available data, it appears that triplet states are ruled out by the
Knight shift anisotropy, and singlet states with line nodes seem to be
ruled out by the temperature dependence of the penetration depth.  If
both of these tentative results stand up to further scrutiny, then the
only remaining candidate state would be the conventional one.
\smallskip}

The caution expressed by AGR has, in a sense, been vindicated by
developments, especially by two key sets of experimental
results.  First, the data
of Ref. \fiory\ were re-analysed\Ref\agrii{J.F. Annett, N.D. Goldenfeld
and S.R. Renn, \journal Phys. Rev. B., &43&2778(91).} and shown, in
fact, to be inconsistent with an exponential temperature dependence;
the actual dependence of the penetration depth on temperature was found
to be quadratic, known to be consistent with an unconventional singlet
state subject to impurity scattering.  
This finding was followed by a series of decisively
accurate and controlled experiments on very clean \Yba\ single crystals
by Hardy, Bonn and co-workers at the University of British Columbia
(see their Chapter in the present Volume for a detailed survey of these
developments)\rlap,\Ref\hardyi{W.N. Hardy, D.A. Bonn, D.C. Morgan, R. Liang,
and K. Zhang, \journal Phys. Rev. Lett. &70&3999(1993).} which
indicated a strong temperature dependence of the penetration depth
below 20 K, consistent with a linear behaviour, and clearly ruling out
the much weaker dependence expected for a nodeless superconductor.
Subsequent experimental and theoretical developments, discussed in
detail below, add support to the interpretation that the superfluid
density has a temperature dependence at low temperatures which is
inconsistent with the simplest s-wave pairing state models, but
consistent in a semi-quantitative manner with the generic predictions
of a \dx2y2\ pairing state.

The intense interest that these discoveries provoked resulted in a
flurry of experimental results which bore more or less directly on the
identification of the pairing state.  The second key set of
experimental results are a series of elegant measurements of phase
interference effects associated with the sign change of the gap
function at the Fermi surface, beginning with the paper (describing the
experiment referred to below as UIUC I) by
Wollman, Van Harlingen, Lee, Ginsberg and
Leggett\rlap.\Ref\wollmani{D.A. Wollman, D.J. Van Harlingen, W.C. Lee,
D.M. Ginsberg and A.J. Leggett, \journal Phys. Rev.  Lett.
&71&2134(93).}  As we discuss in detail below, the observation of these
interference effects can only be reconciled with s-wave pairing under
hypotheses that we regard as either already ruled out or extremely
implausible.

These key developments, augmented by the many other experimental
observations consistent with nodes in the energy gap,
such as photoemission determinations of the
gap function symmetry, have lead to what many, including the present
authors,  believe to be a compelling case for the \dx2y2\ pairing
state.  Nevertheless, it is only fair to point out that there are
still grounds for caution, based upon two experimental results which, if
taken at face value, do not conform in an obvious way to the d-wave
pairing picture.  These are the apparent observation of c-axis
tunneling currents between a conventional superconductor and a cuprate
superconductor\Ref\sun{A.G. Sun, D.A. Gajewski, M.B. Maple and R.C.
Dynes, \journal Phys. Rev. Lett. &72&2267(94).} and the observation of
a non-zero critical current across a closed grain-boundary junction
loop\rlap.\Ref\chaudhari{P. Chaudhari and S.-Y. Lin, \journal Phys.
Rev.  Lett. &72&1084(94).}

The purpose of this Chapter is to discuss these developments in some
depth, highlighting the logical connections between the experimental
findings and their theoretical interpretation.  Rather than attempting
a comprehensive review of all the experimental findings that may be
taken as evidence for or against d-wave pairing, we shall concentrate
on those experiments which, in our view, are most decisive. Also, we
shall not repeat discussion of experiments which were already reviewed
extensively in AGR, unless there have been major changes since 1991.

The layout of this Chapter is as follows. Firstly, in Section 2, we
shall very briefly  review the assumptions which underly
the classification of different pairing states in superconductors.
Then, in Section 3, we show how very general symmetry and 
thermodynamic considerations limit the possible
superconducting states, especially mixed symmetry states such as
$s+d$, $s+id$. In Section 4, we shall discuss the specific
symmetries involved in the high \Tc\ materials. 
In particular we discuss the symmetry implications of
their highly layered structures,  the existence of chains, 
and the orthorhombic distortions of the CuO$_2$ planes. 
In Section 5 we discuss the evidence for sign changes in
the superconducting gap function $\Delta({\bf k})$. This includes
penetration depth and photoemission experiments among others.
Phase sensitive
interference experiments  providing evidence for a macroscopic order
parameter with \dx2y2\ symmetry are discussed in Sections 6 and 7.
Section 6 discusses the general principles underlying this
type of experiment, while Section 7 discusses the specific
experiments which have been performed to date. In
Section 8, we briefly discuss the effects of impurities, since the
extreme sensitivity of unconventional superconductors to non-magnetic
impurity scattering is often cited as an objection to d-wave pairing in
the high \Tc\ materials. Finally, our conclusions present our synthesis
of the overall experimental situation as it currently stands.

{\bf \chapter{Basic Assumptions: Pairing and ODLRO}}

In discussing the symmetry of the superconducting state in the
high \Tc\ materials we want to make as few assumptions
as possible about the nature of the superconductivity or
of its mechanism.  At first sight this might appear to be
especially difficult in the high \Tc\ superconductors since
there is not even a universally agreed upon picture of the
normal state.  If the normal state were generally agreed to be
a Fermi liquid, as is the case in $^3$He and the heavy Fermion materials,
then one could develop the theory of possible superconducting
states arising from attractive interactions among the quasiparticles.
The instabilities of the Fermi liquid to Cooper pairing in various
channels would lead to a system of BCS-like gap equations,
whose solutions would be the various possible pairing states 
of the system.  The difficulty of applying this procedure
to the high \Tc\ superconductors is that it is by no means clear that
the normal state above \Tc\ is a Fermi liquid, or that 
any form of BCS-like gap equation is valid below \Tc.

In fact, it is not necessary to assume that the normal
state is a Fermi liquid in order to understand
the symmetries of possible superconducting states. 
The only assumption which it is necessary to make
is to assume that {\it pairing} and {\it off diagonal long range order} 
(ODLRO) occur in
the superconducting state.
In other words, one must assume that \Tc\ corresponds to
the temperature at which the following correlation function
first becomes non-zero:
$$
 \langle \psi^*_\alpha({\bf r}_1,t_1) \psi^*_\beta({\bf r}_2,t_2)
  \psi_\gamma({\bf r}_3,t_3) \psi_\delta({\bf r}_4,t_4) 
\rangle  \left \{ \eqalign{ & = 0 \ \ T > T_c \cr
                         & \neq 0 \ \ T < T_c \cr }  \right .
\eqn\odlro $$
where ${\bf r}_1$ and ${\bf r}_2$ are taken to be
macroscopically distant from  ${\bf r}_3$ and  ${\bf r}_4$. 
Here $\psi^*_\alpha (\bf{r},t)$ and $\psi_\alpha (\bf{r},t)$ 
are the usual electron creation and annihilation operators,
for an electron of spin $\alpha$ at point $\bf{r}$ and time $t$.
The existence of such ODLRO will ensure the rigidity
of the thermodynamic state under weak applied magnetic fields,
and hence follows flux quantisation, zero resistivity and 
the Meissner effect\rlap.\Ref\yang{C.N. Yang, \journal
Rev. Mod. Phys. &34&694(62).}\  Whether or not
the state above \Tc\ is a Fermi liquid is clearly irrelevant
in considerations solely of properties of the state
below \Tc.  Possibly if other kinds of long 
range order are also present above or below \Tc, then these may 
alter the physics, but this is not the case for the majority of the
non-Fermi liquid normal states proposed to date, such as a gas of
pre-formed pairs\rlap,\Ref\alexandrov{A.S. Alexandrov and N.F. Mott,
\journal Int. J. Mod. Phys. B &8&2075(94).}\
 or the short ranged RVB spin liquid\rlap.\REFS\rvb{
P.W. Anderson, \journal Science &235&1196(87).}\REFSCON\kivelson{
S. Kivelson, D. Rokshar and J. Sethna, \journal Phys. Rev. B
 &35&8865(87).}\refsend

Of course ODLRO could in principle also occur in other forms besides \odlro.
For example, a charge $e$ condensate is not ruled out {\it a priori}\rlap.\Ref\zou{
Z. Zou and P.W. Anderson, \journal Phys. Rev. B &37&627(88).}
However, 
for reasons discussed extensively in AGR\refmark{\agr}
we believe there is overwhelming 
 evidence for a conventional $2e$ ODLRO in the cuprates.
As well as the value of the flux quantum,
the evidence includes
the existence of ac and dc Josephson effects in cuprate-cuprate
junctions and in junctions with ordinary superconductors such as 
Pb and Nb, and the observation of Andreev reflection. Therefore
from now on we shall take it as given that the onset
of superconductivity at \Tc\ corresponds to the formation
of a condensate of electron pairs.  At this point we
do not need to make any assumptions about the nature of the
state above \Tc; for example it could equally well
be a Fermi liquid, a marginal Fermi liquid\rlap,\Ref\littlewood{
P.B. Littlewood, C.M. Varma and E. Abrahams, \journal Phys. Rev. B
&46&405(92).}
a Luttinger liquid\Ref\chakravarty{
S. Chakravary, A. Sudbo, P.W. Anderson and S. Strong, 
\journal Science &261&337(93).}\ or
a gas of bound pairs
which are not Bose condensed\rlap.\refmark{\alexandrov}

The existence of a condensate of electron pairs below \Tc\ also
implies that
the order parameter for
superconductivity corresponds to the appearance (i.e. non-zero value)
of  Gor'kov
type off-diagonal Green functions 
$$  {\cal F}({\bf r}_1,t_1,{\bf r}_2,t_2; \alpha\beta) =
     \langle \psi_\alpha({\bf r}_1,t_1) \psi_\beta({\bf r}_2,t_2) 
\rangle ,  \eqn\gorkov
$$ 
in an appropriate grand canonical ensemble. We shall assume
(apart from one exception discussed below)  that
there exists a Ginzburg-Landau order parameter $\Psi({\bf r}_1,{\bf r}_2;\alpha,\beta)$
identified with the equal time,  $t_1=t_2$, limit of
\gorkov\ (the zero-frequency limit is essentially
equivalent).
The observation of three dimensional critical scaling in the XY
universality class\Ref\kamal{S. Kamal \etal,
\journal Phys. Rev. Lett. &73&1845(94).} and the
existence of a $\lambda$-transition-like
specific heat anomaly near \Tc\ which is consistent with that of
the $^4$He
universality class\Ref\howson{N. Overend, M.A. Howson, and I.D. Laurie,
\journal Phys. Rev. Lett. &72&3238(94).} is evidence
for the existence of a (two component) 
Ginzburg-Landau  order parameter, at least in the 
scaling limit as $T \rightarrow T_c$.  

As soon as the Green function
${\cal F}({\bf r}_1,t_1,{\bf r}_2,t_2; \alpha\beta) $ 
is non-zero, the electron self-energy must also develop 
anomalous, or  off-diagonal,
terms, corresponding to the BCS gap function $\Delta({\bf r}_1,t_1,{\bf r}_2,t_2; \alpha\beta) $.  If the normal state is a Fermi
liquid, then $\Delta$ determines the gap in the quasiparticle 
spectrum in the usual way. However, if the normal state is not a Fermi 
liquid, then there may be no direct correspondence between
$\Delta$ and the quasiparticle spectrum. For example in the
pre-formed pairs scenario the quasiparticle spectrum has become gapped
even above \Tc\rlap.\refmark{\alexandrov}
There is also
no reason to suppose that $\Delta$ obeys a mean field equation
such as the BCS gap equation. 

By the {\it symmetry of the pairing state} we mean
the transformations of the function ${\cal F}({\bf r}_1,t_1,{\bf r}_2,t_2; \alpha\beta) $ 
under the various symmetry operations of the crystal and spin symmetry groups.
Clearly these symmetries can be defined equally well whether or not the
cuprates have a Fermi liquid normal state.
Many of the experimental tests of the pairing state symmetry which we discuss
below (such as the Josephson experiments) are also completely
independent of Fermi liquid assumptions. 
Other tests, such as the evidence for a node in the gap function
from penetration depth data, do have to make some Fermi liquid-like 
assumptions. In particular it is necessary to assume the existence
of a Fermi surface.  In practical terms a ``pseudo'' Fermi surface
should be sufficient, namely a surface 
in momentum space
where the electron distribution function, $n({\bf k})$,
has a cusp, rather than a full Fermi liquid discontinuity\rlap.\refmark{
\littlewood,\chakravarty}\
The positron annihilation and angle resolved 
photoemission experiments\Ref\positron{J.C. Campuzano \etal,
\journal Phys. Rev. B &43&2788(91).}\
both show that  such a Fermi surface exists (at least
in the ab-plane) of the cuprates, whether or not the quasiparticle lifetimes
are Fermi liquid, marginal or Luttinger.

With regard to general symmetry properties, almost all of the non-Fermi liquid
models of the cuprates will behave similarly.
One of the few exceptions to this rule is the anyon theory
of superconductivity\rlap.\Ref\rokhsar{D.S. Rokhsar,
\journal Phys. Rev. Lett &70&493(93).}\
This is because, unlike the marginal or Luttinger
liquid models, this state has a symmetry which is different
from a normal Fermi liquid: broken time reversal symmetry, T.
However, even broken T symmetry in the normal state
does not significantly change the symmetry analysis
outlined in the next two Sections. In particular
the considerations of Section 3 imply that two separate phase
transitions are to be expected in the proposed
$\hbox{\dx2y2}+id_{xy}$ anyon superconducting state, whether or not
T is broken in the normal state.  In any case, to date 
there appears to be no strong signature of a broken T state
in \Yba\rlap.\Ref\laughlin{T.W. Lawrence, A. Szoke, and R.B. Laughlin,
\journal Phys. Rev. Lett. &69&1439(92).}

{\bf \chapter{Mixing of Different Symmetries:  Thermodynamic
Constraints}}

In the previous Section we introduced the general idea of a superconducting
order parameter:
$$ \Psi(\bf{r}_1, \bf{r}_2 ; \alpha \beta) \equiv \langle
\psi_{\alpha}(\bf{r}_1) \psi_{\beta} (\bf{r}_2)\rangle .
 \eqn\pairing$$ 
In this
Section we will discuss to what extent thermodynamic or related
observations enable us to infer constraints on the behaviour  of the
expression \pairing\ under the (exact or approximate) symmetry operations
of the crystal Hamiltonian. In particular we will emphasize
the conditions under which the pairing state belongs to a single
irreducible representation of the symmetry group, and the conditions
under which it becomes a ``mixed'' state involving two or more representations.
As we show below,  simple thermodynamic considerations place
quite strong restrictions on the conditions for such mixed states to occur.
It should be emphasised that the
considerations of this Section are very general and do not in
themselves exclude e.g. that the equilibrium $\Psi$ has nontrivial
transformation properties under the crystal translation group, i.e.
that the translational symmetry is spontaneously broken in the
superconducting state, although in the rest of this Chapter we should
usually assume that it is not.

Consider first the group $G$ of {\it exact} symmetries of the
Hamiltonian for a given crystal.  This is the direct product of the
gauge group $U(1)$, the crystal lattice translation group $T_\ell$ 
and the crystal point
group $H$:  
$$ G=U(1){\otimes}T_\ell{\otimes}H \eqn\group $$ 
As
remarked above, we will normally assume that $\Psi$ transforms as the
identity representation of $T_\ell$, in which case we can simply take $G$ to be
given by $U(1)\otimes H$; however, we may as well keep \group\ for
generality.  We now expand $\Psi$ in the irreducible representations
$\chi(\bf{r}_{1}\bf{r}_{2};\alpha\beta)$ of the group G: 
 $$
\Psi({\bf r}_1, {\bf r}_2; \alpha \beta)=\sum_{l}\sum_{m=0}^{D_l}\psi_{lm}
\chi_{lm}({\bf r}_1, {\bf r}_2: \alpha \beta) \eqn\reps 
$$ 
where $D_{l}$
is the dimension of the $l$-th irreducible representation and the
$\psi_{lm}$ are coefficients which are in general complex.  Note that it
is not assumed, anywhere in this argument, that $\psi$ is constant as a
function of the centre of mass variable $\bf{R} \equiv 
(\bf{r}_{1}+\bf{r}_{2})/2$.

We now proceed, in the spirit of Ginzburg and Landau, to express the
free energy $F$ as a multiple power series in the coefficients
$\psi_{lm}$, using the principle that F must be invariant under all
operations of G (and moreover must be real).  It is immediately clear
that invariance under U(1) implies that all terms containing an odd
number of $\psi_{lm}$'s vanish identically, while the rest must contain
an even number of $\psi_{lm}$'s and $\psi^{*}_{lm}$'s.  Moreover, the
second-order terms must have the form $\sum _{l} \alpha_{l}(T) \sum_{m}
| \psi_{lm} |^{2}$, where the $\alpha_{l}(T)$ are functions of $T$
which, barring pathology, will be different for different { \it l}.  At
fourth order the generic term is
$$ \frac{1}{2}\sum \beta_{l_{1}m_{1} l_2 m_{2}
l_3 m_3 l_4 m_{4}} \psi^{*}_{l_{1}m_{1}} \psi^{*}_{l_{2}m_{2}}
\psi_{l_{3}m_{3}} \psi_{l_{4}m_{4}}.  \eqn\mixing $$
For the tetragonal and orthorhombic symmetry groups relevant to
the high \Tc\ superconductors the relevant representations and
Ginzburg-Landau expansions up to fourth order are well 
known\rlap.\Ref\annett{ J.F. Annett, \journal Adv. Phys. &39&83(90).}\
However, such symmetry analyses usually assume that only 
a single irreducible representation, say $l$, is relevant.
Instead, let us examine terms in the Ginzburg-Landau expansion
which couple two or more representations.

The quartic terms can be grouped, for reasons which will become apparent,
 into what we shall call ``mixing'' and ``non-mixing'' terms.
The non-mixing terms are those in which for each representation
 the $\psi_{lm}$'s
enter to even order, and mixing terms are the rest. 
The non-mixing terms are thus the terms such as
$|\psi_{l_1m_1}|^2
|\psi_{l_{2}m_{2}}|^2$ and
($\psi^{*}_{l_1m_1})^{2}(\psi_{l_{2}m_{2}})^2$ + c.c., while the
mixing terms are of the form
$ \psi^*_{l_1m_1}\psi^*_{l_2m_2}\psi_{l_1m_1}^2 $.
The usefulness of this separation into mixing and non-mixing terms
lies in the fact that in all orthorhombic and
tetragonal superconductors, such as the
cuprates,  simple symmetry arguments imply that the mixing terms
are absent\rlap.\refmark{\annett} \ 
Thus the general form of the Ginzburg-Landau expansion
up to fourth order is:
$$ \eqalign{
F(T) = & \sum_{lm}\left ( \alpha_l(T) | \psi_{lm} |^2
 + {1 \over 2} \beta_{l,m_1,m_2,m_3,m_4} \psi^*_{lm_1}  
\psi_{lm_2} \psi_{lm_3}  \psi_{lm_4}  \right ) \cr
&    +  \sum_{l_1,l_2,m_1,m_2,m_3,m_4} 
  {1 \over 2}   \beta_{l_1l_2 m_1m_2m_3m_4}
\psi^*_{l_1m_1} \psi_{l_1m_2} \psi^*_{l_{2}m_{3}}\psi_{l_{2}m_{4}}
  \cr
&    +  \sum_{l_1,l_2,m_1,m_2,m_3,m_4}
 {1 \over 2}   \kappa_{l_1l_2 m_1m_2m_3m_4}
\psi^*_{l_1m_1} \psi^*_{l_1m_2} \psi_{l_{2}m_{3}}\psi_{l_{2}m_{4}}  ,
} \eqn\gleqn $$
plus terms involving three or four distinct representations, $l$.
For the purposes of the present argument, we may
neglect the sixth- and higher-order terms, which do not affect the
results qualitatively. In fact there are no mixing terms to {\it any}
order\refmark{\annett} unless three or four distinct 
representations are mixed.
General results from non-truncated
Ginzburg-Landau expansions have also been developed by 
Gufan \etal\Ref\gufan{Yu. 
M. Gufan \etal, \journal Phys. Rev B &51&9129(95); and
\journal ibid. &51&9228(95).} for
p- and d-wave superconductors. 

At a sufficiently high temperature all the $\alpha_{l}(T)$ are positive,
and minimisation of the free energy is achieved by setting all
$\psi_{lm}$ equal to zero, i.e.  the system is in the normal phase.  As
the temperature falls, there comes a point, $T^{l_{0}}_{c}$, where
one $\alpha_{l}$ corresponding (e.g.) to $l=l_{0}$, becomes
negative while (in the absence of pathological coincidence) all other
$\alpha_{l}$ remain positive.  We first consider the case (actually not
very likely to be relevant to the high temperature superconductors) where the irreducible
representation $l_{0}$ is multidimensional.  In this case, for $T$ just
below $T^{l_0}_{c}$, some or all of the corresponding $\psi_{l_{0}m}$
will be nonzero; the weights with which the various $m$ are represented
will be controlled primarily by the fourth-order terms, and may in
principle depend on $T$ through the $\beta$'s (or through the omitted
higher-order terms).  Thus it is possible that the ``configuration''
(i.e. the relative weight of the various $m$) undergoes either a
continuous or a discontinuous change below $T^{l_0}_{c}$; an example of
the latter is the A-B transition in superfluid $^{3}$He.

We turn to the case, probably more relevant to the high temperature superconductors, that the irreducible representation
$l_{0}$ is one-dimensional.  The crucial question is:  What is the
condition that below $T^{l_0}_{c}$ some $\psi_{lm}$ corresponding to
{\it other\/} values of $l$ are nonzero?  It is clear that if the
mixing terms are nonzero this can happen in a continuous way (even if
all other $\alpha_{l}$ are positive for all $T$).  But let us consider
the case that the mixing terms vanish, corresponding to the high \Tc\ 
superconductors. For simplicity of notation, we
shall specialise to the case where there is only one relevant irreducible representation
besides $l_{0}$ and it is moreover also one-dimensional (the
generalisations are straightforward).  In this case we note that we can always
minimise the free energy by choosing ${\rm arg}(\psi_{l_0m_0}\psi_{lm})$ 
to be either $0$ or
$\pi$; having done this, and changing the notation for convenience
($l_{0}\rightarrow 1$, etc.), we can write 
$$ \eqalign{ F(T) = &
F_{0}(T) + \alpha_{1}(T)|\psi_{1}|^{2}+
\alpha_{2}(T)|\psi_{2}|^{2}+ \cr
 & \frac{1}{2}\beta_{1}(T)|\psi_{1}|^{4}+
\frac{1}{2}\beta_{2}(T)|\psi_{2}|^{4}+ \kappa
(T)|\psi_{1}|^{2}\cdot|\psi_{2}|^2. } \eqn\glenergy $$ 
For
purposes of illustration we shall choose the temperature dependences of
the coefficients to have the simple Ginzburg-Landau form $$ \eqalign{
\alpha_{1}(T)=  & \ \alpha_{1}(T-T_{c_{1}}), \cr \alpha_{2}(T)= &
\ \alpha_{2}(T-T_{c_{2}}),  \cr T_{c_{2}} \leq & \ T_{c_{1}} \ \  ( {\rm
and\  possibly}\  T_{c_{2}} <0), \cr \beta_{1}(T)= \ & b_{1},\cr
\beta_{2}(T)= \ & b_{2},\cr \kappa(T)= \ & \kappa \cr } \eqn\GLform $$
where $b_{1}$, $b_{2}$, and $\kappa$ are constants, and where stability
requires\ref{In BCS-type theories these conditions are automatically
fulfilled.} that $\beta_{1},\beta_{2}>0$, $\kappa>-
\sqrt{\beta_{1}\beta_{2}}$.  The qualitative results
are independent of this ansatz.

The phase diagram of a system with a free energy of the form \GLform\ is
discussed in detail by Imry\rlap.\Ref\imry{ Y. Imry, \journal J. Phys.
C. &8&567(75).}\ For our purposes it is sufficient to know the following:
For $T$ above the ``upper'' transition $T_{c_{1}}$, it is clear that
$\psi_{1}=\psi_{2}=0$ (normal phase).  For $T$ just below $T_{c_{1}}$,
the free energy is minimised by the choice $$
\psi_{1}=(\alpha_{1}/\beta_{1})(T_{c_{1}}-T) ^{{1}/{2}},\qquad
\psi_{2}=0.  \eqn\fmin $$ At lower temperatures there are various
possibilities, depending on the ratios of the parameters\rlap.\refmark{\imry}  However, it is clear that if $\psi_{2}$ is ever to obtain a
finite value, one of two things must happen:  either it must jump
discontinuously from zero to this value, which evidently corresponds to a
first-order phase transition with an actual discontinuity in the value
of various physical quantities, or there must be a second second-order
phase transition at a temperature $T^*$ given by $$ \eqalign{ T^{*}-
T_{c2}= & \ \lambda(T^{*}-T_{c1}) \cr \lambda= & \  \kappa
\alpha_{1}/\alpha_{2}\beta_{1} \cr } \eqn\tstar $$ This second
possibility requires $\lambda \leq T_{c2}/T_{c1}$.  In this case it is
straightforward to show that the (positive) specific-heat
discontinuity  at $T^*$ is given by the expression $$ \Delta
c^{*}_{v} = (a^{2}_{2}/b_{2})
\frac{(1-\lambda)^2}{1-\kappa^{2}/b_{1}b_{2}} \geq (a^{2}_{2}/b_{2})
\frac{[(T_{c1}-T_{c2})/T_{c1}]^{2}}{1-\kappa^{2}/b_{1}b_{2}}
\eqn\tstartwo $$ Thus, assuming that the ratio
$(a^{2}_{2}/b_{2})/(a^{2}_{1}/b_{1})$ is not pathologically
small, the anomaly at $T^*$ is comparable to that at $T_{c}$ ($\equiv
T_{c1}$) unless $T_{c2}$ is extremely close to $T_{c1}$ (and quite
likely even then).  A second quantity of interest is the change in
slope at $T^{*}$ of the ``total'' order parameter $| \psi |^{2}
\equiv | \psi_{1}|^{2}$ + $| \psi_{2} |^{2}$; crudely
speaking, many physical quantities such as the mean-square energy
gap are likely to be roughly proportional to this.  We find for the
{\it relative} change in slope $\delta^{*}$ at $T^{*}$ the
expression $$ \delta^{*} = \frac {(1-\lambda)(1-\lambda a_{2} /
a_{1})(a_{2}b_{1}/a_{1}b_{2})} {1-\kappa^{2}/b_{1}b_{2}} \eqn\deltastar
$$ In simple BCS-type theories it turns out that $a_{j} \propto 
T^{-1}_{cj}$, and thus (in view of the constraint on $\lambda$) the
quantity $\delta^{*}$ is positive in such theories and of order one
except possibly for $T_{c2}$ very close to $T_{c1}$.

It is clear that the above results, derived for the explicit form \glenergy\
of the free energy, should generalise qualitatively to more
realistic ``non-mixing'' forms, with the sole caveat that if the
temperature $T^{*}$ is very low compared to $T_c$ the anomalies $\delta
c_{v}^{*}$ and $\delta^*$ are likely to be correspondingly reduced.
With this caveat, therefore, we draw the following very important
conclusion:  If the order parameter is a superposition of two
functions from different irreducible representations of the crystal
symmetry group and symmetry considerations forbid ``mixing'' of these
functions in the free energy (as in the cuprates), then {\it  there 
must inevitably be a second
phase transition at some temperature below $T_c$}.  Unless the
relevant irreducible representations have transition temperatures which are very close, both
the entropy and other physical quantities will undergo either
discontinuities or substantial changes in slope.  The almost complete
absence (see Section 5.1), to date, of any suggestion of such phenomena in the cuprate
superconductors is a very strong argument against this  type of
superposition. 

For completeness we should briefly discuss what happens when, owing
perhaps to some small breaking of the symmetry (such as may be constituted,
for some at least of the high temperature superconductors, by the orthorhombic crystalline
anisotropy) the mixing terms are small but not zero.  We first note
that any ``quadratic'' mixing, that is, any term in the free energy of
the form $$ \gamma (\psi_1\psi^{*}_{2} + c.c.) \eqn\mixer $$ can always
be eliminated by a re-diagonalisation of the quadratic terms; however,
the quartic terms will then contain terms of the form $\psi^{3}_{1}
\psi_{2}$ and $\psi^{3}_{2}\psi_{1}$, whose coefficients will be
proportional to $\gamma/ | \alpha_{1}-\alpha_{2} |$ when this is
small.  We note that in most cases these coefficients will not be
strongly $T$-dependent; in particular, in BCS theory, because of the
special form of the $\alpha_{j}(T)$, $| \alpha_{1}-\alpha_{2}|$
is approximately $T(T_{c2}^{-1}-T_{c1}^{-1}$).  There may in
addition be ``direct'' fourth-order mixing terms.  In the following,
let $T^*$ be the temperature at which a second second-order phase
transition would have occurred in the absence of a mixing
term\rlap.\Ref\noteit{If the original second transition were first order, it
would be little affected by a small amount of mixing.}\  For
small mixing, the term in $\psi_{1} \psi^{3}_{2}$ will have little
effect for $T\geq T^*$, so we shall neglect it.  Suppose now that the
coefficient of $\psi_{2}\psi^{3}_{1}$ is of order of magnitude $\xi$.
Then for $T<T_{c1}$ and not too close to $T^{*}$, the effect of the
mixing term is that $\psi _{2}$ is non-zero, increasing as
$T(T_{c1}-T)^{3/2}$:  $$ \psi_{2} \sim \xi \psi_{1}^{3}|
\alpha_{2}^{\prime} (T) \sim
\xi(\alpha_{1}/\beta_{1})^{3/2}(T_{c1}-T)^{3/2}/\alpha^{\prime}_{2} (T),
\eqn\mixertwo $$ 
where $\alpha_{2}^{\prime} (T) \equiv
\alpha_{2}(T)(1-\lambda$).

Well below $T^{*}$, on the other hand, the mixing will have only a
small effect and $\psi_{2}$ will increase approximately as
$[-(\alpha^{\prime}_{2}(T)/\beta_{2})\cdot(1-\kappa^{2}/\beta_{1}\beta
_{2})^{-1}]^{1/2}$.  We may obtain an estimate of the order of
magnitude $\Delta T$ of the crossover region by equating the value of
the first expression at $T^*+\Delta T$ to that of the second
at $T^*-\Delta T$; because $\alpha ^{\prime}_{2}\equiv
\alpha_{2}(1-\lambda)(T-T^{*}$), this gives $$ \Delta
T=\left(\frac{\xi}{\beta_{2}}\right)^{2/3}
\left(\frac{\alpha_{1}\beta_{2}}{\alpha_{2}\beta_{1}}\right)
\left(\frac{T_{c1}-T^{*}}{1-\lambda}\right)\left(1-\kappa^{2}/
\beta_{1}\beta_{2}\right)^{1/3}
\eqn\belowtc $$ Thus for $\xi/\beta_{2}$ non-zero but $\ll 1$ (and
$\lambda$ not too close to 1, etc.) the vestige of the second
second-order transition persists in the form of a sharp kink in the
thermodynamic properties near $T^*$, as we should expect intuitively;
the absence of observation of such behaviour may be used to put limits
on $\xi$ in any specific case of interest.

Finally, what about the quartic terms in the Ginzburg-Landau
expansion which involve three or four distinct representations?
In  this case the ``mixing" terms need not vanish.
 For example, mixing terms such as
$\psi^*_{l_1m_1}\psi^*_{l_2m_2}\psi_{l_3m_3}\psi_{l_4m_4}$
are symmetry-allowed in a tetragonal crystal
with
$l_1=A_{1g}$
$l_2=A_{2g}$, $l_3=B_{1g}$, $l_4=B_{2g}$ (in the notation of
Ref. \annett). 
However, mixing terms of this kind 
do not lead to a continuous 
mixing of representations of the ground state, and thus 
do not change the qualitative observations given above, regardless of how
many $\alpha_l$'s go negative.
Again a
second phase transition below $T^{l_0}_c$ is required
if a mixed symmetry state is to exist at low temperatures.

{\bf\chapter{Classification of Pairing States in the Cuprates}}

The basic principles of the symmetry classification of superconducting
states are well known\rlap,\REFS\volovik{ G.E. Volovik and L.P. Gor'kov,
\journal Zh. Eksp. Theor. Fiz.  &88&1412(85) 
[Sov. Phys. JETP  {\bf 61}, 843 (1985)].
}\REFSCON\ueda{ K. Ueda and T.M. Rice, \journal Phys.
Rev. B &31&7114(85).}\REFSCON\ozaki{ M. Ozaki, K. Machida and T. Ohmi,
\journal Prog. Theor. Phys. &74&221(85).}\REFSCON\blount{E.I. Blount,
\journal Phys. Rev. B &32&2935(85).}\REFSCON\sigrist{M. Sigrist and
K. Ueda, \journal Rev. Mod. Phys. &63&239(91).}\refsend 
and have been reviewed for the high \Tc\ symmetry groups
in Ref. \annett.  However, there are several issues which 
potentially complicate the situation in the high \Tc\ materials.
We have already discussed the possibility that the normal
state is not a Fermi liquid. Another complication
is the strongly two-dimensional nature of the crystals,
with possibly no coherent single electron motion between planes.
Further complications are the existence of chains and other
orthorhombic distortions, and the bilayers and trilayers
present in these materials. 

{}From now on we shall restrict our attention to singlet pairing states since,
as discussed in AGR\rlap,\refmark{\agr} the Knight shift
experiments in \Yba\ imply that the superconducting state has no
unpaired spins at $T=0$.  More recent Knight shift experiments
\REFS\halperin{
W.A. Halperin \etal, \journal Phys. Rev. Lett. &70&3131(93).}\REFSCON 
\kitaoka{Y. Kitaoka \etal,
\journal Physica C &179&107(91); K. Fujiwara \etal,
\journal Physica C &184&207(91); O.M. Vyaslev \etal,
\journal Physica C &199&50(92).}\REFSCON\zheng{G.-q. Zheng \etal, 
\journal Physica B &186&1012(93).}\REFSCON\han{Z.P. Han \etal, 
\journal Physica C &226&106(94).}\
on 
La$_{2-x}$Sr$_x$CuO$_4$ and on the  one-, two- and three-layer
systems,
 Tl$_2$Ba$_2$CuO$_{6+y}$, TlSr$_2$Ca$_2$Cu$_2$O$_{7-\delta}$,  and 
Tl$_2$Ba$_2$Ca$_2$Cu$_3$O$_{10-\delta}$,
all indicate that the spin susceptibility vanishes at
zero temperature (except possibly for a small impurity induced
contribution)  and hence that these are all singlet superconductors.

At first
sight, singlet pairing would imply, by the Pauli principle, 
that the orbital parity
must be even.  However, at this point we should note a possible loophole in
this argument: as pointed out originally by 
Berezinskii\Ref\berez{V.L. Berezinskii, \journal 
ZhETP Pisma &20&628(74) [\journal JETP Lett &20&287(74)].}\
in the context
of superfluid $^3$He, and explored more recently by 
Balatsky and Abrahams\Ref\balatsky{A.V. Balatsky and E.
Abrahams, \journal Phys. Rev. B &45&13125(92).}\ in the
context of heavy-fermion and other superconductors,
states can exist which are odd in
{\it frequency}, so that the Gor'kov Green function
$ {\cal F}({\bf r}_1,t_1,{\bf r}_2,t_2; \alpha\beta) $
is an odd function of $t_1-t_2$. This precludes having a non-zero
pairing amplitude at equal times, $t_1=t_2$, or at zero frequency
and hence the order parameter cannot be defined as in \pairing.
For these odd frequency states, spin singlet superconductivity 
corresponds to odd parity pair pairing,
while triplet pairing states will have even parity, opposite to
the usual situation. Furthermore since the gap function,
$\Delta({\bf k},\omega)$ (or for triplets ${\bf d}({\bf k},\omega)$)
must vanish at $\omega=0$, these superconductors have gapless
excitations at every point on the Fermi surface.
This does not appear to be compatible with the experimental
situation in the cuprates where, although there are gapless excitations
(as evidenced by the temperature dependence of the
penetration depth) much of the Fermi surface appears to
have a well developed energy gap (for example as seen in the
photoemission experiments discussed below).
We shall therefore not consider these odd frequency
states further in this Chapter.

{\bf \section{Possible Pairing States in a Plane Square Lattice}}

        Because of the highly layered nature of the cuprates,
 we first consider the possible pairing states
of the electrons confined to a single CuO$_2$ plane with a square lattice
structure.  Thus we assume until further notice that in Eqn.
\pairing\ ${\bf r}_1$ and ${\bf r}_2$ lie in the same CuO$_2$ plane.

        For the case of a single square CuO$_2$ plane the relevant crystal
point group is simply the group of the square, $C_{4v}$.  The symmetry
operations are:
(a) rotation through $\pi/2$ about the $\left<001\right>$ ($z$) axis, $\hat{R}_{\pi/2}$
 (b) reflection in a  $\langle 100 \rangle$ ($x/y$) crystal 
 axis, $\hat{I}_{axis}$, and 
(c) reflection in a 45$^{\circ}$ or $\langle 110 \rangle $ axis, $\hat{I}_{\pi/4}$. 
Since for an even-parity state the only allowed eigenvalue of
$\hat{R}^2_{\pi/2} $  is $+1$, and quite generally the only
 allowed eigenvalue of $\hat{I}^2_{axis}$ is +1, it immediately follows
 that the even-parity
irreducible representation of the group C$_{4v}$ can be uniquely labeled by the
possible eigenvalues $\pm 1$, of $\hat{R}_{\pi/2} $
and $\hat{I}_{axis}$ ($\hat{I}_{\pi/4}$ is not independent).
These four representations correspond to the tetragonal states
$A_{1g}$, $A_{2g}$, $B_{1g}$, and $B_{2g}$ in
standard group-theoretical notation\rlap.\refmark{\annett} 
These four possible states  transform
under the symmetry operations as the identity, $xy(x^2-y^2)$,
$x^2-y^2$ and $xy$, respectively.  These four states are shown in
Fig. \FIG\reps{The four singlet irreducible representations
possible in a single square CuO$_2$ plane.}\reps.
 (Notice that there are no
E representations  ($d_{xz}$, $d_{yz}$) because
of our assumption of a single CuO$_2$ plane, rather than a full
tetragonal crystal).  
It should be emphasised that these ``representative'' functions
are typical only as
regards the symmetry of the pair state in question, and may give a very
poor indication of the actual nodal structure of the gap function.
For example, a possible $s$
state is  $\Delta({\bf k}) = A + B \cos{4\theta}$, 
($\theta=\tan^{-1}{k_x/k_y}$) which for $|B|>|A|$ has 8 nodes.
(This is sometimes called an ``extended s-state'').

In the following discussion it will be convenient to use a more
informal, and perhaps intuitively more appealing, nomenclature
for these four states.  We shall call any
state which is even under $\hat{R}_{\pi/2} $  an ``s-state'' and 
any which is odd a ``d-state".  To distinguish the two ``$s$''  states
we shall use superscripts $+$/$-$ for states which are
even and odd under $\hat{I}_{axis}$. Thus in this simplified 
notation the four states are just $s^+$, $s^-$, \dx2y2\ and $d_{xy}$,
respectively, with the properties 
summarised in Table 1 and Fig. 1.
(Note that because of the $C_{4v}$ 
symmetry the ``g-wave'' state $xy(x^2-y^2)$ 
is invariant under $\hat{R}_{\pi/2}$, i.e. it has the same
rotational properties as an $s$ state, and is only distinguished
from $s$ by its mirror reflections. Hence we use the nomenclature $s^-$
rather then $g$ for this state.)

\vskip 1cm

\epsfxsize=\hsize
\epsfbox{pairII_table_ps}

A point which will be absolutely crucial to the argument of this
review is that because of the structure of the group C$_{4v}$, all
terms in the
Ginzburg-Landau free energy which involve two of these irreducible representations
must be ``non-mixing'' in the sense of Section 3. 
Mixing type terms involving four of these distinct irreducible representations
are possible in principle, but do
not allow ``mixing'' with a single second-order phase transition.  We
conclude that for single square CuO$_2$ planes the existence of a 
``mixed''
order parameter, i.e. of two or more components transforming according to
different irreducible representations of the symmetry group, (such as
$s^+ + id_{x^2-y^2}$) is incompatible with the observed absence of more than one
phase transition (see Section 5.1).  Unless fluctuation effects are very much more dominant
than they are usually assumed to be in the cuprates (in which case the very
use of a Ginzburg-Landau free energy analytic in the order parameter components might be regarded as
dubious), it seems to us very difficult to avoid this important conclusion.

{\bf \section{Effects of Orthorhombicity}}

       As is well known, the CuO$_2$ planes in most of the cuprates, with the
exception of the Tl- and Hg-based compounds, are not exactly tetragonal but
have an orthorhombic structure; in addition to the slight difference
(usually $ <  2\%$) of the a- and b-crystal axes, the ``buckling'' of the
planes usually picks out a special axis.  In the case of YBCO (both ``1237''
and ``1248'') a more substantial anisotropy of the crystal lattice as a whole
is induced by the presence of chains in the b-direction; although one's
immediate instinct is that most of the ``action" as regards to
superconductivity is likely to be in the CuO$_2$ planes and the chains should
therefore be a relatively minor perturbation, the fact that the penetration
depth is observed\REFS\basov{D.N. Basov
\etal, \journal Phys. Rev. Lett. &74&598(95).}\REFSCON\tallon{J.L. 
Tallon \etal, \journal Phys. Rev. Lett. &74&1008(95).}\refsend\
 to be appreciably anisotropic ($\sim 50\%$) 
in the ab-plane
shows that this ``perturbation'' cannot necessarily be neglected.

The presence of the orthorhombic anisotropy means that the relevant
symmetry group is no longer the group of the square but a subgroup of it.
However, it is important to appreciate that this subgroup is not the same
for all the cuprate superconductors.  In the case of YBCO, the a- and b-
crystal axes become inequivalent, but each remain a twofold axis and a
mirror plane: thus $\hat{I}_{axis}$ remains a symmetry operation while
$\hat{R}_{\pi/2}$ and $\hat{I}_{\pi/4}$ are no longer so. 
The effect is to permit mixing
of what were in a square lattice the $s^+$ and \dx2y2\ states, and 
likewise the
pair $s^-$ and $d_{xy}$; however, there can for example still be no mixing of
\dx2y2\ and $s^-$.  On the other hand, in LSCO and BSCCO it is the two
orthogonal $45^{\circ}$ axes which become inequivalent, while remaining mirror
planes; thus $\hat{I}_{\pi/4}$  is still a good symmetry operation but 
$\hat{R}_{\pi/2}$ and 
$\hat{I}_{axis}$  are not, and the members of the pairs 
($s^+$, $d_{xy}$) and ($s^-$, \dx2y2)
can mix.  These considerations have important consequences for the
structure of the gap function (a ``local" quantity which is defined
separately in each twin domain, see below):  while in YBCO the nodes of the
gap, if it is ``\dx2y2\ like" (cf. below) may now occur at an angle
different from $45^{\circ}$, in LSCO and BSCCO it must still occur at
exactly $45^{\circ}$.

In many of the experiments conducted on the orthorhombic cuprates
to determine the symmetry of the order parameter, the samples have been heavily twinned;
untwinned samples are the exception.  To describe the general nature of the
order parameter within a single twin domain we shall adapt the following terminology:
in YBCO, an order parameter which is even under $\hat{I}_{axis}$ will be called
``$s^{+}$ - like'' or ``\dx2y2\ - like" according as it has
 the same or opposite
sign on the x- and y-axes. 
 Similarly, an odd state will be called ``$s^{-}$
 -like'' or ``$d_{xy}$ -like''
 depending on whether its value at some arbitrary
angle, say $30^{\circ}$, does or does not change sign under $\pi/2$ rotation; 
while in
LSCO and BSCCO an order parameter which is even under $\hat{I}_{\pi/4}$ will be called
``$s^{+}$ like" or ``$d_{xy}$ - like" according as it has the
 same or opposite sign under a $\pi/2$ rotation.

Where we come to discuss the heavily twinned samples used in most
experiments, an extremely important question arises:  how does the order parameter
behave as we cross a twin boundary?  The alternatives for the \dx2y2\
like
state, are illustrated in \FIG\twins{The two possible behaviours of the
 \dx2y2\
order parameter at a twin boundary in YBCO.}
Fig. \twins.  In Fig. \twins (a), the ``+" and ``-" signs
follow the a- and b-axes, while in Fig. \twins (b), they remain oriented with
respect to the ``absolute" North-South-East-West (``NSEW") axes, a behaviour we refer to as
``gyroscopic."  {\it We believe there are extremely strong considerations which
favour the conclusion that the behaviour is gyroscopic}.  In the first place,
the critical current of twinned samples does not, as far as is known,
differ markedly from that of untwinned ones, and this would seem to
indicate that the coupling of the order parameter across twin boundaries, in contrast to
that across grain boundaries, is comparable to that in bulk.  If so, then
since there is no obvious reason for the sign of the coupling between two
neighbouring ``NS" lobes of the order parameter on different sides of the twin boundary to
be different from that between two lobes in the same grain, the gyroscopic
configuration is favoured by a large energy.  This argument could possibly
fail in a strongly orthorhombic crystal such as YBCO, if for example the
coupling across the twin boundary should turn out to be dominated by the
chains\Ref\itery{V.J. Emery, private communication.}\ 
(something for which as far as we are aware there is currently no
evidence either for or against).  However, we believe an even stronger
argument for gyroscopic behaviour 
 (in YBCO) is the observation of a
reproducible phase shift of zero in the UIUC I\refmark{\wollmani} 
and Maryland control
experiments\Ref\mathai{A. Mathai, Y. Gim, R.C. Black, A. Amar and
F.C. Wellstood, \journal Phys. Rev. Lett. &74&4523(95).}
 (see Section 7 below), both of
which were performed on heavily twinned (as well as, in the case of UIUC
I, untwinned) samples.  Were the ``non-gyroscopic" hypothesis correct it
would be impossible to understand this reproducibility.  Although we have
gone through this argument explicitly for the case of \dx2y2\  pairing, it is
clear that similar considerations apply to the $d_{xy}$ and 
$s^-$ states (for the $s^+$
state, the question clearly does not arise).

In light of the above, it is convenient to define a
``twin-averaged" order parameter, that is the order parameter averaged over the two
possible twin-orientations on the assumption that it behaves gyroscopically
across the boundary between them.  It is clear that the possible symmetries
of this ``twin-averaged" order parameter are exactly those already found for the case of
pure tetragonal (square) symmetry.  Moreover, when we come to discuss the
Josephson experiments (Section 7 below), it will turn out that in the
``thermodynamic limit" (by which we mean the limit in which the number of
twin domains in the area of the relevant junction tends to infinity and
there is no systematic bias in favour of one orientation) and with the
``gyroscopic" assumption, it is only this ``twin-averaged" order parameter which is
relevant.  In addition, it is clear that, given the above assumptions,
the bulk Ginzburg-Landau free energy, averaged over the twins, can be expressed
in terms of the twin averaged order parameter, and the arguments about mixing 
then go through exactly as in the tetragonal case.
Thus to a large extent (though not entirely, see Section 7.1) in
the context of the Josephson experiments the whole issue of orthorhombic
anisotropy is irrelevant.  Of course, it cannot be neglected when
analyzing experiments such as ARPES, which probe the ``local" gap and do not
average over a large number of twin domains.

{\bf \section{Bilayer and Trilayer Structures}}

Up to now, we have assumed that all pairing states take place within a
single plane, i.e. that in the expression \pairing\ for the order parameter,
${\bf r}_1$
 and ${\bf r}_2$  lie
in the same CuO$_2$ layer.  We must now face up to the complications
associated with the fact that many of the cuprates, including YBCO, possess
double or in some cases triple CuO$_2$ planes; and that even for single-layer
materials such as Tl 2201\rlap,\Ref\lsco{It is also worth noting that LSCO,
usually regarded as a ``single layer'' material actually has two
non-equivalent (but widely separated) CuO$_2$ layers per unit cell.}
it is not obvious that there cannot be pairing
between electrons in different layers.

Let us first dispose of the latter complication.  What we are
fundamentally interested in throughout this review is the symmetry of the
order parameter in the ab-plane, and given that $\Psi({\bf r}_1,{\bf r}_2;\alpha \beta)
= \langle \psi_\alpha({\bf r}_1) \psi_\beta({\bf r}_2) \rangle$ is
non-zero for ${\bf r}_1$ and ${\bf r}_2$ within the same CuO$_2$ layer, 
we can analyze this
symmetry as in subsection 4.1 and 4.2 above, quite irrespective of
 whether $\Psi$  is
also non-zero for ${\bf r}_1$ and ${\bf r}_2$ 
in different layers.  Thus the only
circumstance which would render the analysis of Sections 4.1 and 4.2 
invalid is
if $\Psi({\bf r}_1,{\bf r}_2;\alpha \beta)$
 were to vanish for ${\bf r}_1$ and ${\bf r}_2$ in the same plane.  While this
hypothesis is not incompatible with the existence of a superconducting
state, we believe its plausibility in single-layer materials is so low that
it is legitimate to neglect it in the present 
context\rlap.\REF\aswell{As a matter of fact, even if it did, it would matter only if the behaviour 
under interlayer exchange were nontrivial, cf. the discussion of the
``symmetric'' bilayer case below.}\  Thus, for such
materials the analysis of Sections 4.1 and 4.2 is adequate as it stands.

For double- or triple-layer materials such as YBCO or BSCCO 2223 it
is less obvious {\it a priori\/} that the hypothesis of ``exclusively inter-layer"
pairing can be excluded.  Let us focus for definiteness on a bilayer
material such as YBCO.  If the order parameter is symmetric with respect to interchange
of the ``layer" indices of ${\bf r}_1$ and ${\bf r}_2$
 (a possibility which actually seems
pathological if there is to be no intra-layer pairing) then it is clear
that the analysis of subsections 4.1 and 4.2 goes through unchanged. 
 If on the
other hand it is antisymmetric (and we continue to assume, as in subsection
4.1, spin singlet even-frequency pairing), then the symmetry with respect to
inversion within the ab-plane must be odd rather than even, and we have to
deal with a set of irreducible representations different from those of
Fig. \reps.  We will not explore this possibility further here, since (a) in
view of the general qualitative similarity between the superconducting
behaviour of various classes of cuprates, it seems to us extremely unlikely
that the pairing state is radically different in one- and two-layer
materials\rlap,\Ref\theposs{The possibility that
 simple even-parity pairing occurs in one-layer
materials and a combination even- and odd-parity pairing in bilayer ones is
difficult to reconcile with the absence of more than one transition in the
latter, cf. Section 3.}
 and (b) a state with odd parity in the ab-plane seems difficult
to reconcile with the existence of a reproducible (a- or b-direction)
Josephson effect with ordinary s-wave superconductors (cf. Section 6, below).
Thus we conclude that within a single CuO$_2$ plane the order parameter is finite and can
be classified as in Fig. \reps.  Given this state of affairs, it is still a
nontrivial question how the order parameter behaves (a) under reflection in the symmetry
plane ``spacing" the two layers (e.g. in the case of YBCO, the plane
containing the Y atoms), and (b) under translation
up the c-axis from one unit cell to the next.  With regard to (b), the
``natural" assumption is that the order parameter is identical both in magnitude and (in
the absence of superflow) in phase from one unit cell to the next, (i.e.
behaves according to the identity representation of the translation group
$T_\ell$), and any assumption different from this would seem to complicate the
picture gratuitously without helping in any way to resolve the apparent
experimental inconsistencies we shall comment on below (indeed, in many
cases it would exacerbate them).  We will therefore make the ``natural"
assumption from now on.  (The same comment applies to the behaviour of the
order parameter within the ab-plane).

Finally, what is the behaviour of the order parameter under reflection in the
``spacing'' plane?  Evidently the irreducible representations are even or odd
under this operation, and thus are ``non-mixing'' in
the language of Section 3.  Thus, by the arguments of that Section, the
absence of more than one phase transition in the bilayer materials is
strong evidence in favour of one and only one irreducible representation being realised.  Now if
it is the odd representation that occurs, it is difficult, if not
impossible, to understand the existence of an (a- or b- direction)
Josephson effect with ordinary superconductors (the argument is essentially
that of Section 6.1, slightly generalised).  Thus we conclude that the
symmetry under reflection is even, and it is then clear that we can for
many (though not all) purposes neglect the bilayer structure entirely, i.e.
treat each bilayer as effectively a single layer.

It is clearly possible to give a similar discussion of trilayer
structures, but since few if any of the crucial experiments have been done
on trilayer materials, there seems no point in doing so here.

{\bf \section{The energy gap}}

       While the existence of the two-particle anomalous average (order parameter)
defined by Eqn. \pairing\ is of course fundamental to the whole of our
discussion, none of the results we have obtained so far depend in any way
on any specific assumptions about the mechanism of its generation; in
particular, they are completely independent of its relation to the
single-particle elementary excitations of the crystal (if such exist).
This point should be emphasised, since an important class of theories of
superconductivity in the cuprates starts from the premise that in the
effectively two-dimensional
systems formed by the individual CuO$_2$ planes no ``single-particle"
excitations in the usual sense of Fermi-liquid theory exist.  If this
should turn out to be the case, then it is not obvious that any of the
microscopic concepts of BCS theory apply to the superconducting states, and
in interpreting the experimental data on (e.g.) specific heat, penetration
length and ARPES one must start again from scratch in the light of the
concepts of the specific alternative theory under consideration.  For the
purposes of the present review, we shall take the point of view that the
apparent signature of a Fermi surface in the normal-phase ARPES
 data\rlap,\refmark{\positron} plus
the qualitative similarity of many of the superconducting-state data to
what is predicted in (generalisations of) BCS theory, make it reasonable to
discuss the microscopic properties in what is essentially a ``BCS-like"
picture.  That is, we shall assume in this subsection that, at least for
the purposes of discussing the superconducting state, a description of the
normal state in terms of a Fermi-liquid type of picture is adequate, and
will then go on in Section 5 to interpret various ``microscopic'' experiments
in light of the description of the superconducting state so obtained.
It should be emphasised that even if the assumptions which underlie the
work of this subsection and Section 5 should eventually turn out to be
false, none of the other results obtained in this review will be affected.

 In a ``Fermi-liquid-like'' picture of the normal state of a typical
``3D'' metallic crystal, the low-lying single-particle excitations are Bloch
waves $\phi_{n{\bf k}}({\bf r})$ characterised by a crystal momentum 
${\bf k}$, a band index $n$ and a spin
index $s$, and have an energy $\epsilon_{n{\bf k}}$ which becomes
arbitrarily sharply defined
as ${\bf k}$ approaches the Fermi surface of band $n$. 
(This would also be true in the marginal Fermi liquid and Luttinger liquid
pictures\rlap.\refmark{\littlewood,\chakravarty}) In the very simplest
microscopic model of superconductivity (essentially the original BCS
picture) one considers only a single band $n$, and one forms (spin singlet) Cooper
pairs by pairing electrons with momentum-spin values ${\bf k} \uparrow$ 
and $-{\bf k} \downarrow$ : see e.g \REF\tinkham{ 
M. Tinkham, {\sl Introduction to Superconductivity}, (McGraw Hill,
1975), ch.~2.} Ref.  \tinkham.
In such a theory the single-particle excitation energies $E_{\bf k}(T)$
are given by the familiar BCS formula
$$
  E_{\bf k}^2(T)  = \epsilon_{\bf k}^2  + | \Delta_{\bf k}(T) |^2 
\eqn\bcs$$
where the quantity $\Delta_{\bf k}(T)$ is usually known as the ``energy gap.''
As discussed in Section 2, a non-zero value for the Gor'kov
off-diagonal Green function \gorkov\ implies the existence of
an anomalous self-energy $\Delta_{\bf k}$ whether or not the
normal state is a Fermi liquid. However this BCS quasiparticle
energy formula \bcs\ is only guaranteed to be  valid
in a Fermi liquid picture.  In this case, 
there is a simple relation between $\Delta_{\bf k}(T)$ 
and the order parameter $\Psi({\bf r}_1,{\bf r}_2)$
(we suppress the spin indices, assuming as always spin singlet pairing):
$$ \eqalign{
 \Psi({\bf r}_1,{\bf r_2};T) = & \sum_{{\bf k}} 
{\Delta_{\bf k}(T) \over 2E_{\bf k}(T)}
 \tanh{(E_{\bf k}(T)/2T)}
\phi_{{\bf k}}({\bf r}_1)\phi_{{\bf k}}({\bf r}_2) \cr
 \equiv & \sum_{{\bf k}} \Psi_{\bf k}(T)
 \phi_{{\bf k}}({\bf r}_1)\phi_{{\bf k}}({\bf r}_2) 
}\eqn\op$$
In the case of the cuprates we have to decide whether to treat the Bloch
waves as three-dimensional or as defined only within the plane
(cf. refs. \REF\andersonzou{P.W. Anderson and Z. Zou,
\journal Phys. Rev. Lett. &60&132(88).} \andersonzou,
\REF\leggettbraz{A.J. Leggett, \journal Braz. Journ. of Physics,
&22&129(92).} \leggettbraz). 
 In the former case we can take over formula \op\ as it stands,
for arbitrary values of ${\bf r}_1$ and ${\bf r}_2$: 
 in the latter we should consider only
values of ${\bf r}_1$ and ${\bf r}_2$
 lying in the same plane or bilayer, etc. (but see
below), and the sum over {\bf k} then goes only over values in the ab-plane.

  For a single band the Bloch wave function and energies must respect
the crystal symmetry, in the sense that 
$\epsilon_{n{\bf k}}=\epsilon_{n\hat{R}{\bf k}}$ and 
$\phi_{n{\bf k}}({\bf r})=\phi_{n\hat{R}{\bf k}}(\hat{R}{\bf r})$  ,
where $\hat{R}$ is any symmetry operation of
the crystal point group.  Since all the irreducible representations
relevant to our problem (i.e. those of Fig. \reps) are one-dimensional, it
follows immediately from \bcs\  and \op\ that whenever 
$\Delta_{\bf k}(T)$  transforms
according to a single irreducible representation of the crystal group, then
$\Psi({\bf r}_1,{\bf r}_2)$ 
transforms according to that same irreducible representation.  Note that this
conclusion is independent of any consideration concerning the details of
the dynamics.  Is the converse also true?  It will certainly be so, if the
condition which determines how $\Delta_{\bf k}(T)$  transforms
 can be written in the form of a
generalised BCS gap equation, i.e. in the form
$$ \Delta_{\bf k}(T) = \sum_{{\bf k}'} V_{eff}({\bf k},{\bf k}')
\Psi_{{\bf k}'}(T). \eqn\gapeqn $$
For, even if $V_{eff}({\bf k},{\bf k}')$
 is itself a function of  $\Delta_{\bf k}(T)$, analyticity
arguments similar to those employed in the discussion concerning the
Ginzburg-Landau free energy, show that it can depend only on 
$| \Delta_{\bf k}(T) |^2$  and must
therefore (in the case of a 1D representation) continue to respect the
crystal symmetry.  It immediately follows that if 
$ \Psi({\bf r}_1,{\bf r}_2)$ transforms
according to a given single irreducible representation then so does $\Delta_{\bf k}(T)$.

It should be emphasised that the above arguments do not prove that
the Fourier transform of the order parameter, $ \Psi({\bf k})$, is
 simply proportional to $\Delta_{\bf k}(T)$ in its
complete dependence on the direction $\hat{\bf k}\equiv {\bf k}/|{\bf k}|$, 
even for a single irreducible representation
(let alone for the ``mixed" case).  However, in the so-called weak coupling
limit in which the effective cutoff energy in the sum over ${\bf k}'$ 
is much greater
than typical values of the gap, the effect of the nonlinearity in Eqn.
\gapeqn\ is small and the two quantities should be approximately proportional.
(cf. the related discussion for the case of O(3)
 symmetry\rlap.\Ref\andersonmorel{P.W. Anderson and P. Morel,
\journal Phys. Rev. &123&1911(61).})
Irrespective of this, we note from \op\ that the nodes of 
$\Delta_{\bf k}(T)$  and $ \Psi({\bf k})$
coincide even when they are not symmetry-determined.  Thus for most
purposes it should not be too bad an approximation to take the two
quantities to be proportional.

When we allow for more than one band to be involved in the process
of Cooper pair formation, things become more complicated.  If we define
operators $c_{n{\bf k}}$ 
for the Bloch waves $\phi_{n{\bf k}}$
 in the standard way, the quantity
$\langle c_{n{\bf k}} c_{n'{\bf k}'}\rangle $ 
 and hence the ``generalised gap function" $\Delta_{nn'{\bf k}}(T)$  may be a
matrix with respect to the band indices. (cf. the related problem of
 pairing in a spin triplet state ($S=1$) when all
the Zeeman substates are represented\rlap.\Ref\balian{R. Balian and N.R.
Werthamer, \journal Phys. Rev. &131&1553(63).})
An extreme case of this
situation is a scenario recently proposed by 
Tahir-Kheli\rlap,\Ref\tahirkheli{J. Tahir-Kheli,
preprint.} in which
the pairing is exclusively ``inter-band," i.e., the diagonal elements of
$\Delta_{nn'{\bf k}}(T)$  are zero.  
In general the formula for the energy spectrum and
hence the experimental properties in such models are considerably more
complicated than in the one-band case.  However, we emphasize that these
complications in no way invalidate the conclusions we have already reached
about the possible symmetries of the order parameter; in fact, the form of
the quantities $\Delta_{nn'{\bf k}}(T)$ must respect 
these conclusions, so that for example
in a square lattice the occurrence of only one phase transition constrains
all the elements $\Delta_{nn'{\bf k}}(T)$
to belong to the same representation.

As a matter of fact, the ``multi-band'' case most likely to be of
practical interest arises in the context of bi- or multilayer systems.
Considering the bilayer case for definiteness, we have two obvious limiting
cases, depending on our hypothesis about the nature of the electronic state
in the normal phase.  If one-electron hopping between the two component
layers of the bilayer is coherent, with hopping matrix element,  $t_\perp$,
 then the bands are split into even and odd bands, with
$\epsilon_{\pm{\bf k}} = \epsilon_{0{\bf k}} \pm t_\perp$. 
Furthermore, if the pairing interaction is sufficiently weak that 
$k_BT_c \ll t_\perp$, then
the pairing matrix will be diagonal in the band indices (although the
equation for the gaps may couple them\Ref\liulevin{D.Z. Liu, 
K. Levin and J. Maly, \journal Phys. Rev. B &51&8680(95).}). 
 In the opposite
limit, where the normal-state hopping is completely incoherent, or coherent
but with a very small matrix element (compared to the effects of the
(intra-layer) pairing interaction $t_{\perp} \ll k_B T_c$), 
pairing is likely to be
diagonal in the layer indices, with a subsequent Josephson-type coupling
between the layers.  The case in which the normal-state inter-layer contact
is completely incoherent is particularly fascinating, since Chakravarty \etal\refmark{\chakravarty} have argued that in this case 
the pair tunneling in the
superconducting state can raise the transition temperature very
substantially.

In the intermediate case ($t_\perp \sim k_B T_c$) 
the gap matrix is likely to be diagonal
neither in the ``layer" nor in the ``even-odd" representation.  We
reemphasise, however, that irrespective of these complications the
conclusions reached earlier still obtain; in particular, in a bilayer
cuprate superconductor with a square lattice and a single phase transition,
even if there is more than one ``energy gap" all gaps must transform under
ab-plane operations according to the same representation\Ref\samerep{
At first sight the model constructed by Liu \etal\rlap.\refmark{\liulevin}
for YBCO
might seem to violate this principle. However, as the authors themselves
 acknowledge, a finite degree of orthorhombic anisotropy
is required for its viability.}\
 (though they need
not, of course, have an identical dependence in detail on {\bf k}).

{\bf\chapter{Evidence for a Sign Change in the Gap Function}}

In this Section, we discuss the experimental evidence for a sign change
in the gap function at the Fermi surface.  These data include a
systematic investigation of the low energy excitations in the
superconducting state, some of which was discussed in AGR.  Here we
focus primarily on the electromagnetic penetration depth and the
photoemission results.  These two probes are perhaps the most direct
measures of the superconducting state, and with a minimum of theoretical
interpretation yield useful information about the pairing state.  The
penetration depth provided perhaps the first strong indication that the
pairing state was not s-wave.  Photoemission directly
probes the symmetry of the gap function, and potentially can disprove a
d-wave scenario.  Indeed, as we shall see, until very recently, it did
seem that the photoemission data were inconsistent with a d-wave pairing
state.

{\bf \section{Electromagnetic Penetration Depth}}

The electromagnetic penetration depth is a powerful probe of the
superconducting state because the data, if taken at face value, directly
reflect the response of the condensate to electromagnetic
perturbations.   Thus, the penetration depth is a measure of the
superfluid density tensor.  In general, a detailed understanding of the
superconducting state is required to interpret correctly penetration
depth data.  Details of many factors, including the coupling between the
electrons and the pairing bosons, electronic structure and disorder can
all quantitatively modify the predictions based upon a simple BCS-type
theory.

In the absence of a detailed microscopic theory, it behoves one to
examine the universal qualitative behaviour of the penetration depth,
which with relatively weak assumptions can be usefully interpreted,
independent of the mechanism of superconductivity.  Two such regimes are
available in the cuprate superconductors.

The first is the asymptotically low temperature regime, which we will
discuss extensively below, and which can directly test hypotheses about
the existence or non-existence of a change in sign of the gap function
over the Fermi surface.  This regime is also available in classic
superconductors.  At low temperatures, it should be possible to
distinguish a superconductor with a non-zero gap function minimum from
one that has nodes by the presence or absence of exponential behaviour
as a function of temperature.  As discussed (\eg) in AGR, different
varieties of nodal structure give rise to a temperature dependent
penetration depth tensor whose components approach their zero
temperature value as a power law function of temperature, with an
exponent that is related (but not in a one-to-one correspondence) to
the pairing state.

The second universal regime is that near \Tc, where fluctuation effects
become important\rlap.\Ref\ffh{D.S. Fisher, M.P. Fisher and D.A. Huse,
\journal Phys. Rev. B &43&130(91).}\  Although it has been generally
accepted for some time that weak Gaussian fluctuations about mean field
theory are observable in the cuprates, it is only recently that genuine
critical fluctuations have been observed\rlap.\refmark{\kamal}\ 
This critical fluctuation
regime is not accessible in other three dimensional superconductors, and
arises in the cuprates because the Ginzburg-Landau correlation length
extrapolated to low temperatures is of order the unit cell dimensions.

{\bf \subsection{Summary of experimental observations}}

During the last 6 years, experimental data have become available on
better characterised samples than previously.  The salient experimental
findings have been reviewed by D.A. Bonn and W.N. Hardy in this Volume, and
the results can be summarised as follows:

\item{(1)} Measurements\refmark{\hardyi,}\REFS\sonier{J.E. Sonier \etal,
\journal Phys. Rev. Lett. &72&744(94).}\REFSCON\maoi{J. Mao \etal,
\journal Phys. Rev. Lett. &51&3316(95).}\refsend\  of the $a-b$ \/ plane
penetration depth $\lambda(T)$ in twinned single crystals of
\Yba\ indicate a linear temperature dependence with a slope of about
4\AA/K, from about 1K to about 30 K.  These results have now been
reproduced by a variety of different groups using different techniques,
including microwave and $\mu$SR techniques.

\item{(2)} Measurements\Ref\vaulechier{L.A. de Vaulchier \etal, preprint
{\tt cond-mat/9504062}.} of $\lambda(T)$ in high quality twinned thin
\Yba\ films epitaxially grown on MgO or LaAlO$_3$ substrates also exhibit
the linear behaviour mentioned under (1).  These measurements agree
quantitatively with those obtained on the single crystals, with the
exception that $\lambda(0)$ was found to be $1750 \pm 160$\AA, slightly
larger than the value of $1400 \pm 50$\AA\  usually quoted for crystals.
Linear behaviour has also been observed on \Yba\ films grown by
evaporation onto a SrTiO$_3$ substrate\rlap.\Ref\ulmii{E. Ulm and T.
Lemberger, {\it Thermal phase fluctuations and the low temperature
behaviour of the ab-plane penetration depth $\lambda(T)$ in \Yba.},
preprint.}

\item{(3)} Measurements\refmark{\vaulechier,}\REFS\fiory{A.T. Fiory
\etal, \journal Phys. Rev. Lett. &52&2165(88); see also the reanalysis
of the data reported in Ref. \agrii.}\REFSCON\ma{Z. Ma \etal, \journal
Phys. Rev. Lett. &71&781(93).}\REFSCON\leeapl{J.Y. Lee and T.R.
Lemberger, \journal Appl. Phys. Lett.
&62&2419(93).}\REFSCON\leeohio{J.Y. Lee \etal, \journal Phys. Rev. B
&50&3337(94).}\refsend\  on lower quality twinned thin \Yba\ films yield
a temperature dependence for $\lambda$ that varies as $T^2$.  In Ref.
\vaulechier, the films whose temperature dependence was quadratic were
found to have a value for $\lambda(0)$ of 3600\AA, whereas in the other
references cited, where a value can be deduced, it is of order 1700\AA.
The nominally pure films exhibited a quadratic temperature dependence at
temperatures below 25K, but between 25K and 40K the temperature
dependence matched that measured on single crystals\refmark{\hardyi}
very well\rlap.\refmark{\leeohio}\  The authors of Ref. \leeohio\ were
able to plot the coefficient of the quadratic term against $\lambda(0)$
and found agreement with the d-wave theory of Hirschfeld and
Goldenfeld\rlap.\Ref\hirschfeldi{P.J. Hirschfeld and N. Goldenfeld,
\journal Phys. Rev. B &48&4219(93).} 

\item{(4)} The anisotropy of the penetration depth\refmark{\basov,}
\REFS\zhang{K. Zhang
\etal, \journal Phys. Rev. Lett. &73&2484(94).}
\refsend\  and surface
resistance\refmark{\zhang} of untwinned \Yba\ crystals has been measured
using microwave cavity perturbation techniques\refmark{\zhang} and far
infra red spectroscopy\rlap.\refmark{\basov}\ It is found that the
temperature dependence of the principal components
of the penetration depth tensor (i.e. in both the $a$
and $b$ directions) is similar to that found in the average penetration
depth observed in twinned crystals.  For present purposes, the quantity
that is of interest is $\sigma_{00}$, the limiting value of the real
part of the complex conductivity, extrapolated to zero frequency.  A
surprising prediction is that in a d-wave superconductor, $\sigma_{00}$
is independent of the scattering rate, and takes a value of
approximately 0.3 times the dc conductivity at \Tc.  The observed values
for the residual conductivity (that is, the conductivity extrapolated to
zero temperature) are consistent with this prediction.

\item{(5)} Twinned single crystals of \Yba\ similar to those which give
a linear temperature dependence for $\lambda(T)$ at low temperatures
were systematically doped with Zn and Ni
impurities\rlap.\Ref\bonndope{D.A. Bonn \etal, \journal Phys. Rev. B
&50&4051(94).}\  These impurities substitute for the Cu(2) atoms (in
the CuO$_2$ planes) so that
the sequence of crystals studied were
YBa$_2$(Cu$_{1-x}$A$_x$)$_3$O$_{6.95}$, where A stands for Zn or Ni, 
$x=0.0015$ and $0.0031$
for Zn and $x=0.0075$ for Ni.  The Zn-doped crystals experience a strong
\Tc\ suppression\rlap,\Ref\chien{T.R. Chien \etal, \journal Phys. Rev.
Lett. &67&2088(91).} $\pd T_c/\pd x\approx -1260$K, whereas the Ni-doped
crystals (and thin films)\Ref\sumner{M.J. Sumner \etal, \journal Phys.
Rev. B &47&12248(93).}  experience much less of a suppression in \Tc:
$\pd T_c/\pd x\approx -390$K.  This is seemingly at odds with the
observation\refmark{\sumner} that the dc resistance at \Tc\ is increased
to roughly the same extent by both Ni and Zn.  The influence on the low
temperature behaviour is quite dramatic.  The Zn
doping\refmark{\bonndope} causes the penetration depth to exhibit a
quadratic variation with temperature below a crossover temperature
$T^*$, and a linear variation with temperature above $T^*$.  That is, if
one considers only the temperature interval over which the nominally
pure crystals exhibit a linear temperature dependence, this interval
becomes divided into two regions.  The Ni doping shows very little
effect on the penetration depth, although the barely discernible trend
is qualitatively similar to that for the Zn doping.  These data could
be sensibly fit to an interpolation formula derived from the theory of
impurity scattering in d-wave
superconductors\rlap.\refmark{\hirschfeldi}

\item{(6)} Ni and Zn dopants were added to \Yba\ films\Ref\ulm{E.R. Ulm
\etal, \journal Phys. Rev. B &51&9193(95).} but in much greater
concentrations than in the single crystal experiments described above.
The values of $x$ for both Ni and Zn ranged from 2\% to 6\%,
corresponding to a substitution of 3\% to 9\% of the plane copper
atoms.  As with the single crystals, it was found that Ni and Zn have
approximately the same scattering rate, as determined from infrared
reflectivity measurements, but that Zn depresses \Tc\ about three times
more rapidly than Ni.  The penetration depth was determined from kinetic
inductance measurements and the variation of $\lambda(0)$ with doping
concentration analysed and compared with a theory of Kim
\etal\Ref\kim{H. Kim \etal, \journal Phys. Rev. B &49&3544(94).} which
is an extension of Ref. \hirschfeldi\ to high concentrations.  The
agreement is not very good, with the reduction in the zero temperature
superfluid density $n_s(0)$ being severely underestimated by as much as
a factor of 10 (Ni at 6\% doping).

\item{(6)} The c-axis penetration depth has been measured\Ref\maoc{J.
Miao \etal, \journal Phys. Rev. B &51&3316(95).} and a linear
temperature dependence obtained.

\item{(7)} Near \Tc, the penetration depth diverges as $\lambda(T)
\propto (1-T/T_c)^{-y}$ with $y\approx 0.33$.  This behaviour has been
observed in \Yba\ single crystals\refmark{\kamal,}\Ref\anl{S. Anlage
\etal, {\it Fluctuations in the microwave conductivity of \newline\Yba\ single
crystals in zero dc magnetic field}, {\sl Phys. Rev. B\/} in press.} and
La$_{2-x}$Sr$_x$CuO$_{4\pm\delta}$ thin films\rlap.\Ref\jaccard{Y.
Jaccard \etal, \journal Physica C &235-240&1811(94).}

\item{(8)} The low temperature 
behaviour\Ref\anlageND{S.M. Anlage \etal, 
\journal J. Superconductivity &7&453(94).} of $\lambda (T)$ for
thin films and single crystals of the 
Nd$_{1.85}$Ce$_{0.15}$CuO$_{4-\delta}$ compound exhibits an exponential
temperature dependence with $2\Delta(0)/k_BT_c \approx 4.1$.  This
material seems to be exceptional, because it is the only
known cuprate which appears to be an s-wave superconductor.  Presumably
this is associated with the fact that it is electron-doped as opposed
to hole-doped.

{\bf\subsection{Evidence for one complex order parameter}}

The observed critical behaviour of the penetration depth is consistent
with the three dimensional (3D) XY universality class.  In Ref. \kamal,
the authors checked this interpretation in a number of ways, apart from
the value of the exponent for the penetration depth.  First, the
observed exponent was not altered by the presence of Zn impurities;
generically, disorder would be expected to affect the universality class
of the phase transition, but in the case where the specific heat
exponent is negative, such as the 3D XY model, the Harris criterion
predicts that weak disorder should be irrelevant in a renormalisation
group sense.  Secondly, Kamal \etal\ used hyperscaling relations to
calculate the correlation length in the c-axis; thus they were able to
verify that the system was indeed in a regime where the fluctuations
would be expected to be three dimensional, showing that the analysis was
self-consistent.

The observed behaviour has several implications.  First, no evidence for
a second transition is observed.  Gross thermodynamic measurements on single
crystals away
from \Tc\ have failed to indicate any anomalies or discontinuities.  The
measurements of Ref.  \kamal\ exhibit scaling phenomena over two decades
of reduced temperature $t\equiv (T_c-T)/T_c$ over the range $0.001 < t <
0.1$, setting a bound on how close to the upper value of \Tc\ a putative
second transition would have to be, assuming that it did indeed affect
the penetration depth by causing an anomaly in the superfluid density.
In light of the thermodynamic arguments in Section 4, {\it we
conclude that an order parameter consisting of a mixture of
irreducible representations does not occur}.  Secondly, the observation
of 3D XY scaling strongly indicates that {\it the order parameter is a
single 
complex number}  and hence is a one dimensional irreducible
representation.

{\bf \subsection{Evidence for nodes}}

The observation of linear temperature dependence of $\lambda$
at low temperatures is
certainly a striking departure from the s-wave prediction of a flat or
exponential temperature dependence, and is stronger evidence for a
d-wave interpretation than the earlier observations of quadratic
temperature dependence.  The latter could conceivably have been an
artifact of several proposed mechanisms, but within the context of an
s-wave pairing scenario: these mechanisms include trapped
vortices\rlap,\Ref\hebardii{A.F. Hebard \etal, \journal Phys. Rev. B
&40&5243(89).} vortex pair nucleation at defects\Ref\hebardiii{A.F.
Hebard \etal, \journal Phys. Rev. B &44&9753(91).} and low frequency
phonons\rlap.\REFS\klimo{G.V. Klimovich \etal, \journal JETP Lett.
&53&399(91).}\REFSCON\gant{V.F. Gantmacher \etal, \journal Physica C
&171&223(90).}\refsend\ However, none of these mechanisms was shown to
account quantitatively for the experimental evidence for the quadratic
temperature dependence observed in thin films.

The linear temperature dependence observed more recently is in accord
with the arguments of Ref. \agrii, which showed that {\it both\/}
principal components of the a-b plane penetration depth tensor should grow
linearly with temperature if \Yba\ were a singlet unconventional
superconductor.  There are two main scenarios in which this
interpretation could be incorrect.  First, the theoretical prediction
concerns the asymptotic low temperature behaviour of the penetration
depth.  It is conceivable that the gap function is strongly anisotropic,
but with a gap minimum that is not zero but nonetheless very small
compared with $k_BT_c$, for example of order 1mK.  In this case, the
asymptotic low temperature behaviour would be exponential, but would not
be observed until $T \sim 1$mK.  For temperatures greater than this, but
still much less than $T_c$, the temperature dependent penetration depth
would resemble that of a genuine d-wave superconductor.  This scenario
can never be entirely eliminated (by penetration depth measurements
alone), even by taking data at much lower temperatures.  An s-wave
multilayer proximity effect coupling model of \Yba\ has been
presented\rlap,\Ref\klemmliu{R.A. Klemm and S.H. Liu, \journal Phys. Rev.
Lett. &74&2343(95).} which is able to fit the nominally pure data of
Ref. \bonndope\ by appropriate choice of the parameters in the model.
However, this model predicted that the temperature dependence of
$\lambda_c(T)$ should be strongly exponential, in clear disagreement
with subsequently measured data of Ref. \maoc.

The second scenario is the proposal that phase fluctuations may be
important in the high temperature superconductors\rlap,\Ref\itery{V.J.
Emery and S.A. Kivelson, \journal Nature &374&434(95).} and could
account for the linear temperature dependence of the penetration
depth\rlap.\REFS\roddick{E. Roddick and D. Stroud, \journal Phys. Rev.
Lett. &74&1430(95).}\REFSCON\coffey{M.W. Coffey, \journal Phys. Lett. A
&200&195(95).}\REFSCON\itii{V.J. Emery and S.A. Kivelson, \journal Phys.
Rev. Lett. &74&3253(95).}\refsend  However, it is hard to see how this
proposal could account for the observed crossover to a quadratic
temperature dependence in the presence of impurities.  In addition, the
theory generically predicts that $d\log\lambda(T)/dT$ extrapolated to
zero temperature should increase and be proportional to $\lambda(0)^2$
as this latter quantity is increased by impurity doping.  However, the
experimental result\refmark{\ulmii} is that the zero temperature
extrapolated value of $d\log\lambda(T)/dT$ decreases with increasing
$\lambda(0)$.  Moreover, the observed trend is qualitatively consistent
with the node scenario, because isotropising or filling in of the
density of states will tend to reduce the temperature dependence of
$\lambda(T)$.  Thus, it seems very unlikely that phase fluctuations are
making an important contribution to the low temperature penetration
depth, so they cannot account for the observed behaviour.

Probably the strongest grounds for associating the observed low
temperature behaviour of the penetration depth with nodes in the gap is
the behaviour in the presence of small amounts of doping, discussed in other
contexts in Section 8.  Assuming that
one can use the standard Abrikosov-Gor'kov theory for impurity
scattering in superconductors\REFS\gorkov{A.A. Abrikosov and L.P.
Gor'kov, \journal Zh. Eksp. Teor. Fiz. &39&1781(60)\ [\journal Sov.
Phys. JETP &12&1243(61)].}\REFSCON\skalski{S. Skalski, O.
Betbeder-Matibet and P.R. Weiss, \journal Phys. Rev.
&136&A1500(64).}\refsend\ and that the unitary scattering limit is
relevant, it is straightforward to show that the penetration depth in a
superconductor with line nodes (in three dimensions) gives rise to a
penetration depth that crosses over from linear to quadratic temperature
dependence\rlap.\refmark{\hirschfeldi,}\Ref\carbotte{For related
calculations of this effect, see C.H. Choi and P. Muzikar, \journal
Phys. Rev. B &37&5947(88); R. Klemm \etal, \journal Z. Phys. B
&72&139(88); M. Prohammer and J.P. Carbotte, \journal Phys. Rev. B
&43&5370(91).}\ The crossover is accompanied by an increase in the zero
temperature value of the penetration depth, which can also be
computed\refmark{\hirschfeldi,\kim} and observed by extrapolation
of the data.  The theory predicts
that for small impurity concentrations, the limiting low temperature
behaviour should be quadratic, crossing over to a linear temperature
dependence at higher temperatures but still in the regime $T \ll T_c$.
The observed crossover temperatures $T^*$ are in reasonable agreement
with estimates made from the theory.  For example, the theory predicts
that $T^*\approx 0.83\sqrt{\Gamma\Delta_0}$, where $\Gamma$ is the impurity
scattering rate, which is estimated from conductivity measurements for
the 0.31\% Zn-doped single crystals of Ref. \bonndope\ to be 5K, and
$\Delta_0$ is the gap function amplitude, taken to be of order
3$k_BT_c$; thus $T^*$ should be of order 31K, which compares favourably
with the observed value of 28K for this concentration.  Although the
theory does not include a variety of effects which may be important,
such as taking proper account of the electronic structure along the c-axis,
van Hove singularities and localisation effects, it is expected to be a
reasonable indicator of the main trends, and should be useful in a
semi-quantitative way.  

Another question is the use of the unitary limit in the scattering
calculation.  The motivation for this was two-fold: first to ensure that
the impurities did not rapidly reduce \Tc\ to unobservably low values,
and second because of the observation\Ref\uher{A comprehensive review of
the data is given by C.  Uher, in {\sl Physical Properties of High
Temperature Superconductors III}, edited by D.M. Ginsberg (World
Scientific, New Jersey, 1992), Chapter 3, p. 159.} of a crossover in the thermal
conductivity to linear behaviour below about 300mK.  Not taking the
unitary limit would ensure that \Tc\ would decrease too rapidly with
doping to be consistent with experiment; but with the unitary limit in
place, there is no reason to believe that the theory would give a proper
account of phenomena near \Tc: for example, inelastic scattering is
neglected.  Nevertheless, the theory focuses on the most important
physics at low temperatures, where it was intended to be used.

The doping dependence of the penetration depth contrasts sharply with
what one would expect for a state without nodes, for which the
addition of impurities would be expected to increase the minimum of the
gap function\REFS\bork{L.S. Borkowski and P.J. Hirschfeld, \journal
Phys. Rev. B &49&15404(94).}\REFSCON\fehr{R. Fehrenbacher and M. Norman,
\journal Phys. Rev. B &50&3495(94); \journal Physica C
&235-240&2407(94).}\refsend\ leading to a more obvious exponential
behaviour.  Thus, even if for some accidental reason, the nominally pure
samples had a gap function that went to zero or nearly vanished at the
Fermi surface (thus simulating a d-wave superconductor from the point of
view of exhibiting a linear temperature dependence), the doped samples
would exhibit an exponential temperature dependence in
disagreement with experiment.

{\bf \subsection{Conclusion}}

What do these observations and the fitting with simple theoretical
models enable us to conclude about the pairing state?  First, unless
there is a very weak secondary transition below the threshold of
detection, we can rule out a state with two order parameters.  Second,
the superconducting state of \Yba\ and at least one of the La materials
is a one-dimensional representation in the universality class of the 3D
XY model, consistent with a $s$, $d$ or mixture of $s+d$ states in an
(untwinned)
orthorhombically-distorted tetragonal copper-oxide plane.   The linear
temperature dependence down to about 1K places an upper bound on any
minimum in the gap function, and the observed crossover to quadratic
behaviour, even at relatively high levels of doping, are consistent with
nodes, but do not completely remove the possibility that the minimum in
the gap function is accidentally very small in the nominally pure
samples, so that any exponential temperature dependence occurs at
temperatures well below those at which measurements have been made.
In the $s+d$ state that is permitted in the orthorhombically-distorted
tetragonal plane, the node positions are shifted away from the Brillouin
zone diagonal; the theoretical predictions are unchanged, except that
there is the logically allowed possibility that at high doping the gap
function will have become so isotropic that it no longer changes sign at
the Fermi surface, a possibility that cannot occur in a pure d-wave
state.   The fact that \Tc\ can be driven to zero with enough electron
irradiation\Ref\giap{J. Giapintzakis \etal, {\sl Determination of the
symmetry of the superconducting pairing state and formation of a low
temperature normal metallic state in YBCO by electron irradiation},
Illinois preprint P-95-09-066.} indicates that this possibility does not
seem to occur, at least in \Yba.  Finally, it is reasonable, but of
course unproven, that the quadratic temperature dependence observed in
BSCCO films is a reflection of disorder imposed on an underlying d-wave
state, just as is the case with YBCO.

{\bf \section{Angle Resolved Photoemission}}

Angle resolved photoemission spectroscopy (ARPES) is potentially one of 
the most powerful techniques for elucidating the superconducting gap
function over the whole Brillouin zone. Unlike other probes of the
energy gap, if the gap function has nodes then the ARPES experiment
should be able to pinpoint exactly where on the Fermi surface
the nodes lie.  This should clearly distinguish superconducting states
which have ``accidental'' nodes, such as extended s-wave states, from
those which have symmetry determined nodes, such as \dx2y2. Unfortunately
the interpretation of the photoemission data is not quite as straight forward
as would appear at first sight, and this has led to some controversy
about which pairing state is actually indicated by the experiments. 
First we shall discuss the current state of the experiments, and then
we shall make some comments on the theory of photoemission from
superconductors and some possible interpretations of the experiments.

{\bf \subsection{Experiments}}

Photoemission only became a viable probe of the  the superconducting gap after
the discovery that Bi$_2$Sr$_2$CaCu$_2$O$_8$ could be cleaved in
vacuum to produce a good quality surface with a sharp photoemission
Fermi edge.
The first measurements, by Imer \etal\rlap,\Ref\imer{J.M. Imer \etal,
\journal Phys. Rev. Lett. &62&336(89).} showed a significant temperature
dependence of the spectra at the Fermi edge, between 105K and 15K. The 
changes were not just the sharpening of the Fermi edge expected from the
Fermi-Dirac distribution but the clear pile up of states
below the BCS gap. They found that the spectrum was roughly
consistent with a BCS density of states convolved with
instrumental broadening.  
They estimated the gap to be $30 \pm 5$ meV,  
assuming a Gaussian broadening with a $20$meV full-width-half-maximum.
In contrast to these results for Bi$_2$Sr$_2$CaCu$_2$O$_8$, photoemission
measurements on \Yba\ show relatively poor evidence for a 
superconducting energy gap. Indeed only rare surface cleaves
show a sharp metallic Fermi edge at 
all\rlap.\REFS\liu{R. Liu \etal, \journal
Phys. Rev. B &46&11056(92).}\REFSCON\tobin{J.G. Tobin \etal,
\journal Phys. Rev. B &45&5563(92).}\REFSCON\schroeder{N. Schroeder
\etal, \journal Phys. Rev. B &47&5287(93).}\refsend
For this reason we shall concentrate mostly on the
Bi$_2$Sr$_2$CaCu$_2$O$_8$ spectra here. 
Very recently  ARPES measurements of the superconducting
gap in \Yba\ have also become available\rlap,\Ref\schabel{M.C. 
Schabel \etal,  preprint. } and these agree quite closely with the
BSCCO data. A more extensive review of photoemission
results in high \Tc\ superconductors is also available 
elsewhere\rlap.\REFS\shenI{ Z.X. Shen and D.S. Dessau, \journal Physics 
Reports &253&1(95).}\REFSCON\shenII{Z.X. Shen, W.E. Spicer, D.M. King,
D.S. Dessau and B.O. Wells, \journal Science &267&343(95).}\refsend

Angle resolved photoemission experiments on Bi$_2$Sr$_2$CaCu$_2$O$_8$
were first measured by 
Manzke \etal\Ref\manzke{R. Manzke \etal,
\journal Europhys. Lett. &9&477(89).} and
Olson \etal\rlap,\Ref\olson{C.G. Olson, \journal Science &245&(89).}
who reported energy gaps of $30$meV and $24$meV respectively.
The first systematic study of the angular dependence of the gap
appeared to support an isotropic energy 
gap\rlap,\Ref\olsonII{C.G. Olson \etal, \journal Solid State 
Commun. &76&411(90).} although
this interpretation was based upon only a limited set of sampling points
in the Brillouin zone.  The first evidence for gap anisotropy was
obtained by Dessau \etal\Ref\dessau{D.S. Dessau \etal,
\journal Phys. Rev. Lett. &66&2160(91).} and by 
Wells \etal\Ref\wells{B.O. Wells \etal, \journal
Phys. Rev. B &46&11830(92).} who reported that the gap decreased by
roughly a factor of two on going from the $\Gamma$M line to the 
$\Gamma$X line in the zone. From the leading edges of the
spectra, the gap function $\Delta({\bf k})$ is much bigger nearer 
the $\Gamma$M line than the $\Gamma$X line.  Fitting the spectral shape,
Wells \etal\ reported that the gap was about $15$meV along $\Gamma$X
compared to $24$meV along $\Gamma$M.
Note that because the Bi$_2$Sr$_2$CaCu$_2$O$_8$ crystal axes are rotated
by $45^{\circ}$ relative to the Cu-O bond directions, 
the $\Gamma$X direction corresponds to the nodal point 
of a \dx2y2 energy gap, using the usual nomenclature
appropriate to the  square Brillouin zone of a single isolated CuO$_2$ plane.

The first evidence that the gap might actually vanish at the \dx2y2
nodal line was
reported by Shen \etal\Ref\shenII{Z.X. Shen \etal, \journal
Phys. Rev. Lett. &70&1553(93).}\  They showed that for points in
the zone along $\Gamma$Y the gap was very small, and the data 
could be consistent with a gap node at this point. A gap node would
be expected along $\Gamma$Y if Bi$_2$Sr$_2$CaCu$_2$O$_8$ were
a \dx2y2 superconductor.  Shen \etal\ found
that the gap at $\Gamma$Y inferred from the data
was strongly sample dependent.  At many of the experimental
$\Gamma$Y points 
$\Delta$ was as small as $2$meV, while at others it was as large as 
$12$meV. 
Interestingly the gap remained highly anisotropic, even
in an oxygen depleted sample which had a reduced \Tc\ of $78$K
and a much smaller gap at $\Gamma$M of $12$meV. Shen \etal\ argued
that their data supported a \dx2y2\ order parameter or a mixed symmetry
state with a large \dx2y2\  component, and that it was qualitatively
incompatible with extended s-wave states.   
The recent results of Schabel \etal\refmark{\schabel}
show a predominantly \dx2y2\ gap anisotropy in YBCO which is essentially
identical to the behaviour found in BSCCO.

Subsequently the debate has focused on whether or not the gap is truly
zero in the $\Gamma$X and $\Gamma$Y directions. 
Ding \etal\Ref\ding{H. Ding, 
\journal Phys. Rev. B &50&1333(94).} found that the gap
was small and consistent with zero along both $\Gamma$X and $\Gamma$Y.
Kelley \etal\Ref\kelley{R.J. Kelley \etal, \journal
Phys. Rev. B &50&590(94).} found that the
gap was non-zero but only $1-2$meV along $\Gamma$Y, while it was larger, 
$4-8$meV, along $\Gamma$X.  The $\Gamma$X and $\Gamma$Y lines would be
equivalent in a 
tetragonal crystal, but are rendered inequivalent in Bi$_2$Sr$_2$CaCu$_2$O$_8$
by the slight difference in a and b lattice constants of $5.4095$\AA
and $5.4202$\AA\rlap.\Ref\bordet{ P. Bordet \etal, \journal Physica C 
&156&189(89).}  The a and b axes 
 are also inequivalent because of the incommensurate
superstructure in the Bi-O 
planes\rlap.\Ref\petricek{V. Petricek, Y. Gao, P. Lee and
P. Coppens, \journal Phys. Rev. B &42&387(90).}

Ma \etal\Ref\maetal{J. Ma \etal, \journal Science &267&862(95).} measured
the temperature dependence of both the $\Gamma$X and $\Gamma$M gaps. They found
strong coupling BCS-like behaviour near \Tc\ for the
large gap ($14-18$meV) at $\Gamma$M, but a much weaker temperature
dependence for the smaller gap (of $7-11$meV) at $\Gamma$X. For
quite a wide range of temperatures below \Tc\ the gap at $\Gamma$X is
indistinguishable from zero, only becoming clearly non-zero below about
$0.8 T_c$.  This behaviour appears to suggest that
a model with more than one order 
parameter component might be required, for example one in which 
the order parameter
at \Tc\ has pure \dx2y2\ symmetry, while a second
order parameter component (such as $s$ or $d_{xy}$) becomes non-zero 
below \Tc.

Ding \etal\Ref\dingIII{ H. Ding \etal, \journal Phys. Rev. Lett. 
&74&2784(95).} measured the angular dependence of the photoemission
spectrum at a large number of points on the Fermi surface.
They found  non-zero gap values along both $\Gamma$X and $\Gamma$Y, with
values similar to those reported by Kelley \etal\refmark{\kelley}\ 
Furthermore, they also claimed that the gap was actually a local maximum
at both $\Gamma$X and $\Gamma$Y, and that it went to zero
at about $10-15^{\circ}$ around the Fermi surface from the \dx2y2\ gap
node. This surprising result would be incompatible with even mixed symmetry
states such as $\hbox{\dx2y2}+s$ and $\hbox{\dx2y2}+id_{xy}$, but seemed to indicate
an extended s-wave state with no \dx2y2\ component.  However
a subsequent 
polarisation analysis by Norman \etal\Ref\norman{M.R. Norman, M. Randeria,
H. Ding, J.C. Campuzano and A.F. Bellman, preprint.} has suggested that
that non-zero gap observed at $\Gamma$X could be due to
Umklapp scattering events caused by the incommensurate modulation.
They argued that in the absence 
of this Umklapp scattering the gap in the $\Gamma$X quadrant would be identical
to that in the $\Gamma$Y quadrant.
Hence the gap is very small (of order 2meV or less)
at the \dx2y2\ node point in both the $\Gamma$X and $\Gamma$Y 
quadrants. Indeed, apart from one data point, the $\Gamma$Y
gap is zero
within experimental error near the \dx2y2\ nodal point.

If these sometimes conflicting results are not disturbing
enough, photoemission 
experiments also present one further puzzle. This is that the total
area of the photoemission spectrum appears to be temperature dependent.
This was first reported by Dessau \etal$^{\dessau}$ who observed that
the total spectral area appeared to decrease (increase) as a function
of temperature along $\Gamma$M($\Gamma$X). Ma \etal$^{\maetal}$
measured the detailed temperature dependence of this change in the spectral
area.  In fact there are no sum rules which constrain this occupied
spectral area (unlike the total occupied+unoccupied spectral area).
It is only in weak coupling BCS theory that the occupied spectral
area remains unchanged below \Tc. Thus
the change in spectral area could be due to strong coupling effects,
or other many-body effects such as quasiparticle renormalisation
or non-Fermi liquid effects. It is not clear at present how the anisotropy
of the area change relates to the gap anisotropy.

Finally, the work of Kelley \etal\Ref\kelleyII{R.J. Kelley \etal,
\journal Phys. Rev. Lett. &71&4051(93).}
indicated that a polarisation
analysis of the photoemission spectra at $\Gamma$X and $\Gamma$Y
was incompatible with either s-wave or \dx2y2 pairing, and implied either a
$d_{xz}+d_{yz}$ or $p_{x}+p_{y}$ state. The interpretation of the 
spectra appears to be flawed, however, since the presence or absence
of a peak in the spectrum was taken to indicate the presence or absence
of a condensate. No such simple relationship exists between the photoemission
spectral shape and the condensate.

{\bf \subsection{Interpretation and Theory}}

In the above discussion, we have taken the various gaps quoted by the
experimental groups at face value. However, at least some of the
 discrepancies
between quoted results appear to be due to different methods of determining
the gap function from the experimental photoemission spectra. 
For this reason it is worth reexamining the theoretical basis of the 
photoemission experiments, in order to
put the various experimental claims in a proper context.

The many body theory of photoemission was formulated by Schaich and
Ashcroft\rlap,\Ref\schaich{ W.L. Schaich and N.W. Ashcroft,
\journal Phys. Rev. B &3&2452(71).} who showed that the photoelectron current
was given by a  Kubo response theory, in terms of a three-current correlation
function.  In practice one neglects effects such as interactions between
the photoelectron and the hole it leaves behind in the solid,
yielding a simpler  formula due to Pendry\Ref\pendry{J. B. Pendry, in
{\sl Photoemission and the Electronic Properties of Surfaces}, ed. 
B. Feuerbacher, B. Fitton and R. Willis (Wiley, New York, 1978).}
$$ I({\bf K}, E_i+\hbar \omega) \propto {\rm Im}
  \left<{\bf K},z| G^+ H' G_h H'^+ G | {\bf K},z\right> .\eqn\photem $$
Here $| {\bf K},z\left.\right>$ is an electron state with momentum $\hbar{\bf K}$
parallel to the surface at the plane of the detector, $z$, outside 
the surface.
$G$ is the propagator for the photoelectron with energy $E_i+\hbar\omega$.
$H'$ is the electron-photon interaction Hamiltonian 
$$ H' = \int d^3r {\bf A}({\bf r},t)\cdot{\bf J}({\bf r}) \eqn\hint $$
where ${\bf J}({\bf r})$ is the current operator.  $G_h$ is the propagator
for the hole which is excited inside the solid, at energy, $E_i$.  
In the Pendry theory the propagators $G$ and $G_h$ are fully dressed 
by the appropriate self-energy, so many-body effects are included, as well
as the full surface electronic structure\rlap.\Ref\sriva{R.H. Williams,
G.P. Srivastava and I.T. McGovern, in {\sl Electronic Structure of Surfaces},
M. Prutton (ed.) (Adam Hilger, Bristol, 1984).}

In the analysis of photoemission data of the high \Tc\ superconductors
several critical assumptions are usually made:
\item{(1)} 
It is assumed that the photoelectron
propagator, $G$, is unchanged by the superconductivity, because of its
high energy, and so the effects of superconductivity are entirely contained
in the hole propagator $G_h$.  
\item{(2)} The photoelectrons are assumed to be
excited sufficiently far inside the surface that the electronic structure and
superconducting order parameter are essentially bulk-like.  
\item{(3)} 
Any dependence of the matrix elements of the current operator with
momentum around the Fermi surface, or with initial state energy, is neglected.
With this assumption the angle resolved photoemission spectrum intensity 
becomes
proportional to the single particle spectral weight function, 
$A({\bf k},E_i)$.
\item{(4)}  For the cuprates, 
the electronic structure is assumed to be essentially two-dimensional,
and so the observed spectrum as a function of the parallel momentum
${\bf K}$ is assumed to be indicative of the full 
Fermi surface and gap function, neglecting any possible
band dispersion and gap modulation along the c-axis.
\item{(5)} Various fitting schemes are used to extract the angle-dependent
gap function $\Delta({\bf k})$ from the spectra, using either the leading 
edge
of the spectrum or a fit to a strong coupling theory lineshape. 
It seems that differences in fitting procedures, as well as variation among
samples (e.g. oxygen content) accounts for the differences in numerical
gap values found by the various groups. It is also important 
to point out that these fits
neglect possible angular dependences in other relevant quantities, such
as the quasiparticle lifetime, $\tau({\bf k})$. An angular
dependent $\tau({\bf k})$ could lead to increased sharpening
of the spectral edge at some points in the zone, hence leading to
systematic errors in the extracted values of $\Delta({\bf k})$.
One should therefore be very wary of quoted measurement error bars, since these do not take into account possible systematic errors.

While each of the above points may have its justifications, together they
raise several questions about whether the experiments are indeed determining
the gap parameter, $\Delta({\bf k})$, with the precision usually implied.
 Such uncertainties may explain why quoted gap values from
different groups sometimes have non-overlapping error bars. Of course varying 
sample and surface qualities may also be a factor, as is clear from 
the work of Shen \etal\refmark{\shenII}\  In particular they
bear on the key question of whether $\Delta({\bf k})$ vanishes at
the \dx2y2\ nodal lines, or is merely small there.  At the present time, 
in our opinion, all that can be said is that the photoemission data
appears to be {\it consistent} with a pure \dx2y2 order parameter.
The experiments {\it cannot rule out} a small non \dx2y2\ component,
 such as $s$ or
$d_{xy}$, or a highly anisotropic s-wave state
such as that of Ref. \REF\temmerman{
W.L Temmerman, D. Szotek, B.L. Gyorffy and O.K. Andersen,
{\sl Phys. Rev. Lett.}, to be published.}\temmerman.
 Any such anisotropic gap
must vary in magnitude by a factor of at least two (and probably
more like a factor of 10)  between $\Gamma$M,
$\Gamma$X and $\Gamma$Y.   

More work on the theory of photoemission in superconductors is certainly
warranted. The prospect
 that the detailed spectral shapes for $A({\bf k},E_i)$
may be used to determine the McMillan $\alpha^2F$ function is very
exciting\rlap.\Ref\arnold{
G.B. Arnold, F.M. Muller and J.C. Swihart, \journal Phys. Rev. Lett.
&67&2569(91).} However, the observed spectrum is
highly convolved, with 
instrumental broadening in both energy and  momentum in addition to the strong
coupling self-energy effects\rlap.\Ref\kimlovich{G.V. Klimovich,
\journal Zh. Eksp. Teor. Fiz. &105&306(94) [\journal JETP &79&169(94)].}
It seems unlikely that any deconvolution process will be able to extract the
full angle-dependent 
strong coupling $\Delta({\bf k},\omega)$ with an accuracy of more 
than $\pm 1-2$meV, given
the typical instrumental energy broadening of $15-20$meV.
This would be sufficient to eliminate pure \dx2y2 pairing
if the gap were found to be
non-zero within this error bar at the nodal lines. However it is unlikely 
ever to be possible to completely eliminate the possibility of
small ($<2$meV) non-zero gap values at the
\dx2y2\ nodes using ARPES experiments.

\section{Neutron Scattering}

The possibility that neutron scattering could be used to
probe the pairing state in unconventional superconductors was first
suggested, to our knowledge, by Lu in 1992\rlap.\Ref\jplu{J.P. Lu,
\journal Phys. Rev. Lett. &68&125(92).}
Because the neutron scattering intensity is essentially
the spin susceptibility, 
$$S({\bf q},\omega)= \left ( 1+{1 \over e^{\hbar\omega/k_BT}-1} \right )
{\rm Im}\chi({\bf q},\omega), \eqn\sqomega $$
neutron scattering  
obtains information which is very similar to NMR $1/T_1$
measurements.  However Lu pointed out that
neutron scattering could provide a very clean cut test of
unconventional superconductivity because of the experimental
freedom to tune precisely the wave vector and frequency, unlike NMR
which is wave-vector integrated and in the limit $\hbar \omega
\rightarrow 0$.
The neutron scattering intensity $S({\bf q},\omega)$
should be significantly affected by
superconductivity provided $\hbar \omega < \Delta$. 
In unconventional superconductors with nodes in the gap function,
Lu predicted that sharp ${\bf q}$ dependent resonances
would occur in $S({\bf q},\omega)$ because of scattering 
between states at the gap nodes, as illustrated in
\FIG\figlu{Resonances in
 neutron scattering intensity, $S({\bf q},\omega)$
for a \dx2y2\ superconductor.
For $\hbar \omega < k_B T_c$ the intensity
$S({\bf q},\omega)$ should show sharp peaks at the wave vectors
drawn. Taken from Ref. \jplu.} Fig. \figlu.  
In principle, by measuring the positions of the resonance peaks 
in $S({\bf q},\omega)$ one could unambiguously determine
precisely where on the Fermi surface the gap nodes occur, and hence 
the pairing state.  
Lu pointed out
that the conditions for the appearance of these
resonances: $ T < \omega < T_c < \Delta$ , were experimentally
realisable  in the high \Tc\ superconductors.

Experimental neutron scattering work on the cuprate
superconductors has been active ever since their discovery.
See, for example, the excellent review by Birgeneau and
Shirane in Volume 1 of this series\Ref\birgeneau{
R.J. Birgeneau and G. Shirane, in {\sl Physical Properties of
High Temperature Superconductors vol 1}, D.M. Ginsberg (ed.)
(World Scientific, Singapore 1989), Chapter 4, p. 151.}  or any of the more recent
reviews\rlap.\REFS\shirane{G. Shirane, \journal Physica B &215&1(95).}\REFSCON\bourges{G. Bourges \etal, \journal
Physica B &215&30(95).}\refsend  Much of this work has
focused on the magnetic properties of the antiferromagnetic
phases of the materials, and on the antiferromagnetic
fluctuations (paramagnons) in the normal metallic state. It has been
difficult to obtain spectra in the superconducting state because of the
 necessity of growing large enough single crystals which are
doped homogeneously. Nevertheless superconducting state
neutron scattering spectra are now 
available, for example, 
in La$_{1.85}$Sr$_{0.15}$CuO$_4$  
with $T_c=37$K\Ref\yamada{K. Yamada \etal, \journal
Phys. Rev. Lett. &75&1626(95).}
and for \Yba\ crystals with  $T_c$  of
93K\rlap.\Ref\fong{ H.F. Fong \etal, \journal Phys. Rev. Lett.
&75&316(95).}  The neutron scattering results
from the 214 and 123  materials are quite different
from each other,
in both the normal and the superconducting state.  In both materials
there is evidence of strong antiferromagnetic spin fluctuations
in the normal state,
showing up at peaks in $S({\bf q},\omega)$ near 
${\bf q}=(\pi/a,\pi/a)$.
However in the 214 materials these fluctuations are incommensurate, peaking
at wavevectors ${\bf q}=(\pi/a,\pi(1\pm\delta)/a)$ 
and ${\bf q}=(\pi(1\pm \delta)/a,\pi/a)$ 
with
$\delta \sim 1/4$, while in 123 they remain commensurate
at $(\pi/a,\pi/a)$.  In 123, as a function of frequency and temperature 
$S({\bf q},\omega)$
shows evidence for a gap of order $3.5k_BT_c$.
However this gap is usually believed to be  
a magnetic gap in the paramagnon spectrum, i.e. a spin gap,
 rather than a superconducting gap.
In 214, by contrast, there is no such gap and there is
evidence for low energy excitations at least down to 
$1-2$meV\rlap.\REFS\thurston{T.R. Thurston \etal, \journal
Phys. Rev. B &46&9128(92).}\REFSCON\mason{T.E. Mason \etal,
 \journal Phys. Rev. Lett.
&71&919(93).}\REFSCON\matsuda{ M. Matsuda \etal, \journal
Phys. Rev. B &49&6959(93).}\refsend

Let us now consider in detail the superconducting state neutron
scattering in the 214 materials, and the constraints
it places on the superconducting pairing symmetry.
As already mentioned, $S({\bf q},\omega)$ is strongly peaked
at the four incommensurate vectors 
${\bf q}_{\delta}=(\pi/a,\pi(1\pm\delta)/a)$
and $(\pi(1\pm \delta)/a,\pi/a)$. These vectors clearly do
{\it not\/} connect the gap nodes of a d-wave superconductor
as in Fig. \figlu, since these
 would be of the form $(1\pm\epsilon)(\pi/a,\pm\pi/a)$. Instead the
incommensurate peaks must be associated with nesting vectors
of the Fermi surface\refmark{\mason} and the peaks
are clearly a normal state property since they are
present both above and below $T_c$.  On cooling the sample into the
superconducting state the incommensurate peaks
remain at the same position, ${\bf q}_\delta$, and have essentially
unchanged width in ${\bf q}$\rlap.\refmark{\mason}
Thurston \etal\refmark{\thurston} and 
Matsuda \etal\refmark{\matsuda}
found that at the
peak maximum  ${\rm Im}\chi({\bf q}_\delta,\omega)$ remained independent
of temperature below \Tc\ for  $1.5  \leq \omega \leq 6$meV. 
In contrast Mason \etal\refmark{\mason} found a gradual
decrease in intensity below \Tc.
Despite this disagreement, in either case, it is clear that
the spectra are {\it inconsistent with an isotropic gap}.
The considerable intensity in ${\rm Im}\chi({\bf q}_\delta,\omega)$ 
at low frequencies and below \Tc\ implies the presence of excitations 
with energies well below the expected isotropic BCS gap of
$\Delta \sim 10$meV.   Mason \etal\refmark{\mason} argued that
these low energy excitations could not be taken as direct  evidence
for \dx2y2\ pairing, because the wave vector dependence of $S({\bf q},\omega)$
 was the
same in the normal and superconducting states.

The most recent experiments on La$_{1.85}$Sr$_{0.15}$CuO$_4$ by
Yamada \etal\refmark{\yamada} have clarified this picture
somewhat. They found that
the incommensurate peak intensity,  ${\rm Im}\chi({\bf q}_\delta,\omega)$,
was essentially independent of $T$ for $\hbar\omega \geq 4.5$meV,
and decreased below $T_c$ for $\hbar\omega \leq 3$meV. This implies that
there {\it is} an energy gap at the wave vector ${\bf q}_\delta$ and that it
is close to
$3.5$meV. Since the neutron scattering intensity at wave vector
{\bf q} depends on electronic excitations from ${\bf k}$ to 
${\bf k}+{\bf q}$\rlap,\refmark{\jplu} one can define a {\bf q}
dependent effective gap as
$$ \Delta_{eff} = \min_{\bf k} | E_+({\bf k}+{\bf q}) - E_-({\bf k})|  ,
\eqn\deltaeff$$
where as usual 
$E_\pm({\bf k})=\pm \sqrt{(\epsilon({\bf k})^2 + |\Delta({\bf k})|^2)}$.
Yamada \etal\ noted that if one assumes a \dx2y2 gap function
$$
\Delta({\bf k})=(\Delta_0/2) (\cos{k_x a} - \cos{k_y a}),\eqn\dgap $$
then the effective excitation gap
at the incommensurate vector ${\bf q}_\delta$ will be about $3.5$meV
if $\Delta_0 = 10$meV.  This value of $\Delta_0$ would be consistent
with several theoretical predictions based upon models of
\dx2y2\ pairing\rlap.\refmark{\yamada}

Of course, these observations could not 
be said to completely rule out an anisotropic s-wave state, whether
with gap nodes or not. Any gap function with $\Delta_{eff}=3.5$meV
at ${\bf q}_\delta$ should also give similar behaviour. 
What would be needed to distinguish clearly \dx2y2\ pairing
from the anisotropic s-wave states is a more detailed
analysis of the neutron scattering intensities precisely
at the gap node wave vectors (as shown in Fig. \figlu),
rather than at the nesting vectors ${\bf q}_\delta$.
In this case one would have to see $\Delta_{eff}=0$ 
for a \dx2y2\ pairing state. Mason \etal\refmark{\mason}
did scan along a line ${\bf q}=(1\pm\epsilon)(\pi/a,\pi/a)$
in the superconducting state, where one might expect to see \dx2y2\
gap nodes, but the intensity was rather featureless. If the
sharp resonances predicted by Lu in Ref. \jplu\  are present,
then they have a relatively low spectral weight compared to
the strongly enhanced antiferromagnetic scattering around ${\bf q}_\delta$. 
Future experiments with higher resolution may be able
to search more closely for these peaks. It should also be noted that
Lu's theoretical predictions did not include the effects of strong
electron-electron interactions, or of disorder. It is quite
likely that disorder would smear out the sharp ${\bf q}$ structure
he predicted, although one should still see $\Delta_{eff}=0$
in a \dx2y2\ superconductor.

As mentioned above, the neutron scattering data in the YBa$_2$Cu$_3$O$_{6+x}$
compounds  differs considerably from that in La$_{2-x}$Sr$_x$CuO$_4$.
The structure factor $S({\bf q},\omega)$ has a broad, commensurate,
peak around ${\bf q}_{AF}=(\pi/a,\pi/a)$.  In contrast to the lanthanum
cuprate there does not appear to be a continuum of low energy
excitations at this wave vector, but  rather there exists
an energy gap, $E_g$\rlap.\refmark{\bourges} The gap is
around $33-35$meV in fully oxygenated samples, and decreases
as the oxygen content is reduced, roughly following
$E_g \sim 3.5 k_B T_c$.  The exact nature of this
gap is controversial, but it is generally believed
to be associated with a gap in the spin wave spectrum which
develops because of the loss of long range antiferromagnetic order
in the metallic state. It is apparently not directly associated with
the superconducting gap, although it is generally of the same
order of magnitude.
Interestingly, if we assume that \Yba\ is a \dx2y2\ superconductor
with a gap given by eq. \dgap\ with $\Delta_0=25$meV
as found in photoemission\rlap,\refmark{\schabel}
then the effective excitation gap $\Delta_{eff}$ in Eq. \deltaeff\
for the antiferromagnetic wave vector ${\bf q}_{AF}$ is
expected to be around $30$meV,  for the Fermi surface 
and gap function found
in Ref. \schabel.  
Unfortunately it appears probable\refmark{\bourges} 
that  for YBa$_2$Cu$_3$O$_{6+x}$ there are two
distinct gaps, one magnetic and one superconducting, which
have similar magnitudes at ${\bf q}_{AF}$.  This considerably
complicates the interpretation of the neutron scattering data
and makes an unambiguous identification of the pairing state
difficult.

A second feature of the neutron scattering spectra ---
a resonance
at $\hbar\omega=41$meV ---
 is more
clearly associated with the superconductivity\rlap.\Ref\mook{H.A. Mook
\etal, \journal Phys. Rev. Lett. &70&3490(93).}\ 
This peak in $S({\bf q},\omega)$
above the ``spin gap" appears only in optimally oxygenated samples
at ${\bf q}_{AF}$.  It is quite sample dependent\refmark{\shirane}
and  occurs at an oxygen content of O$_{6.9}$-O$_7$. The peak
intensity increases suddenly below \Tc\ and then saturates
for $T \ll T_c$, clearly following the
opening up of the superconducting gap.
A detailed study of this resonance
was carried out by Fong \etal\refmark{\fong}\ They observed that
the peak had both magnetic and non-magnetic (phonon) contributions,
and it was the sudden appearance of the magnetic scattering 
below \Tc\ which led to the increase in intensity.
They argued that the appearance of magnetic
scattering in this channel was strong evidence for a sign change in the
gap function, such as one expects in \dx2y2\ pairing. 
The argument is based upon the BCS coherence factor for 
neutron scattering exciting electrons from ${\bf k}$ to ${\bf k}+{\bf q}$.
It is most clearly demonstrated by considering the weak coupling
BCS spin susceptibility given by Lu\rlap:\refmark{\jplu}
$$
 \chi({\bf q},\omega)  =   \sum_k  {1 \over 4}  \left\{2\chi_1 +
 \chi_2 + \chi_3\right\},\eqn\nigelchi
$$
where 
$$\eqalign{
\chi_1 & \equiv \left ( 1 + {\epsilon({\bf k}+{\bf q})\epsilon({\bf k}) +
\Delta({\bf k}+{\bf q})\Delta({\bf k}) \over E({\bf k}+{\bf q})E({\bf k})}
\right  ) \left (
{ f({\bf k}+{\bf q})-f({\bf k}) \over \omega - [E({\bf k}+{\bf q})
-E({\bf k})] + i\Gamma } \right ) \cr
 \chi_2 & \equiv \left ( 1 - {\epsilon({\bf k}+{\bf q})\epsilon({\bf k}) +\Delta({\bf k}+
{\bf q})\Delta({\bf k}) \over E({\bf k}+{\bf q})E({\bf k})} \right )
\left ({ 1-f({\bf k}+{\bf q})-f({\bf k}) \over \omega - [E({\bf k}+{\bf q})
+E({\bf k})] + i\Gamma } \right ) \cr
 \chi_3 & \equiv\left ( 1 - {\epsilon({\bf k}+{\bf q})\epsilon({\bf k}) +\Delta({\bf k}+
{\bf q})\Delta({\bf k}) \over E({\bf k}+{\bf q})E({\bf k})} \right )
\left ({ f({\bf k}+{\bf q})+f({\bf k}) -1 \over \omega + [E({\bf k}+{\bf q})
+E({\bf k})] + i\Gamma } \right )
\cr} \eqn\jpluchi $$

In the $T \rightarrow 0$ limit, the first term dominates,
and either $E({\bf k}+{\bf q})$ is above the Fermi level
and $E({\bf k})$ is below the Fermi level, or vice versa.
In either case the product $E({\bf k}+{\bf q})E({\bf k})$
is negative. Precisely on the Fermi surface $\epsilon({\bf k}+{\bf q})=
\epsilon({\bf k})=0$, and thus the prefactor for the scattering
process is:
$$    1 - {\Delta({\bf k}+{\bf q})\Delta({\bf k}) \over 
 | \Delta({\bf k}+{\bf q})\Delta({\bf k})| } .\eqn\prefactor $$
Clearly this vanishes if $\Delta ({\bf k}+{\bf q})$ and
$\Delta({\bf q})$ have the same sign, as in s-wave pairing,
but is non-zero if they have different signs, as in
\dx2y2\ pairing.   Unfortunately, since the exact nature of this
$41$meV peak remains uncertain, this argument must be treated with
some caution. In particular the resonance appears
to be associated with the bi-layer structure of 
\Yba\rlap,\refmark{\mook}
which leaves open the question of which Fermi surface sheets 
correspond to the initial and final states ${\bf k}$ and ${\bf k}+{\bf q}$.
An s-wave gap which changed signs between the two Fermi surfaces
might also be compatible with the existence of the $41$meV 
peak\rlap.\REFS\bulutscalapino{ N. Bulut, D.J. Scalapino and
R. Scalettar, \journal Phys. Rev. B &45&5577(92).}\REFSCON\lichtenstein{
A.I. Lichtenstein, I.I. Mazin and O.K. Anderson, \journal
Phys. Rev. Lett. &74&2303(95).}\REFSCON\dahm{T.Dahm and
L. Tewordt, \journal Physica C &253&334(95).}\refsend

In summary, the neutron scattering data have not proved as
simple a test of unconventional pairing in the cuprates as originally
thought. The difficulty arises from the combination of several
factors, most importantly the interaction between the 
particle-hole excitations of the superconductor and the 
magnetic fluctuations present in both normal and superconducting states.
This implies that weak coupling BCS calculations\rlap,\refmark{\jplu}
will not capture the full physics of ${\rm Im}\chi({\bf q},\omega)$;
instead calculations are necessary which explicitly take into
account the strong correlations. Many groups have
carried out such calculations using a variety of techniques\REFS\monthoux{
P. Monthoux and D.J. Scalapino,  \journal Phys. Rev. Lett. &72&1874(92).
}\REFSCON\si{Q. Si, Y. Zha, K. Levin and J.P. Lu, \journal
Phys. Rev. B &47&9055(93).
}\REFSCON\dahmprl{ T. Dahm and L. Tewordt, \journal Phys. Rev. Lett.
&74&793(95).}\REFSCON\onufrieva{ F. Onufrieva, \journal
Physica B &215&41(95).}\refsend but it is beyond our scope to
discuss them in detail. However, good  
qualitative, and sometimes
quantitative, agreement with the neutron scattering data
can be obtained in both the normal and superconducting states
assuming a single band Hubbard model or bilayer Hubbard model.
Furthermore if this calculated spin susceptibility 
is used as the pairing interaction, usually in an Eliashberg type formalism,
a self-consistent picture of superconductivity appears to
emerge. In a single layer this pairing state must always be
\dx2y2, while in the bilayer both \dx2y2 and s-wave solutions appear
with similar values of \Tc.
This s-wave state has a gap function which changes sign between the two
Fermi surface sheets\rlap.\refmark{\bulutscalapino - \dahm}
At present the netron scattering appears to {\it be consistent} with
a \dx2y2\ pairing state in both La$_{1.85}$Sr$_{0.15}$CuO$_4$ and
\Yba\, but {\it cannot unambiguously rule out} some of these
other alternatives.

In the future we hope that the neutron scattering data will
put more precise limits on the possible pairing states. 
One key issue is whether or not there is a true gap in the spectrum
$S({\bf q},\omega)$, or whether there is a tail of low energy
excitations corresponding to the d-wave gap nodes.
It is surprising that such low energy excitations have not been seen
in neutron scattering in \Yba\, unlike La$_{1.85}$Sr$_{0.15}$CuO$_4$,
especially when there is plenty of other evidence for gapless
behaviour as seen in Raman scattering, infra-red response
penetration depth and so on. It is possible that the relevant
region of the ${\bf q}$, $\omega$ space (${\bf q} \sim (1\pm \epsilon)
(\pi/a,\pi/a)$, $\hbar\omega \ll \Delta_0$) has not been searched
thoroughly enough on good quality samples, or that the intensity of
these excitations is just much weaker than the strong antiferromagnetic
fluctuations near $(\pi/a,\pi/a)$ and they are lost in the background.
Mook \etal\refmark{\mook} did see a weak  tail of low energy
excitations below the ``spin gap'', but other groups generally
consider these to be part of the background\rlap.\refmark{\bourges}
A clarification of this issue would considerably strengthen the claims
that the neutron scattering supports the \dx2y2\ pairing hypothesis.

{\bf \section{Raman Scattering}}

Raman scattering provided some of the first strong
evidence for nodes in the superconducting gap in \Yba,
as reviewed by AGR. The data available by 1988 
clearly showed that the continuum of electron-hole pair excitations
extended well down below the superconducting ``gap'', which
appeared as a broad edge in the spectra. The 
interpretation of the continuum as due to electron-hole pair excitations
was clear cut because of the Fano lineshapes of several of the
phonon modes\rlap.\Ref\cooper{S.L. Cooper, M.V. Klein, B.G. Pazol,
J.P. Rice and D.M. Ginsberg, \journal Phys. Rev. B &37&5920(88).}\
Monien and Zawadowski\Ref\monien{H. Monien and A. Zawadowski,
\journal Phys. Rev. Lett. &63&911(89).}\ 
argued (based on a specific microscopic model)
that this spectrum was consistent with a $d_{xy}$ state among others,
but not with \dx2y2\ or s-wave.

Since then, the superconducting state
Raman measurements have been extended from \Yba\
to, for example,
Bi$_2$Sr$_2$CaCu$_2$O$_{8+\delta}$\rlap,\Ref\staufer{T. Staufer,
R. Nemetschek, R. Hackl, P. Muller and V. Veith,
\journal Phys. Rev. Lett. &68&1069(92).}\ again showing
the continuum of excitations extending well below the nominal gap.

Clearly this continuum is {\it prima facie\/} evidence against
an isotropic gap, such as isotropic s-wave or $d_{x^2-y^2}
+ \alpha i d_{xy}$ with $\alpha \sim 1$, and in this sense is
consistent with the photoemission results described above.
In order to ask whether the data place stronger constraints on
the gap function it is necessary to predict the Raman spectra
for various anisotropic gaps, such as \dx2y2, or anisotropic
s-wave states. Such a calculation has recently been 
provided\rlap.\REFS\devereaux{ T.P. Devereaux and D. Einzel,
\journal Phys. Rev. B &51&16336(95).}\REFSCON\devereauxi{
T.P. Devereaux, D. Einzel, B. Stadlober, R. Hackl, D.H. Leach,
and J.J. Neumeier, \journal Phys. Rev. Lett. &72&396(94).}\refsend
The results showed very good quantitative agreement for
the Raman spectra {\it including its polarisation dependence}
for \dx2y2\ while both isotropic and anisotropic s-wave
were qualitatively inconsistent. 
Of particular interest is
the observed $\omega^3$ low frequency power law 
of the $B_{1g}$ spectrum (note here $B_{1g}$ etc. labels the
symmetry of the excitations measured, and is not
the pair state symmetry), compared to the linear power law
for $A_{1g}$ and $B_{2g}$ spectra. 
It is important to realise that this change in power law
is not just a density of states effect, but arises from the
BCS-like coherence factors in the Raman scattering intensity.
Thus the calculated \dx2y2\ spectra is completely different from
the spectra calculated with the
$\Delta({\bf k}) \sim | d_{x^2-y^2} |$ gap function
proposed by Chakravarty \etal\refmark{\chakravarty}\
even though it has exactly the same density of states.
In our view these results not only strongly support the 
overall picture emerging from photoemission that $| \Delta({\bf k})|
\sim | k_x^2 - k_y^2 |$ but also {\it force us to the conclusion
that $\Delta({\bf k}) \sim k_x^2-k_y^2$}. 
It is also interesting to note that the calculation
clearly distinguishes \dx2y2\ from $d_{xy}$ states, since the
absolute orientation of the crystal axes is known
and the $B_{1g}$ and $B_{2g}$ spectra are inequivalent.

{\bf\chapter{Quantum Phase Interference: General Principles}}

	In the last couple of years a new class of experiments has been
developed which looks {\it directly} at the symmetry of the order parameter
as a function of its arguments\rlap.\Ref\wolffot{We do not
attempt to discuss the so-called Wohlleben effect, because its
observation in Nb (P. Kostic \etal, to appear in {\sl Phys. Rev. B}) 
suggests that it is not peculiar to unconventional superconductivity.}\  
Experiments of this type were
originally proposed in the context of heavy-fermion superconductors
such as UPt$_3$ which were suspected of forming pairs in a p-state, but
were never successfully conducted (or indeed as far as we know
attempted) there; this is probably not an accident, given 
that there is an important difference between the p- and d-wave cases.
The principle of the experiment is based on the Josephson effect, so we
start with a brief review of the latter.

Imagine a pair of classic (isotropic BCS)
superconductors 1 and 2 joined by some ``weak link,'' e.g. a
thin, clean layer ($\sim 10$ \AA) of oxide (``ideal tunnel oxide
junction'').  The order parameter in each of the bulk superconductors has no
interesting dependence on relative coordinate and may for the moment be
characterised in each case by a single complex number
$\Psi_{1}$($\Psi_{2}$).  Now the junction can transmit, in some
intuitive sense, not just quasiparticles but also Cooper pairs, and one
might therefore imagine that there could be a term in the free energy
which involves both $\Psi_{1}$ and $\Psi_{2}$.  The lowest-order such
term of interest must, by gauge invariance, be of the form 
$$
E^{(2)}_{J}=-{\rm const.}(\Psi^{*}_{1}\Psi_{2} + c.c.)\equiv -J \cos  \Delta
\varphi, \eqn\josephson 
$$ 
where $\Delta \varphi \equiv$ arg
($\Psi_{1}/\Psi_{2}$)  is the relative phase of the Cooper pairs on the
two sides and $J$ is a function of $T$ which turns out to be related to the
critical supercurrent $I_c$ of the junction by the formula 
$$ \eqalign{
| J |= & \ I_{c}\varphi_{0}/2 \pi \cr (\varphi_{0} \equiv h/2e =
& \ \hbox{flux quantum}). \cr} \eqn\josephsontwo $$ 
It is usually assumed
(cf. below) that $J$ is positive, (and this is almost certainly the case
for an ideal tunnel oxide junction), but in principle there can be
types of weak link for which it is negative\rlap,\Ref\bulaevskii{L.N. 
Bulaevskii \etal, \journal Solid State Commun. &125&1053(78).} 
in which case the junction
is called a ``$\pi$-junction'' (since the equilibrium value of the
relative phase $\Delta \varphi$ is $\pi$).  The existence of a coupling
energy of the form \josephson\ has a number of well-known consequences, most of
them already pointed out in Josephson's original 1962 paper:
\item{(1)} Under zero dc voltage the junction
can carry a dissipationless current (supercurrent) of any magnitude up
to the critical current I$_c$ (Eqn. \josephsontwo).  
\item{(2)} Since the
time-dependence of $\Delta \varphi$ is given by the Josephson relation
$$ d\Delta \varphi/dt  = 2eV/\hbar\equiv
2\pi V/\varphi_{0},\eqn\joseppp$$
under a finite dc voltage the junction will
carry an ac supercurrent at frequency $2eV/\hbar$., which can be
detected by its resonance with a superimposed ac field (``Shapiro
steps''). 
\item{(3)} If a uniform magnetic field B is applied in the plane of the junction,
then the total critical current through the latter shows a ``Fraunhofer"
modulation as a function of the quantity $\Phi_J = B A_{eff}$, where 
$ A_{eff}$  is the
``effective area" of the junction:
$$  I_c(\Phi_J) = I_c \left| { \sin{(\pi\Phi_J/\varphi_0)} \over 
 (\pi\Phi_J/\varphi_0)}\right|  . \eqn\fraunhofer $$ 
The ``effective area" $A_{eff}$  is given by the product of the length
perpendicular to the field times the quantity ($d + \lambda_1 + 
\lambda_2$), where  $d$ is
the thickness of the junction and $\lambda_1$, $\lambda_2$ 
 are the London penetration depth
of the two bulk superconductors.
 \item{(4)} If the junction is inserted in a bulk superconducting
ring (thickness $\gg$ London penetration depth: \ ``rf SQUID''
configuration), the quantity $\Delta \varphi$ must satisfy the
condition $\Delta \varphi$ = 2$\pi\Phi/\varphi_{0}$ (modulo $2\pi$),
where $\Phi$ is the total flux trapped through the ring.  In the limit
of a ``strong'' junction ($J \gg\varphi_{0}^{2}/2L$, where $L$ is the
ring self-inductance)  minima with respect to $\Phi$ occur close to 
integral multiples of $\varphi_{0}$, so the trapped flux is {\it quantised}
in units of $\varphi_0$.  
\item{(5)} Most importantly in the present context,
suppose we incorporate two junctions in a ring, as part of a 
circuit, and trap a flux through the ring (``dc SQUID''
configuration).  Then the {\it difference} between the phase
differences across the individual junctions will be $2\pi
\Phi/\varphi_{0}$, and since the elements act in parallel the total
critical current will have maxima at integral values of
$\Phi/\varphi_{0}$, including zero, and minima at half-integral
values.  Note that this conclusion rests crucially on two implicit
assumptions: 
 
\itemitem{(a)} there are no sources of phase difference
``around the ring'' other than the trapped flux, and
\itemitem{(b)} the junctions
are both ``normal'' (or both ``$\pi$,'' but not mixed).

\noindent Note also that the flux which enters the result is the
{\it trapped} flux, which may not necessarily be equal to that
applied externally as the ring may set up its own circulating
currents.

We next consider the case of a junction where one or both of
the bulk superconductors involved shows ``unconventional'' pairing.
The possible forms of the bilinear term in the free energy analogous to
\josephson\ are now strongly constrained by symmetry considerations: The
relevant expression must be invariant under any transformation under
which the Hamiltonian describing the junction is invariant (Principle
A)\rlap.\refmark{\annett,}
\Ref\yip{S-K. Yip, O.F. de Alcantara Bonfim and P. Kumar,
\journal Phys. Rev. B &41&11214(90).}\ For
example, if the junction Hamiltonian \josephson\
is invariant under rotation in
spin space, it cannot give rise to bilinear coupling between a spin
triplet order parameter on the one side and a spin singlet one on the other.
Independently of this, if the junction is invariant under inversion
then an odd-parity order parameter cannot couple to an even-parity one.  This latter
result has immediate application to the case of UPt$_3$ and possibly
other heavy-fermion systems:  although it is very unlikely that the
junction is strictly invariant under parity at a microscopic level, it
is likely to be much more so after macroscopic averaging, and therefore
any overall Josephson coupling between an odd-parity superconductor and
a conventional one is likely to be (a) small and (b) random in sign
(that is, the predominant coupling of the (single) order parameter of the
conventional superconductor is as likely to be to the ``incoming'' lobe
of the p-wave order parameter as to the ``outgoing'' one).  It is therefore not
surprising (if one believes UPt$_3$ is indeed odd-parity) that there
has been no reliable observation of a Josephson effect between this
material and any conventional superconductor: moreover one would
predict that even if we were to learn how to make junctions which give
a finite effect, the SQUID experiments proposed for this system would
in real life give random and hence not very convincing results.

By contrast, the predictions for Josephson contact between a d-wave
superconductors and an s-wave one are clear-cut.  First consider the
case where the junction is parallel to the ab-plane, and assume that
the junction itself is invariant under $\pi$/2 rotation.  Then
{\it no} d-wave state can show a bilinear coupling to the
s-wave state and one should get no Josephson supercurrent (or more
accurately no lowest-order one, see below).  In addition, even in
orthorhombic symmetry no coupling can occur between the $s^-$ or $d_{xy}$
states and the s-state of the conventional superconductor  (if the
junction is invariant under reflection in the orthorhombic crystal
axes).  By contrast, if the plane of the junction is perpendicular to
the ab-plane then there is no symmetry argument which forbids coupling
of the s-wave order parameter predominantly to (e.g.) the d-wave lobe
which is
oriented perpendicular to the junction, so that the existence of a
finite Josephson current between YBCO and classic superconductors
in this geometry (which has been long known) is perfectly compatible
with d-wave pairing in the former.  (It should be emphasised that the
above symmetry arguments constrain only the terms in the free energy
which are {\it bilinear} in the $\Psi$'s.  In most cases one
cannot exclude the possibility that effects of higher order in the
$\Psi$'s may give rise to a finite Josephson supercurrent.  However, in
such cases the periodicity observed in e.g. the ac Josephson radiation
or the flux trapping behaviour  (or the Fraunhofer diffraction pattern,
cf.  below)  would be $\varphi_{0}/n$, where $n$ is an integer greater
than 1.  Most of the experiments to be discussed have explicitly
checked that the periodicity is indeed $\varphi_{0}$ and thus
discounted this possibility, so we will assume in the following that
any Josephson effect observed is indeed bilinear).

Let us now consider the general case.  In the following we shall, in
common with most if not all of the existing literature on this subject,
assume that the dependence of the order parameter on the center-of-mass coordinate
${\bf R}$ is important only in so far as it affects the overall phase.  More
specifically, we shall assume that the (spin singlet) order parameter
$\Psi({\bf r}_1,{\bf r}_2)$ can for our purposes be written in the form
$$ \Psi({\bf r}_1,{\bf r}_2) = F({\bf \rho}) f({\bf R}), \eqn\factor $$
where ${\bf \rho}$ is the relative coordinate ${\bf r}_1-{\bf r}_2$.
The quantity $F_{\bf k}$ is defined as  just the Fourier transform of
$F({\bf \rho})$.  We emphasize that the assumption \factor\
is made only for the
purposes of simplifying the notation in the ensuing argument; those results
below which follow from pure symmetry arguments are completely unaffected
by a relaxation of the assumption, and the (non-symmetry-based) results of
microscopic theory are likely to be only weakly sensitive to it.

Consider a given Josephson junction between bulk superconductors 1 and 2,
and suppose that the order parameter in superconductor 1 (2) may in general be a
combination of various irreducible representations $j$($j$'). 
 (In the case of
an orthorhombic crystal, we take those to be the irreducible representations of the tetragonal
group.)  For each $j$ or $j'$ we choose a basis function $Y^{(1)}_j({\bf k})$,
$Y^{(1)}_{j'}({\bf k})$,
which, as well as transforming appropriately, reflects the
detailed ${\bf k}$-dependence of the order parameter, so that we can write
$$ \eqalign{
 F^{(1)}_{\bf k} = & \sum_j \psi^{(1)}_j Y^{(1)}_j({\bf k}) \cr
  F^{(2)}_{\bf k} = & \sum_{j'} \psi^{(2)}_{j'} Y^{(2)}_{j'}({\bf k}) .
} \eqn\foneftwo $$
Then, from gauge invariance, the lowest-order Josephson
coupling energy can be written in the form
$$ \eqalign{
E^{(2)}_J = & - {\rm Re} \sum_{jj'} A_{jj'} \psi^{(1)}_j \psi^{(2)}_{j'} \cr
   =  & - {\rm Re} \sum_{jj'} A_{jj'} | \psi^{(1)}_j| |
 \psi^{(2)}_{j'} | \cos\left(\varphi^{(1)}_j - \varphi^{(2)}_{j'}\right)
} \eqn\josephsonagain $$
where the quantity $A_{jj'}$  will in general depend on the properties of the
junction.

In the application of our analysis to circuits including more than one
Josephson junction a crucial role is played by the choice of phase
convention for the $Y_j({\bf k})$.  In our opinion by far the simplest and
most natural choice is so that for any one bulk superconductor the form of
$Y_j({\bf k})$ is real and is defined {\it relative to absolute space}: 
 e.g. for a
crystal oriented with axes along the NS and EW directions in ``absolute''
(laboratory) space, we define a \dx2y2-like basis function so that its $+$
signs are always on the NS lobes and the $-$ signs on the EW ones.  (For a
twinned crystal we apply this definition to the ``twin-averaged" order parameter, see
Section 4).  Note that even in cases where various bulk regions are made of
the same superconductor (e.g. YBCO), if the crystal axes are differently
oriented the definitions of the $Y_j({\bf k})$ will be different in different
regions and the relative phase convention may be arbitrary; this however
causes no difficulty provided we keep the accounting straight.

With the above choice of phase convention, the quantity $A_{jj'}$ 
may depend not
only on the ``intrinsic" properties of the junction {\it but also on its
orientation}, and in particular may be negative for some combinations $j,j'$.
For example, if we define the phase convention for a \dx2y2\ state as above,
and consider a junction with an ordinary s-wave superconductor with the
junction plane (edge) in the NS-direction, then since the natural
assumption is that the exterior s-wave couples more strongly to the lobe of
the order parameter which is perpendicular to the junction than to the ``parallel'' one,
it is plausible that $A_{sd}$ is in this case negative.  We
see more generally that with our convention, for a single irreducible 
representation in each
bulk superconductor the lowest-order Josephson coupling can always be
written in the form
$$ E_J^{(2)} = - J \cos{\Delta\varphi} \eqn\josephsonyetagain $$
where however $J$ may have either sign.  In the recent literature junctions
for which $J$ is negative are sometimes referred to as ``$\pi$-junctions'';
however, this seems to us very confusing, since the fact that $J$ is negative
has nothing to do with the intrinsic properties of the junction as such but
is entirely a consequence of our choice of phase convention.  We urge that
in future the term ``$\pi$-junctions'' be reserved for junctions of the type
discussed in Ref. \bulaevskii,
or more generally for those junctions which induce
half-quantum flux quantisation in a single-junction SQUID ring with uniform
bulk crystal orientation.  To make quite sure that there is no ambiguity,
in the rest of this review, we shall use the term 
``intrinsic $\pi$-junctions'' for such a case.

In the case where more than one irreducible representation occurs in one or both
superconductors we need to invoke a very important general consideration
(cf. Section 3):  In the absence of pathology, the relative phase of the
different $\psi_j$  
occurring in each of the bulk superconductors is uniquely fixed by the
energetics (see the form of the fourth-order terms in Eqn. \gleqn).  Let us
then choose one particular irreducible representation $j,j'$ on each side, and (with the general
phase convention above) define the phases  $\varphi_j(\varphi_{j'})$ 
 of the other components relative to
$ \psi_{j0}(\psi_{j'0})$.  We then define,
conventionally, the ``total'' phase difference $\Delta\varphi$ 
  across the junction as the
difference between the phases of $ \psi_{j0}$ and $\psi_{j'0}$.  
 Then we can write the
lowest-order Josephson coupling energy in the generally valid form
$$ E_J^{2} = -J \cos{(\Delta \varphi + \alpha)} \eqn\equationsixsix $$
with $J,\alpha$ real and
$$ J e^{i\alpha} \equiv \sum_{jj'} A_{jj'} | \psi_j | \cdot |
\psi_{j'} | \exp{i(\varphi_j-\varphi_{j'})} .\eqn\equationsixseven $$
We note explicitly that the quantity $\alpha$ need not be $0$ (or $\pi$). 
 For example,
if (contrary to the arguments of Section 4) the twin-averaged order parameter of YBCO
is of the $s + id$ form, then we expect that for certain junctions $\alpha$
would
vary continuously from $0$ to $\pi/2$ as the ``d-wave'' component is increased.

In the context of a determination of the symmetry of the order parameter the crucial
question is:  What do we know for sure about the coupling constants 
$A_{jj'}$ in
\equationsixseven, and what may we reasonably surmise?  
It is essential to appreciate
that the only totally reliable knowledge of the $A_{jj'}$
 which is independent
of a particular microscopic model is that which follows from symmetry
arguments.  We will assume that when averaged over scales large compared to
a typical ``correlation'' length, which is in general of an atomic order of
magnitude or, at least, very much smaller than the transverse dimensions of
the junction, the properties of the latter itself are invariant under
appropriate symmetry operations ($\pi/2$ rotations about the normal, 
reflection
in the plane defined by the normal and a bulk crystal axis, etc.).  This
then immediately implies that the Josephson energy $E^{(2)}_J$ must be
invariant under the appropriate operations when applied simultaneously to
the junction and the bulk superconductors, and this will in general
determine (a) which of the $A_{jj'}$ must be zero (``principle A''), and 
(b) to
some extent, how these quantities transform when the orientation of the
junction is changed.  It is clear that when the bulk samples are heavily
twinned (and we assume no correlation in the twinning across the boundary),
then in the thermodynamic limit (only!) these arguments
can be applied to the $A_{jj'}$ defined in terms of the ``twin-averaged''
 order parameters.
We already saw an example of (a):  if the plane of a junction between a
tetragonal cuprate superconductor and (e.g.) Pb is normal to the c-axis, and
the properties of the junction are invariant under $\pi/2$ rotation, then
$A_{s,d_{x^2-y^2}}$ must be zero. 
 An example of (b) is exploited in the ``type-II"
experiments of Section 7:  for two junctions between Pb and a tetragonal
cuprate with normals lying in the ab-plane and at right angles the sign of
$A_{s,d_{x^2-y^2}}$  must be different.  
We will exploit some rather more complicated
applications of these principles in Section 7.

Can we obtain any further information about the coupling coefficients 
$A_{jj'}$
over and above what is implied by symmetry?  To do so, we clearly need (a)
a microscopic model of the formation of Cooper pairs in bulk, and (b) a
microscopic model of the processes in the junction which give rise to the
Josephson coupling.  As regards (a), while a few 
calculations\Ref\muthukumar{Eg. V.N. Muthukumar and G. Baskaran,
\journal Mod. Phys. Lett. &B8&699(94).} have been
done for the case of ``exotic'' (non-Fermi-liquid-based) models of the bulk
superconductivity, to the best of our knowledge those have all assumed
simple s-wave symmetry; since the main relevant differences of those models
from the standard BCS one appears to lie mainly in the energy-dependence of
the single-particle propagators, it seems unlikely that the behaviour  of the
$A_{jj'}$  in (anisotropic versions of) such models would be quantitatively
different from what it is in BCS-type theories.  However, we should caution
that while these existing calculations have produced a Josephson coupling
with the conventional sign, we can see no compelling {\it a priori\/} argument to
exclude the possibility that in such models ``intrinsic $\pi$-junctions'' (in the
sense defined above) might be the norm, between two cuprate superconductors
and/or between a cuprate and a classic superconductor such as Pb.
As regards point (b), while the literature of the last thirty years
contains hundreds or perhaps thousands of papers which work out the details
of the Josephson effect in the various different known types of junction
(tunnel oxide barriers, SNS junctions, microbridges, point contacts...) the
vast majority of these not only deal with s-wave superconductors but
neglect the details of the directional properties of tunnelling matrix
elements (as is legitimate in the s-wave case) so that the generalisation
to unconventional pairing is not obviously 
trivial\rlap.\Ref\yiplastminute{See (e.g.) S.K. Yip, 
\journal J. Low Temp. Phys. &91&203(1993)\ and 
\journal Phys. Rev. B &52&3087(1995).}
One case where one can write down such a generalisation is that of 
a simple tunnel oxide junction
described by the simple Bardeen-Josephson Hamiltonian
$$ \hat{H}_T = \sum_{{\bf k}{\bf q}\sigma\sigma'}
 ( T_{{\bf k}{\bf q}\sigma\sigma'} a^+_{{\bf k}\sigma}b_{{\bf q}\sigma'}
+ {\rm H.C.} ) 
\eqn\bardeenjosephson$$
where $a^+_{{\bf k}\sigma}(b^+_{{\bf q}\sigma'})$  create an electron 
in Bloch-wave state ${\bf k}({\bf q})$ 
and spin $\sigma(\sigma')$ in bulk superconductor 1 (2).  
For a simple BCS-like picture
of the bulk state of both superconductors the calculation of the
second-order Josephson coupling energy is a straightforward generalisation
of the original one of Ambegaokar and Baratoff\Ref\ambegokar{
V. Ambegokar and A. Baratoff, \journal Phys. Rev. Lett. &10&486(63).} 
for the s-wave case:
for simplicity of notation we state it 
under the simplifying
assumptions\Ref\relaxation{Relaxation of these assumptions seems
extremely unlikely to affect qualitatively the direction-dependence 
which is the prime object of our attention here.}
 of zero temperature and a matrix element
$ T_{{\bf k}{\bf q}\sigma\sigma'} = T_{{\bf k}{\bf q}} \delta_{\sigma\sigma'}$ 
which is proportional to a unit matrix in spin space
and is moreover time-reversal invariant $ T^*_{{\bf k}{\bf q}} 
=  T_{-{\bf k}-{\bf q}} $
 but is otherwise arbitrary:
$$ E^{(2)}_J = - 2{\rm Re} \sum_{{\bf k},{\bf q}}  
\frac{| T_{{\bf k}{\bf q}} |^2  F^{(1)}_{\bf k} F^{(2)*}_{\bf q}}{E^{(1)}({\bf k})
+ E^{(2)}({\bf q}) },
\eqn\tkqsquared $$
where $F_{\bf k}^{(j)}\equiv \Delta^{(j)}({\bf k})/2E^{(j)}({\bf k})$, $j=1,2$.

For the case of s-wave pairing, Ambegaokar and Baratoff showed that the critical
current $I_c\equiv 2\pi E_J/\varphi_0$ which follows from Eqn. \tkqsquared\
is related to the normal state resistance $R_n$ of the junction by the formula
$$
I_cR_n = \frac{\pi\Delta^{(1)}}{2e}f(\Delta^{(1)}/\Delta^{(2)}),
\eqn\tonylastmin
$$
where $f$ is a specified function of the gap ratio, taking the value one when its
argument is unity.  We will refer to the right hand side of Eqn. \tonylastmin\
as the ``Ambegaokar-Baratoff" value of the product $I_cR_n$, and denote it
by $(I_cR_n)_{AB}$.

	Expression \tkqsquared\
 respects the symmetry principle we have
called ``principle A.''  We can therefore rewrite it in the form \equationsixseven,
 with
the $A_{jj'}$  given by the formal expression
$$ A_{jj'} \propto  \int {d\Omega \over 4 \pi}
\int {d\Omega' \over 4 \pi} | \tilde{T}(\hat{\bf n},\hat{\bf n}') |^2
\tilde{Y}_j(\hat{\bf n}) \tilde{Y}_{j'}(\hat{\bf n}') \eqn\ajjprime$$
where $ \tilde{Y}_j(\hat{\bf n}) $   and $\tilde{T}(\hat{\bf n},\hat{\bf n}')$
are respectively the
energy-averaged value of $Y_j({\bf k})$
     and a quantity which, while in general not simply related
to   $T_{{\bf k}{\bf q}} $, respects its symmetries and its 
qualitative features (in
particular, if     $T_{{\bf k}{\bf q}} $  is sharply peaked for  
${\bf k}$      and  ${\bf q}$     normal to
the junction plane, then so is  $\tilde{T}(\hat{\bf n},\hat{\bf n}')$).

As already stated, the expression \tkqsquared\ (and hence \ajjprime) 
has been
derived for the anisotropic case only for a simple oxide tunnel
junction described by the Hamiltonian \bardeenjosephson. 
 However, it
seems plausible to guess that it would apply more generally, with however
$T_{{\bf k}{\bf q}} $
replaced by some ``effective'' matrix element which, for example, in
the case of an SNS junction would depend strongly on the properties of the
normal-metal layer (including its temperature).  Unfortunately, for most
types of junction both our theoretical and our experimental information on
the detailed angular dependence of  $| T_{{\bf k}{\bf q}} |^2$ 
(or its average) is
rather poor.  For an SNS junction, and to a somewhat lesser extent for a
simple tunnel oxide barrier, it seems very plausible that
$| T_{{\bf k}{\bf q}} |^2$ 
(hence also $\tilde{T}(\hat{\bf n},\hat{\bf n}')$) 
is strongly peaked in the normal
direction (${\bf k}$ and ${\bf q}$ both close to the junction normal 
$\hat{\bf n}_0$). If this is the case, then for those types of 
junction  $A_{jj'}$   is a direct
measure of the product  $Y_j(\hat{\bf n}_0)Y_{j'}^*(\hat{\bf n}_0)$.
However, in the
case of junctions formed by grain boundaries (which not only tend to be
``jagged'' but also to act as sinks    for various chemical impurities etc.)
it would seem optimistic to expect that the
  $A_{jj'}$  have this simple form.

Finally, if we not only make the assumption of the last paragraph but
assume that the \dx2y2\   irreducible representation has literally that form (i.e.
$Y_j(\hat{\bf n})= \hat{n}_x^2 - \hat{n}_y^2 $),
 we obtain the expression written down by Sigrist
and Rice\Ref\sigristrice{M. Sigrist and T.M. Rice, \journal
J. Phys. Soc. Japan &61&4283(93).}
 for the coupling energy of two differently oriented YBCO
grains each described by a \dx2y2\
state: 
$$  E^{(2)}_J \propto \cos{2\theta_1}\cos{2\theta_2} 
 \cos{(\varphi_1-\varphi_2)} \eqn\sigristriceansatz
$$
 where  $\theta_1$  and  $\theta_2$   are the angles made by the
 positive lobe  of
the order parameter in superconductor 1 and 2 respectively with the boundary normal.
While this ``Sigrist-Rice'' ansatz for the coupling energy has of course the
correct symmetries, we would caution that it may be dangerous to take it
too seriously as the exact form of   $A_{dd}(\hat{\bf n})$, 
in particular in the case of a grain boundary junction.

To complete our analysis, we need to consider the behaviour  of the order parameter in
the bulk superconductor, whether the latter be classic or
unconventional.  We will postpone for the moment possible complications
associated with twin boundaries, corners, etc., and assume the
superconductor in question to be homogeneous (though not necessarily
infinite in spatial extent).  We now invoke a very important general
consideration:  in all models of the pairing in the high temperature superconductors currently
considered, the only unbroken {\it continuous} symmetry is the
U(1) gauge symmetry.  Thus to deform the components of the order parameter
continuously in any way other than by an overall phase rotation will
cost a bulk free energy density of order $\alpha^{2}F_{c}$, when
$F_{c}$ is the bulk thermodynamic condensation free energy and $\alpha$ is a
number which is in general of order unity; while for special cases
$\alpha$ could be much smaller than 1, it is very unlikely to be of the order of
the inverse sample size.  On the other hand, the ``bending'' energy
density associated with variation of the components in space over a
distance $L$, turns out to be quite generically of the order of
$F_{c}(\xi(T)/L)^{2}$, where $\xi(T)$ is the correlation
length:  in BCS-type theories this is of order $\hbar v_{F}/\Delta(T)$,
and quite generally is expected to be of the order of the pair size
except close to $T_{c}$.  Thus any deviation of the order parameter from its
equilibrium value (such as, for example, a ``soliton'' in which the
lobes rotate {\it geometrically} in $\bf{k}$-space from one
axis to another) will have a typical length scale of order
$\xi(T)/\alpha$ and cost an energy per unit area of order $\alpha
F_{c} (T) \xi(T)$.  The upshot of this is that for any reasonable
sample size the creation energies for such solitons, etc., will be so
enormous relative to $k_BT$ that their probability of existence in
thermodynamic equilibrium is essentially zero except very close to
$T_{c}$; and while in principle this kind of ``defect'' could be
generated during the quenching through the phase transition, it seems
likely in the cases of physical interest that it would be topologically
unstable and disappear as soon as $T$ falls appreciably below $T_{c}$.
Thus the {\it only} kind of variation of the order parameter over
macroscopic distances which we need to consider in the bulk
superconductor is the overall phase variation corresponding to the U(1)
gauge symmetry.  (The case of ordinary s-wave superconductivity is a
rather trivial special case of this general result).

{}From rather general considerations we can say that if the overall phase
$\varphi(\bf{R}$) varies in space, the associated ``bending'' free energy
density has the form 
$$ F =\rho_{s}(T) \frac{\hbar^2}{8m} ({\bf
\nabla}\varphi-2e{\bf A}({\bf R})/\hbar)^{2} 
\eqn\bendingenergy $$
 and the
supercurrent is given by $$ {\bf J}_{s}=\rho_{s}(T) \frac{\hbar e}{2m}
({\bf \nabla}\varphi- 2e{\bf A}({\bf R})/\hbar), \eqn\supercurrent $$ where the
superfluid density $\rho_{s}(T)$ defined by \bendingenergy\ may in general be a
tensor in Cartesian space.  In any given geometry the equilibrium form
of the order parameter must be found by minimising the sum of (a) the Josephson
energies \equationsixsix\ associated with any junctions in the circuit (b) the
``bending'' energy \bendingenergy, and (c) the usual electromagnetic energy 
$$
F_{em}=\frac{1}{2\mu_0}  \int ({\bf \nabla}\times {\bf A}({\bf r}))^{2}
d^3{\bf r}.
\eqn\electromagneticenergy $$

For an arbitrary geometry this minimisation problem may be quite
complicated (cf. below).  However, in most cases (though not all) of
experimental relevance a major simplification is possible.  The problem
contains three characteristic energies:  
\item{(1)} The energy
$E_{kin}$ which would be necessary to produce one quantum of
circulation (with $\bf{A}=0$) {\it with flow over the whole bulk
volume}.  
\item{(2)} The self-inductance energy $E_{ind}$ of the
corresponding circulating current (note that for most experimentally
interesting geometries, this is not particularly sensitive to the
actual distribution of the current, and in any case we are interested
only in orders of magnitude).  
\item{(3)} The phase-locking energy $E_{J}$
of any Josephson junctions involved.

Using \bendingenergy\ and standard results of electrical circuit theory, we 
obtain the order-of-magnitude estimates $$ E_{kin}\sim \rho_{s}
\frac{\hbar^{2}}{2m} \frac{A_{B}}{L} , \qquad E_{ind} \sim
\frac{(h/e)^{2}}{\mu_{0}L} ,\qquad  E_{J} \sim I_{c}(\hbar / e),
 \eqn\estimates
$$ where $A_{B}$ is the cross-sectional area of the bulk
superconductor(s), and $L$ is the circumference of the circuit.  
The condition for $E_{kin}$ to be large compared to
$E_{ind}$ is $A_{B} \gg \lambda_{L}^{2}(T)$ where $\lambda_{L}(T)$ is
the bulk London penetration depth; this condition is fulfilled for
essentially all geometries of experimental interest.  The condition
$E_{kin}\gg E_{J}$ is generally fulfilled, in particular when the
cross-sectional area of the junction is small compared to $A_{B}$;
however, in certain special ``two-dimensional'' geometries $E_{J}$ may be
comparable to or even larger than $E_{kin}$.  Now, if $E_{kin}$ is
indeed the largest energy in the problem, it will be important to get
rid of it by forcing the current to flow only over a strip of thickness
of order $\lambda_{L}(T)$ on the surface of the bulk superconductor
(the usual Meissner effect).  If furthermore, the thickness of the bulk
is itself $\gg \lambda_{L}(T)$ (a condition which is, of course,
compatible with but not guaranteed by the condition $A_{B} \gg
\lambda_{L}^{2}(T)$), then we will be able to argue as in the standard
account of flux quantisation that since the current is screened out
within the bulk, the phase difference $\Delta
\varphi$ accumulated {\it within the bulk regions} as we go around a
circuit is given by $$ \Delta \varphi= \frac{2e}{\hbar} \int \bf{A}
\cdot \bf{dl} =2 \pi\Phi/\varphi_{0}, \eqn\circuitphase $$ where $\Phi$
is the total flux through the circuit (it is assumed here that the
contribution of the junction regions to the flux is negligible).
Moreover, for a single circuit the sum of the energies $E_{kin}$ and
$E_{ind}$ then has the simple expression ($\Phi_{ext}$=externally
applied flux, $L$=self-inductance of circuit) $$
E_{ind}(\Phi)=\frac{1}{2}LI^{2}= (\Phi - \Phi_{ext})^{2} /2 L.
\eqn\selfinductance $$

	We now choose a particular component   $j_0$ 
 of the order parameter (it does not matter
which) in each bulk region and define the phase difference   $\Delta\varphi_i$
 across
the $i$-th junction as described above.  Since the total phase difference
accumulated by the phase as we go around the complete circuit must be
$2n\pi$, we have from \circuitphase\ the central result
$$ 
\sum_i^{(c)} \Delta\varphi_i = -2\pi \Phi/\varphi_0 \ (+ 2n\pi) ,\eqn\sumphi$$
where the $(c)$
indicates that we must sum over all the junctions in the
circuit as we traverse it in a consistent sense.  Further, the total energy
of the (closed) circuit is given by the expression
$$ E_{tot}(\Phi) = (\Phi - \Phi_{ext})^2/2L - \sum_i E_i^{(J)}(\Delta\varphi_i),
\eqn\totalcircuit $$
where   $E_i^{(J)}(\Delta\varphi_i) $ is the Josephson coupling energy of the
$i$-th junction.  It should be noted that Eqn. \totalcircuit\ applies only for a
circuit with no external leads (or in the (unphysical) case of exact
symmetry, cf. Section 7.3).  We refer to the combination of eqns. \sumphi\ 
and \totalcircuit\ as ``principle B."

{}From here on the analysis proceeds essentially as in the case of a normal
circuit with only s-wave order parameters and ``normal'' junctions.  Let us
focus in particular on the case (realised in most experiments) that the
critical currents of all the junctions involved are large compared to
$\varphi_0/L$.  Then the second term in \totalcircuit\
 dominates, and the value of
{\it total} flux  $\Phi$     through the circuit is simply
$(\varphi_0/2\pi)\Delta\varphi_{tot}$
where $\Delta\varphi_{tot} \equiv \sum_i \Delta\varphi_{i}$
 is the value which
minimises the sum of the Josephson couplings.  Thus in this limit the
current circulating in the ring is given by the simple expression
$$ I = \left (\Phi_{ext} - (\varphi_0/2\pi) \cdot \Delta\varphi_{tot}
\right ) / L
\eqn\iequals$$
of which the s-wave result  ($\Delta\varphi_{tot}=0$)   is of course a special
case.  Moreover, in the case of a two-junction ring set up with external
leads in the ``dc-SQUID configuration", the critical current,
which with our sign convention for the   $\Delta\varphi_i$   is
$I_{c1} \sin{\Delta\varphi_1}$, is easily seen to be periodic in the
trapped flux  $\Phi$   with period  $\varphi_0$, and maxima at
$\Phi = \varphi_0(n + \Delta\varphi_{tot}/2\pi)$.
For the special case $I_{c1}=I_{c2}$ the formula is
$$ I_c = 2 I_c \left| \cos{\left ( {\pi\Phi \over \varphi_0} - {\Delta
\varphi_{tot} \over 2} \right )} \right| \eqn\ictwojunctions $$
but for unequal junctions the minima are finite.  Note that it is the
 {\it total
trapped flux}  $\Phi$, not the external flux $\Phi_{ext}$, which 
appears in this formula:  we return to this point in Section 7.3.

The upshot of all this is that either a direct measurement of the
circulating current as a function of  $\Phi_{ext}$, or 
(with caveats) 
a measurement, in the 2-junction case, of the parallel critical current as a
function of  $\Phi$, will determine the quantity  $\Delta\varphi_{tot}$, 
and hence
allow us to infer something about the $E_i^{(J)}$  and thus the symmetry of
the order parameter.  Two important points should be noted:  (1) unless time-reversal
symmetry is spontaneously broken (which requires, as a minimum, the mixing
of two different irreducible representations) the quantity $\Delta\varphi_{tot}$ must be either 
$0$ or
$\pi$; (2) experiments of this type cannot by themselves ever exclude the
presence of components of the order parameter which, for symmetry or other reasons, do
not contribute to the individual $E_i^{(J)}$.

{\bf \chapter{Quantum Phase Interference: Experiments}}

	At the time of this writing (late 1995) there are about a dozen
experiments in the literature which attempt to exploit the Josephson effect
systematically to determine the symmetry of the order parameter.  With one exception
they are all on YBCO, usually pure but in one case also Pr-doped.  It is
convenient to classify them into three groups:
\item{I.} Experiments involving a single Josephson junction between YBCO
and a conventional superconductor such as Pb, in the ``c-axis" geometry.
These exploit only principle A.
\item{II.} Experiments involving a SQUID-type circuit or something related,
the bulk ingredients of the circuit again being YBCO and a
conventional superconductor.  These experiments exploit (directly) only
principle B.
\item{III.} Experiments on a circuit whose bulk elements are entirely YBCO (or
some other high temperature superconductor), with the different bulk regions having different
crystal orientations.  As we shall see, these experiments in some sense
exploit both of the principles A and B.

In reviewing the experiments we shall focus primarily on the application
to them of the general principles enunciated in Section  6  above,
with particular attention to the possible complicating effects of the
orthorhombic structure of YBCO.  For those considerations
which are specific to the particular experimental geometry or method used,
such as the possibility of  unwanted trapped flux, we refer the
reader to the original papers and the subsequent exchanges in the
literature.

{\bf\subsection{Type I experiments}}

To date there have been experiments reporting a finite Josephson
current by two groups:  Sun 
\etal\refmark{\sun} followed by Katz 
\etal\Ref\katz{A.S. Katz, A.G. Sun, R.C. Dynes and K. Char, 
\journal Appl. Phys. Lett. &66&105(95).} being one, and Iguchi and 
Wen\Ref\iguchi{I. Iguchi and Z. Wen, \journal Phys. Rev. B. 
&49&12388(94).} being the other.  We will refer to
those as the UCSD and Tsukuba experiments respectively.  It should be
mentioned that a number of groups had previously tried and failed
to observe a Josephson current in this geometry\rlap.\refmark{\ag}

In the UCSD experiments the authors etched single (but heavily twinned)
crystals of  YBa$_2$Cu$_3$O$_{7-\delta}$   and 
       Y$_{1-x}$Pr$_x$Ba$_2$Cu$_3$O$_{7-\delta}$  so
as to expose faces parallel to the ab-plane, and fabricated junctions by
evaporating on to the exposed surface first about $10$\AA\ of Ag (to prevent
formation of too thick an insulating barrier) and then a $1\mu$m thick Pb film.
Since there is probably a thin material insulating barrier formed at the
YBCO surface, the resulting junction is probably best described as of SINS
(superconductor-insulator-normal-superconductor) type.

For our purposes the salient result of this experiment is that a finite
critical current    $I_c$   was obtained between the YBCO and the Pb.
However, the magnitude of the product   $I_cR_n$    was considerably less
than the (s-wave) ``Ambegaokar-Baratoff'' value    (see
Section 6), which is calculated for a Pb-YBCO junction as 8.0 meV.
In fact, the ratio $I_cR_n/(I_cR_n)_{AB}$
varied from about 0.025 to 0.11, being
highest for the pure YBCO samples.  Since the effective 
dimensions\Ref\dimensions{Recall (cf. Section 6)  that in the context of the
Fraunhofer modulation the effective ``thickness'' of the junction 
includes the bulk Pb penetration depth, which is of order $500$\AA,
so that uncertainties in the thickness of the Ag layer itself
are unimportant.}  of the
junctions were reasonably well known, it could be checked that the
Fraunhofer modulation pattern obtained in a magnetic field applied in the
plane of the junction is not only of the conventional type but corresponded
to a field scale defined by the usual flux quantum    $h/2e$     (and not,
e.g.,  $h/4e$), thus establishing that the Josephson effect seen is
indeed the usual lowest-order one (see Section 6).  No anomalous
Fraunhofer patterns were reported.

The Tsukuba experiments employ a geometry which, at least at the
macroscopic level, is similar to the UCSD one:  they used c-oriented YBCO
and Pb films as the electrodes, while the barrier was either ``natural'' or
an MgO 
layer (and sometimes an   Ag      overlayer) a few nm thick.  In this
experiment too, finite critical currents were observed, with however a
value\Ref\vaules{We take $(I_cR_n)_{AB}$ to be 8meV, as calculated
by Sun \etal\refmark{\sun}} 
of $\alpha = I_cR_n/(I_cR_n)_{AB}$
 in the range           $0.01 - 0.025 $,
thus smaller than in the UCSD experiment.  In addition, the authors report
two intriguing further features:  First, the critical current is a
nonmonotonic function of temperature; it rises steeply below the 
    \Tc\   of Pb, reaches a maximum at around 5K and then decreases until
at 2K it is about half the maximum value.  Secondly, in some cases the
Fraunhofer pattern is ``split,'' i.e. shows a minimum at zero applied field
(see Fig. 2 of Ref. (\iguchi)); as we shall see below, similar behaviour  is
seen in the UIUC II experiment.  However, at low temperatures the critical
current appears to be only weakly dependent on the field.  The Fraunhofer
patterns seen are often asymmetric under reversal of the field or the
current or both.

Let us first try to interpret these data on the hypothesis of simple
 s-wave pairing in the YBCO.  Clearly the UCSD data are qualitatively
consistent with this assignment; while the value of 
$\alpha$     is surprisingly
small, there are various scenarios which could explain it, e.g.
normal-metal ``shorts'' which contribute to $R_n^{-1}$ but not to $I_c$.
Indeed, it is not totally implausible that the even smaller value of $\alpha$
seen in the Tsukuba experiments could be accounted for in this way.
However, the Fraunhofer data in the latter are more difficult to explain on
an s-wave hypothesis:  while the obvious explanation of the frequent
asymmetry of the patterns is as some kind of trapped-flux effect, it seems
difficult to account for the ``split'' patterns in this way without invoking
a rather implausible degree of coincidence.  However, this does not seem
to us a conclusive argument against the s-wave scenario.

Now let us ask whether the data are compatible with any assignment other
than simple s-wave.  For pedagogical clarity let us first pretend that
the crystal symmetry of YBCO is pure tetragonal, so that the competing
possibilities are   $s^-$,  \dx2y2\    and        $d_{xy}$.  Now the Cooper
pairs in Pb are known with a high degree of confidence to form in a
simple s-state, i.e. are even under $\pi/2$ rotation and under inversion in
a crystal axis; further, it seems a natural assumption that the junctions
themselves are isotropic in the ab-plane, at least where averaged over a
few atomic distances (but see below).  Given this state of affairs, and the
experimentally verified fact that the Josephson effect is seen in lowest
order, principle A immediately rules out\Ref\rulesout{
This is strictly true only in the ``thermodynamic limit,'' i.e. when the
averaging scale is negligible compared to the sample dimensions, however
this condition is likely to be very well  approximated in practice.}\
all YBCO pairing symmetries other
than   $s^+$.

It should be emphasised that this conclusion (drawn for the case of
tetragonal symmetry, which is of course not realised for real-life YBCO) is
a great deal more robust than one might think at first sight.  For example,
it is tempting to try to think up mechanisms by which the bulk d-wave order parameter of
YBCO might have some s-wave component mixed into it as we approach the
surface, so that this component could couple to the Pb order parameter and give a finite
lowest-order Josephson effect.  However, any such effect must still come
from a bilinear coupling of the bulk order parameters 
in the Pb and the YBCO, so that
the general symmetry principle still rules it out.

Once the orthorhombic anisotropy of real YBCO is taken into account, the
situation becomes a good deal more complicated.  (For the moment we still
assume that the surface is smooth.)  Indeed, it is immediately clear that
an {\it untwinned} single crystal of orthorhombic YBCO with \dx2y2\
pairing can show a lowest-order Josephson effect in the UCSD/Tsukuba
geometry with an s-wave superconductor such as Pb\rlap;\Ref\suchaspb{
Note however that pure $s^-$ and $d_{xy}$ are still excluded.}
 in fact it can even
have a value of  $\alpha$ of order unity.
However, the UCSD and
(presumably, though it is not stated) the Tsukuba samples were heavily
twinned.  The subsequent analysis depends crucially on the behaviour  assumed
for the order parameter on crossing a twin boundary.  If it behaves ``non-gyroscopically''
as defined in Section 6 above, then the critical currents of the
different twin domains add up in phase and the $\alpha$-value 
for the whole junction
will be essentially the same as for an untwinned sample.  However, we
argued in Section 6 that there are compelling reasons to believe that
the order parameter behaves ``gyroscopically," i.e. 
keeps its $+$ and $-$ signs relative to
``NSEW."  In this case neighbouring twins will tend to interfere
destructively, and one would expect that the $\alpha$-value for the macroscopic
junction will be of order  $N^{-1/2}\alpha_0$, where $\alpha_0$  is the
 ``untwinned''
value, and $N$ is (at least for the moment:  but see below) the number of
twin domains in the area of the junction.

The crucial question now is:  Is the value of $N$ which is likely to have
been realised in the UCSD and Tsukuba experiments small enough that the
predicted $\alpha$ would be as large as observed?  In the 
Tsukuba experiment this seems
not implausible:  the largest value of $\alpha$
 observed was about 0.025, and if
we assume that the twinning of the sample was comparable to that in the
UCSD experiment and use the estimates of the latter authors, there should
have been     $\approx 10^3 - 10^4$    twin domains in the junction area. 
 Thus, provided $\alpha_0$
is of order 1, reconciliation is possible.  Actually there is in any
case a complication connected with the nature of the sample surface which
may help here:  see below.

It is more difficult to explain the UCSD results in this way; the number
of twin domains as quoted by the authors is comparable, but the values of
$\alpha$, as we see, are larger, up to 0.11. 
 Thus even with     $\alpha_0 \sim 1$    the
predicted value of $\alpha$ is too small by at least a factor of 4, unless the
estimate of $N$ is for some reason badly wrong.

A further complication concerns the microscopic nature of the nominal
``ab" surface.  In the above argument it was assumed to be microscopically
flat, so that all tunnelling is literally in the c-direction; however, in
real life this is unlikely to be so.  Indeed, the authors of the Tsukuba
experiment note explicitly that their surface carries a large number of
``micrograins'' (protrusions) whose height and lateral dimension they
estimate at 2-5 nm and 10-100 nm respectively; moreover, they postulate
that all or most of the tunnelling actually takes place across the {\it side}
faces of these micrograins and is therefore in the a- and/or b-direction.
Rather similar phenomena may occur in the UCSD experiment, with a/b
tunnelling occurring across the vertical sides of the etch pits.  Does
this affect the arguments used above?

It should be strongly emphasised that {\it  provided}
 (a) all surface inhomogeneities are on average isotropic
with respect to the laboratory (``NSEW") axes
 and (b) the order parameter behaves gyroscopically across twin boundaries, the
generic symmetry argument  {\it still} rules out states other than $s^+$ 
in the thermodynamic limit (area of junction   $\rightarrow \infty$). 
 The only
difference is that the ``thermodynamic limit'' may be less easy to attain.
Suppose for example it turns out that (perhaps because of the contribution
of the chains) tunnelling in the b-direction is appreciably easier then in
the a-direction.  The effect is (a) to make  $\alpha_0$  
  of order 1, (rather than
the smaller value it is likely to have for a truly planar ab-interface),
and (b) to replace, in the calculation of $\alpha$, the factor        
$N^{-1/2}$  by the
excess number of ac-interfaces normal to (say) the NS axis, which should be
of order  $N_r^{-1/2}$  where  $N_r$   
  is the total number of ``rugosities''
(micrograins and/or etch pits) in the junction area.  Since  $N_r$  could well
be an order of magnitude or so smaller than $N$, this may make it easier to
reconcile the observed value of $\alpha$ with a d-wave scenario. 
 While this may
be an important effect in the Tsukuba experiment\rlap,\Ref\tsukubaexpt{
The value of $N_r$ is not quoted, but we estimate it to be as small as 100.} 
its relevance to the
UCSD results is thrown in doubt by the reported observation that a
systematic study of the effects of different degrees of etching showed no
systematic dependence of $I_c$  on the latter.

We should finally note that the Tsukuba authors cite the observation of
``split'' Fraunhofer patterns as evidence for a d-wave scenario, interpreting
them as the result of the interference of tunnelling through the a- and b-
faces of a single micrograin (cf. the discussion, below, of the UIUC II
experiment).  However, it seems to us that (since  $N_r$, while smaller
than $N$, seems unlikely to be of order two or three!) such an interpretation
would require a high degree of coincidence of the dimensions of the
different micrograins contributing to  $I_c$, which seems difficult to
reconcile with the quoted spread of values.  Thus, this feature of the
Tsukuba experiments, as well as the anomalous temperature-dependence of
$I_c$\rlap,\Ref\ictdependence{It should be observed that
the latter was also seen for a-axis oriented films, so it
may not have much directly to do with the order parameter symmetry.} remains to us puzzling.

In sum, the UCSD experiment is clearly compatible with a simple s-wave
scenario and appears, at least, very difficult to reconcile with a
\dx2y2\
picture (or indeed any unconventional one).  As to the Tsukuba experiment, while
the existence of a critical current of the observed magnitude is compatible
with either d- or s-wave behaviour  (and if anything would seem to favour the
former), other features of the experimental results appear puzzling on any
symmetry assignment. If we are prepared to disregard the thermodynamic
argument against ``mixing'' either experiment could of course accommodate
a real or imaginary mixture of
$s^+$ and \dx2y2.

{\bf \subsection{Type-II experiments}}

To date there are four experiments of type II in the literature; they are
reported in Wollman \etal\rlap,\refmark{\wollmani}\
 Brawner and Ott\rlap,\Ref\brawner{D.A. Brawner and H.R. Ott,
\journal Phys. Rev. B &49&12388(94).}
 Mathai \etal\refmark{\mathai} and
Wollman \etal\Ref\wollmanii{D.A. Wollman, D.J. Van Harlingen,
J. Giapintzakis and D.M. Ginsberg, \journal Phys. Rev. Lett.
&74&797(95).}
  We shall refer to them as respectively the UIUC I, ETH,
Maryland and UIUC II experiments.  The circuit arrangements used in the
first three are essentially the same, the differences lying mainly in the
techniques used to infer the device's current-flux relation and the
precautions taken against the possible effects of trapped flux; the UIUC II
experiment uses a modification of the circuit arrangement.  All four type-II
experiments were done on   YBa$_2$Cu$_3$O$_{7-\delta}$ , with $\delta$ in
 the range
$0-0.1$, and all are interpreted by their authors as incompatible with any
s-wave order parameter (i.e. as favouring a d-state, though not necessarily
pure \dx2y2).  In addition, the Maryland group claim to exclude
violation of time-reversal invariance at any level above about 5\%.

The prototype geometry of these experiments is that of UIUC I (see 
\FIG\uiucfig{Geometry of the UIUC I experiment of
Wollman \etal\refmark{\wollmani}.} Fig. \uiucfig).
For pedagogical clarity we start by discussing an idealised version, in
which the circuit (including the two junctions) is exactly symmetric under
reflection in the diagonal axis.  The analysis is then extremely
straightforward and is simply a generalisation of that given originally by
Geshkenbein \etal\Ref\geshkenbein{V.L. Geshkenbein,
A.I. Larkin and A. Barone, \journal Phys. Rev. B &36&235(87); 
cf. V.L. Geshkenbein and A.I. Larkin, \journal Pis'ma Zh. Eksp. Teor.
Fiz. &43&306(86) [\journal JETP Lett. &43&395(86)].} 
for the case of p-wave pairing:  cf. Sigrist
and Rice\rlap.\refmark{\sigristrice}  We first assume for the 
moment that the YBCO pairing state
corresponds to a single irreducible representation, and 
note that the mere fact that a finite critical current is
observed across either junction individually then implies, by principle A,
that this representation must be invariant under reflection in a crystal
axis; this implies that as regards the bulk YBCO the operations of
rotation through $\pi/2$ and inversion in a $45^{\circ}$ axis are equivalent
 and we may
use either in our argument.  We choose for reasons of subsequent
convenience to consider $\pi/2$ rotation; in any case, the relevant pairing
symmetries are only  $s^+$ and      \dx2y2.

	We can now apply the analysis of Section 6 directly. 
 If the pairing state  is $s^+$,
then the circuit is just the standard symmetric dc-SQUID circuit
described in textbooks of superconducting electronics; the quantity
$\Delta\varphi_{tot}$
is zero and all the usual results hold.  If on the other hand the
pairing state is \dx2y2,
then the symmetry under $\pi/2$ rotation implies that the
couplings $A_{jj'}$
across the two junctions ($j=s$, $j'=d_{x^2-y^2}$)
have opposite sign: $A^{(2)}_{sd} = - A^{(1)}_{sd} $. 
Thus   $\Delta\varphi_{tot}=\pi$, and the current-flux
relation has an offset of half a flux quantum.  This is the basic principle
used in all the type-II experiments.

Let us now discuss the realistic case.  This is complicated by (a) the
orthorhombic bulk asymmetry of YBCO, (b) the lack of identity of the
junctions, and (c) the asymmetry of the bulk external circuit (and the
points of attachment of the junctions, etc.).  In addition one might ask
(d) how the above argument, which implicitly assumes that the faces to
which the junctions are attached are exactly perpendicular to the crystal
axis, would be affected if this is not rigorously the case.  In view of the
fact that the authors of all the type-II experiments report or infer a
current-flux relation corresponding to      $\Delta\varphi_{tot}=\pi$,
we
shall discuss only the question of whether any of the above considerations
allow this result to be compatible with the assignment of a single YBCO
pairing state other than \dx2y2.
For the moment we ignore complication (d) and assume a single irreducible
representation; then the above application of principle A goes through and
the only alternative to      \dx2y2\ is the simple   $s^+$  state.  Do
considerations (a)-(c) allow a loophole for this assignment?

It is immediately clear that consideration (a) is completely irrelevant,
i.e. it in no way changes the predictions for an 
s-state\Ref\considerationa{The converse statement is not {\it prima facie\/} true: had the experiments favoured 
$\Delta\varphi_{tot}=0$ rather than $\Delta\varphi_{tot}=\pi$, advocates
of the d-wave scenario might have tried to invoke the orthorhombic
anisotropy (unsuccessfully we believe) to explain them.} 
(many existing dc
SQUIDS use classic superconductors whose crystal symmetry is not
tetragonal!).  Consideration (b) is relevant, in the sense that were a
single experiment to come out in favour of the conclusion that
$\Delta\varphi_{tot}=\pi$, one
might try to reconcile it with an s-wave scenario by postulating
that one of the junctions involved was normal and the other an
(``intrinsic'') $\pi$-junction.  However, quite apart from the fact that to our
knowledge no experimentalist has ever reliably seen an intrinsic
$\pi$-junction, the UIUC I experiment in particular relies on statistics
obtained with a very large number of different pairs of junctions, and it
stretches credulity that the preponderance of ``normal-$\pi$'' pairs in the
nominally random ensemble could be sufficient to explain the results.

Complication (c) is by far the most annoying in practice.  For a strictly
symmetric ``s-wave'' dc SQUID circuit (meaning that not only the inductances,
but also the critical currents of the two junctions are
identical) the maxima and minima of the critical current occur rigorously
at integral and half-odd-integral values of  $\Phi_{ext}/\varphi_0$
respectively,
and this remains true even in the presence of finite self-inductance
effects.  (This statement follows essentially from the conditions of
time-reversal invariance and periodicity of   $I_c$     in  
$\Phi_{ext}$  and the 
experimental observation that  $I_c(\Phi_{ext})$ is monotonic 
over a half-cycle).
However, an asymmetry in any of the circuit parameters will in general
invalidate this result, since (e.g.) even for $\Phi_{ext}=0$,
by the time the critical current is reached there will in general be a
finite current {\it circulating} in the ``ring,''
 and this will produce its own flux.  
Thus a naive measurement of the dependence of  $I_c$  on the
``applied" flux will in general give totally misleading results.  We refer
the reader to the original papers (cf. also Ref. 
\REF\vanharlingen{D.J. Van Harlingen, \journal Rev. Mod. Phys.
&67&515(95).} \vanharlingen) for the various
ingenious experimental procedures which have been devised to circumvent
this problem; one of them, that of UIUC II, will be discussed further
below.

Next we turn to complication (d), and the related question of the possible
occurrence of ``mixed'' representations:  note that although in Section
we have given strong {\it a priori\/} arguments based on thermodynamic
considerations for excluding this possibility, it is still of interest to
subject it to direct experimental test.  We first note that if we ignore
for the moment consideration (d), i.e. assume all junction planes are
rigorously normal to the crystal axes, etc., then none of the existing
type-II experiments can exclude an arbitrary admixture of the $s^-$  or
$d_{xy}$ 
states, since those are odd under reflection in the crystal axes and
hence by principle A cannot contribute to the single-junction coupling
energy.  (However, as suggested in Ref. \REF\beasleylew{
M.R. Beasley, D. Lew. and R.B. Laughlin, \journal Phys. Rev. B &49&12330(94).}
\beasleylew, future experiments using
faces cut at $45^{\circ}$  could in principle do so.)

Next, what are the consequences of a very small misalignment of the
(macroscopic) junction plane with the crystalline axes?  In principle, if
the misalignment is accurately known, this might put some constraints on
the  $s^-$    and  $d_{xy}$  admixture, but those are likely to be exceedingly
weak.  A more interesting question is whether such misalignment provides a
loophole for the simple s-wave ($s^+$) scenario.  It is clear that the
answer is no, unless the state is so extremely anisotropic that it actually
changes sign (eight times) as we go around the Fermi surface (perimeter).
Even with this extreme assumption it would appear to require an almost
pathologically violent variation of the order parameter and/or the tunnelling matrix
elements with angle.  We shall therefore not discuss this theoretical
possibility further here.

The most interesting ``mixed'' possibility, if we are prepared despite the
thermodynamic counter-argument of Section 4 to consider such 
states\rlap,\Ref\contuerargument{Actually, for (orthorhombic) YBCO 
$s^+- d_{x^2-y^2}$  mixing is forbidden by this argument only in
the thermodynamic limit (cf. Section 4.2).}
is a mixture of  $s^+$ and    \dx2y2\   with some definite relative
phase:  schematically, the (twin-averaged) order parameter would be
$$ \Psi \sim d_{x^2-y^2} + \beta e^{i\varphi} s^+ .\eqn\sdmixture$$
This ansatz has the attraction that it could apparently explain the UCSD
experiment (in which the  \dx2y2\  component is irrelevant).  (It
is more commonly written schematically, for   $\varphi=\pi/2$, as ``$s+id$''.)
It may be seen that such a state leads, in the ``standard'' type-II
geometry, to a total phase shift  $\Delta\varphi_{tot}$  
given by the expression
$$\Delta\varphi_{tot}=\tan^{-1}\left(\frac{-2\gamma
\sin\varphi}{1-\gamma^2}\right),
\eqn\tanchi  
$$
where $\gamma\equiv \beta A_{ss}/A_{sd}$.
All the type-II experiments indicate that  $\Delta\varphi_{tot}$
is close to $\pi$, and
in particular the Maryland experiments are reported as indicating that
$\Delta\varphi_{tot}$
differs from $\pi$  by no more than about 5\%.  
Since there seems no good reason
to believe that the quantities   $| A_{ss} | $    and    
 $| A_{sd} | $     are different in order
of magnitude, this would then indicate that the {\it imaginary} part of any
s-wave admixture cannot be much greater than this.

We finally review a couple of miscellaneous features of the type-II
experiments which are relevant to points raised elsewhere in this review.
First, both the UIUC I and the Maryland experiments included ``control''
experiments in which the two junctions, rather than being attached in the
standard ``corner'' geometry of Fig. \uiucfig, were attached to the same edge
(UIUC I) or the opposite edge (Maryland) of the YBCO crystal.  In both
experiments the data indicate that $\Delta\varphi_{tot}$  is equal or close to
zero in this geometry.  Since all the Maryland samples and some of the UIUC
I samples were heavily twinned, this can be reconciled with the \dx2y2\
assignment required by the ``corner'' data if and only if the order parameter
behaves gyroscopically across a twin boundary; we already used this
conclusion heavily in Section 4.

Secondly, we note that the UIUC II experiment was done not on a true SQUID
geometry, but on a single junction attached at
the corner of the YBCO plate, so that the current flowing across the ``NS"
and ``EW" edges was roughly equal.  Under these conditions a simple
extension of the theory of the Fraunhofer modulation quoted in Section 6
leads, for an s-wave state, to the same expression \fraunhofer\ as for an
``edge'' junction, while for a   \dx2y2\ state it predicts a
symmetrical but ``split'' Fraunhofer peak (i.e. the critical current is a
minimum at  $\Phi=0$).  For the case of exact symmetry with respect to
the diagonal through the corner the formula which replaces \fraunhofer\ is
$$  I_c(\Phi) = I_c \left| { \sin{(\pi\Phi/2\varphi_0)} \over 
 (\pi\Phi/2\varphi_0)} \right|  . \eqn\replacesfraunhofer $$ 
A major advantage of this arrangement over that of UIUC I is that the
effective self-inductance of the ``circuit'' is much smaller and thus we can
essentially replace $\Phi$ in the Fraunhofer formula by the external flux
$\Phi_{ext}$ which is directly measured.  The UIUC II experiment shows a split
Fraunhofer pattern which is incompatible with formula \fraunhofer\ and
reasonably close to \replacesfraunhofer; the residual discrepancy can be
reasonably accounted for by the lack of exact reflection symmetry.

To sum up, all the type-II experiments indicate that the order parameter of YBCO, or at
least a substantial component of it, has      \dx2y2\   symmetry; they
are incompatible with an $s^+$  component such that $A_{ss} > A_{sd}$,
and in addition the Maryland experiment excludes an {\it imaginary}
admixture at a level much greater than 5\%\rlap.\Ref\admixture{
We note that this applies to the ``plane-averaged'' order parameter. The Maryland experiment
does not by itself exclude the possibility of (e.g.) ``$s+id$'' on one layer of
a bilayer and ``$s-id$'' on the other.}
  No experiment of this class
can {\it by itself} (i.e. without appealing to the thermodynamic argument of
Section 4.1) exclude an arbitrary admixture of $s^-$   and/or $d_{xy}$.

{\bf\subsection{Type-III experiments}}

As mentioned above, type-III experiments involve a circuit whose bulk
elements are entirely composed of some high-temperature superconductor
epitaxial films, separated by grain boundaries so that different
bulk regions have different crystal orientations.  To date there are three
such experiments on YBCO in the literature which are reported in refs.
\chaudhari,
\REF\tsuei{C.C. Tsuei, J.R. Kirtley, C.C. Chi, L.S. Yu-Jahnes, A. Gupta
T. Shaw, J.Z. Sun and M.B. Ketchen, \journal Phys. Rev. Lett. &73&593(94).} 
\tsuei\
 and \REF\tcsuh{J.H. Miller Jr., Q.Y. Ying, Z.G. Zou, N.Q. Fan, J.H. Xu,
M.F. Davis and J.C. Wolfe, \journal Phys. Rev. Lett. &74&2347(95).}
\tcsuh; 
those will be referred to as the IBM I, IBM II and
TCSUH experiments respectively.  Very recently, the IBM II experiment has
been repeated on Tl
 2201\rlap.\Ref\kirtleytltt{J.R. Kirtley \etal, preprint.}

In the IBM II experiment (Fig. \FIG\kirtleyfig{The geometry of the
IBM II experiment, as reported in Ref. \tsuei.} 
\kirtleyfig) a ring was constructed out of three
different bulk regions of (heavily twinned) YBCO, with the crystal axes and
grain boundary orientations as indicated in the Figure.  The system was
placed in zero external field and the flux generated by circulating
currents spontaneously induced in the ring was measured; those were found to be
finite and to correspond closely to the result expected if the ring is
``spontaneously'' generating a half-odd-integral number of flux quanta.  In
a control experiment\rlap,\refmark{\tsuei} rings containing zero or two grain boundaries were
found to show integral trapped flux,
as did also a three-junction experiment\Ref\kirtnature{J.R. Kirtley \etal,
\journal Nature &373&225(95).} with the junctions differently oriented in
relation to one another and the crystal axes (the so-called ``zero-ring"
experiment).
In a more recent experiment by the
same group\rlap,\Ref\kirtleyhole{J.R. Kirtley, C.C. Tsuei, M. Rupp, J.Z. Sun,
L.S. Yu-Jahnes, A. Gupta, M.B. Ketchen, K. Moler and M. Bhushan, preprint.}
the hole in the ring was in effect 
removed, and vortices
carrying a half-quantum of flux were found to occur at the ``tricrystal
point'' where the three grain boundaries meet (see Fig. \kirtleyfig).

The TCSUH experiment used the geometry shown in Fig. \FIG\tcsuhfig{
Geometry of the TCSUH experiment, after Ref. \tcsuh.} 
\tcsuhfig.  This is
conceptually quite close to the geometry of the more recent
IBM experiment, and in
fact one would expect that under certain circumstances, in an expanded
version of this geometry (or one with a hole in the middle), a
``half-quantum'' vortex would be generated around the tricrystal
point.  In fact the
dimensions are too small (in the principal experiment) for this to happen,
and rather than look for spontaneously induced flux the authors measured
the critical current of the device as a function of the external flux
applied to the microbridge region.  As they remark, it is possible to
regard this experiment as analogous to UIUC II, (with the role of the Pb in
the latter being played by region III).  As in UIUC II, a ``split''
Fraunhofer pattern was observed, i.e. the critical current is a (local)
{\it minimum} for zero applied flux.  In a second experiment using a much wider
microbridge ($40\mu$ rather than $3\mu$) the Fraunhofer pattern was of the
``normal'' type.  It should be noted that unlike in the IBM II experiments,
the characteristic angles are not close to multiples of $30^{\circ}$ (the
significance of this will become clear below).

Finally, in the IBM I experiment 
the Josephson critical current was measured between a
hexagon of YBCO separated from an outside YBCO region by grain boundaries
and oriented at approximately $45^{\circ}$ with respect to it (see Fig. 
\FIG\chaudharifig{Geometry of the IBM I 
experiment\rlap.\refmark{\chaudhari}} \chaudharifig).  The
measurement was first made for the complete hexagon and then repeated after
the various hexagon faces (grain boundaries) had been ``chopped away'' one by
one by laser ablation (leaving barriers which are essentially completely
insulating).  The salient results for our purposes are that (1) a finite
critical current is observed for the complete hexagon, and (2) this current
decreases, as the various faces are chopped away, approximately in
proportion to the remaining length of face.

In the discussion of the implications of these experiments implicit
assumptions have often been made which are not a consequence of
basic symmetry considerations.  Let us start our own discussion by stating
a few conclusions which rely only on very general principles.  In doing so,
we assume that the applied fluxes, critical currents etc which the authors
quote are the ``true'' ones (e.g. there are no trapped-flux effects).  
We further assume 
that the junctions are
all in the thermodynamic limit with respect to the twinning and are
statistically identical, and moreover (for the moment) that ``defrustration''
can be neglected.
First,
it is clear that neither the IBM II nor the TCSUH results are compatible
with the conjunction of (a) an order parameter which is everywhere real and of the same
sign (i.e. a nodeless  $s^+$  state) with possibly small admixtures of other
symmetries, and (b) junctions which are all of ``normal'' type (i.e. not
(intrinsic) $\pi$-junctions).  This is because under these circumstances it
is straightforward to verify that we can choose our phase conventions so as to make
all the Josephson coupling energies $E_J$ positive, and the circuit in question
will then behave exactly like a ``textbook'' circuit made of (say) Al, and
will show neither a spontaneous flux nor a split Fraunhofer pattern.  Since
we regard the occurrence in nature (let alone in these experiments!) of
(intrinsic) $\pi$-junctions as problematical, we would regard this conclusion
as effectively disposing of the possibility of a real nodeless order parameter, and in
particular of a nodeless    $s^+$    state.

Secondly, while the IBM I experiment is clearly compatible with (and
suggests) an s-wave order parameter, we want to emphasise that, 
{\it taken by itself} and
without additional thermodynamic or other considerations, it {\it does not}
rule out the possibility of a d-state 
(cf. the remark below Eqn. (2) of Ref. \chaudhari). 
Indeed, any experiment using a circuit in which the
bulk components are YBCO only which can be explained by an s-wave order
parameter can equally well be accounted for by the hypothesis that the order parameter
is of the form  $d_{(x+iy)^2}$
(i.e. $\Delta ({\bf k}) \sim ( k_x + i k_y)^2$). 
(In essence this is because the only way,
in this case, in which the order parameter depends on direction is a phase {\it  which  is
independent of the crystal axis orientation} and therefore cancels between
the order parameters on the two sides of the junction.)  While, in view of the evidence
for rather strong crystal lattice effects, we would not regard the
$d_{(x+iy)^2}$
state itself as particularly plausible, it is clear by continuity
that a mixture of the two different irreducible representations
   \dx2y2\     and    $d_{xy}$      of the form
$$ d_{x^2-y^2} + i \alpha d_{xy} \eqn\dplusid $$
with $\alpha$ close to 1, would also replicate the
 s-wave behaviour  at least to a
good approximation.  We note that such a state is not excluded by the
type-II experiments; in fact the only compelling arguments against it
are  the thermodynamic considerations  reviewed in Section 4, and the 
existence of a node in the gap function as described in Section 5.

The further conclusions drawn by the authors of the cited papers
\refmark{\chaudhari,\kirtleytltt} appear to
us mostly to rest on the explicit or implicit assumption of a very specific
form of the Josephson coupling as a function of the orientation of the
crystal axes along 
the two sides, namely the Sigrist-Rice expression
 \sigristriceansatz\ 
or, at least, the ``forward-scattering'' ansatz.  For the reasons given in
Section 6, we would therefore regard these arguments as suggestive
rather than compelling.

However, provided we disregard the small deviation ($\sim 3^{\circ}$) of the actual
angles in these experiments from (what we take to be) the ``design'' angles,
it is still possible to draw some conclusions from symmetry arguments
alone.  Let us for example consider the three junctions of the IBM II
experiment:  for clarity we omit the hole in the ring and bring them
together to a point as in the later experiments of this group.
The crucial point is that, to the extent that the actual angles
approximate to the design angles, the three grain boundaries {\it plus}
 the bulk
crystal domains on either side are related to one another by simple
symmetry operations, and are therefore ``equivalent'' provided we do the
accounting of the axes right.  Suppose first that only a single irreducible representation is
represented in the (twin-averaged) YBCO order parameter.  There it is clear
that $\Delta\varphi_{tot}$  must be zero or $\pi$
 according as to whether this irreducible representation is even or
odd under the product of $\pi/2$ rotation and inversion in a crystal axis, i.e.
under inversion in a $45^{\circ}$ axis (note that this conclusion does not assume
that the area of the three junctions is the same).  Thus, the experimental
observation that   $\Delta\varphi_{tot}$  is (close to) $\pi$    
limits the possible single
irreducible representation to \dx2y2\ or $s^-$.
{\it The IBM II experiment by itself, without further assumptions
about tunnelling matrix elements cannot
distinguish between these two possibilities.}

If, despite the thermodynamic arguments, we allow for more than one irreducible representation,
things become a good deal more complicated:  since the individual grain
boundaries have no special symmetry relative to both the bulk regions, any
of the four possible irreducible representations in one bulk region can couple to any in the
other, giving 16 independent possible coupling constants $A_{jj'}$
(note that the matrix  $A_{jj'}$ is not in general symmetric).  The
value of   $\Delta\varphi_{tot}$ 
is then determined by minimisation of the total
Josephson energy
$$ E_J = - \sum_i \sum_{jj'} A^{(i)}_{jj'} \psi_j \psi_j^* 
\eqn\totaljosephson$$
where $i$ labels the various junctions, and in general $\Delta\varphi_{tot}$ 
need not be $0$ or $\pi$.
The observation that   $\Delta\varphi_{tot}$ is in fact close to $\pi$
  then merely tells
us, crudely speaking, that the Josephson coupling is dominated by irreducible representations
which change sign under inversion in the $45^{\circ}$ axis, i.e. by \dx2y2\
and/or  $s^-$        ; it cannot exclude finite admixtures of    $s^+$  
 and $d_{xy}$.

A similar analysis can be applied to the IBM I experiment, with the
slight simplification that the rather higher degree of symmetry puts some
restrictions on the individual    $A_{jj'}$.  We see immediately from Fig.
\chaudharifig\
that since the complete hexagon is invariant under simultaneous reflection
of the inside grain in a crystal axis and the external one in a $45^{\circ}$ axis,
it would follow from an argument similar to that applied to the IBM II
experiment that if the order parameter corresponds to a single irreducible representation it must be even
under the product of these two operations, or equivalently even under $\pi/2$
rotation--that is, it must be either  $s^+$  or  $s^-$.  It is convenient,
however, to split this symmetry argument into two parts:  supposing that
the only irreducible representation represented is odd under the product of   $\hat{I}_{axis}$
 and $\hat{I}_{\pi/4}$,
then (a) faces 1 and 4 must individually give zero contribution to
the Josephson current, and (b) faces 5 and 6 (and 2 and 3) must cancel one
another.  As to point (a), the fact that face 1 alone gives a non-zero
current (see Fig. 2 of Ref. \chaudhari) seems to rule out $s^-$, leaving
only $s^+$.

As regards point (b), an important observation has been made by 
Millis\rlap:\Ref\millis{A.J. Millis, \journal Phys. Rev. B &49&15408(94).}
The argument for cancellation depends critically on the assumption that
there are no currents in the ``bulk'' inside the grain (so that in the absence of
a vector potential the phase is constant throughout this grain as in Eqn.
\circuitphase). That is, it requires, {\it inter alia}, that in 
\estimates\ the kinetic
energy   $E_{kin}$  is much larger than the Josephson energy $E_J$.  Millis
points out that this is probably not so in the actual experiment, and that
assuming the order parameter is d-wave, the system will then try to ``defrustrate'' by
allowing the phase to vary throughout the internal 
region\rlap.\Ref\defrustrate{A similar argument can be applied to the TCSUH wide-bridge experiment, explaining why a normal Fraunhofer
pattern is seen\rlap.\refmark{\tcsuh}}  Unfortunately,
while this argument can certainly explain how the total Josephson coupling
can be nonzero even if the pairing state is pure  \dx2y2,
it does not by itself account for the fact that the contribution from faces
1 and 4 appears to be finite (and more generally that the total critical
current is found to be simply proportional to the layer exposed).  Millis'
own tentative explanation, which is based on the possible ``jaggedness" of
the grain boundaries which form the junctions, seems to us to require that
the scale of this jaggedness is so great that we are not in the
``thermodynamic limit'', which seems rather unlikely.  In view of the
poorly understood nature of the grain boundary, one cannot entirely
exclude explanations of this general nature.

Once we consider the possibility of more than one irreducible representation, there are few
useful conclusions we can draw from the IBM I experiment; in
particular, as already noted, it cannot exclude a state of the form
$d_{x^2-y^2}+i\alpha d_{xy}$ with $\alpha$
fairly close to 1.

Finally, we note that the analysis of the TCSUH experiment, in which
there appears to be no particular equivalence between the various grain
boundaries, apparently cannot be carried out as above, but would require a
specific hypothesis about the form of the Josephson matrix elements.
The same applies to the ``zero-ring" experiment of 
Kirtley \etal\refmark{\kirtnature}

{\bf\subsection{Summary}}

To summarize our conclusions concerning the Josephson experiments:  (1)
If, relying on the thermodynamic argument of Sections 3 and 4.2, we
allow only a single irreducible representation of $C_{4v}$  for the
(twin-averaged) order parameter, then the conclusions which appear, {\it
prima facie}, to follow from the various Josephson experiments are, on
symmetry grounds alone, not mutually compatible:  under this hypothesis
the UCSD experiment allows only  $s^+$, the IBM I experiment only
$s^+$ or $s^-$, the IBM II experiment only   \dx2y2\  or $s^-$ and the
four type-II experiments only \dx2y2.  The only obvious way to reconcile
the various experiments is to assume that neither in the UCSD nor in the
IBM I experiment was the ``thermodynamic limit'' realised, but we
have seen above that this is not obviously plausible.  (2) If we allow
more than one irreducible representation, then to explain all the
experiments we require as a minimum a mixture of  $s^+$ and \dx2y2, with
some $s^-$  and/or  $d_{xy}$      component as an ``optional extra.''
However, the Maryland experiment excludes an $s^+ + \hbox{\dx2y2}$
mixture with relative phase close to $\pi/2$ (assuming a reasonable
amplitude for each component).  Hence the most plausible assignment
would seem to be a real mixture of $s$ and d.  This would be compatible
with the IBM I experiment if the contribution of the d-wave
component is largely washed out by scattering at the grain boundary.  We
emphasize that it is the twin-averaged order parameter which must have
this character, and that therefore the mechanism for producing it can
have nothing to do directly with the orthorhombic asymmetry of YBCO,
which as we have mentioned above is (given the gyroscopic assumption) in
the Josephson context largely irrelevant.

{\bf\chapter{Effects of Impurities}}

One of the most commonly-expressed doubts about the reality of d-wave
pairing in high temperatures superconductors is the presumed extreme
sensitivity of a d-wave superconductor to dirt.  It is natural to
suppose that a small amount of impurity would generate enough s-wave
scattering as to completely destroy the superconducting state even at
the nominal levels of dirt surely present in even the best samples of the
high temperature superconductors.  This naive argument does not fully
take into account the very short coherence length of the cuprates, and
needs to be investigated quantitatively before any conclusion can be
drawn.  Fortunately, an elegant series of experimental measurements have
recently been completed, which shed considerable light on this important
issue.

We have chosen to discuss several experiments below, which have the
feature that their interpretation is to a large extent
model-independent.  Perhaps the strongest assumptions are the
correctness and applicability of the standard theory of disordered
superconductors, about which more will be said below.  The experiments
fall into two classes: impurity
doping\rlap,\refmark{\sumner,}
\REFS\ishidanmr{K. Ishida \etal, \journal Physica B
&186-188&1015(93).}\REFSCON\kimohio{J.-T. Kim \etal, \journal Phys. Rev.
B &49&15970(94).}\refsend usually with Zn or Ni, and planar oxygen
displacement by electron irradiation\rlap.\REFS\basovrad{D.N. Basov
\etal, \journal Phys. Rev. B &49&12165(94).}\REFSCON\giapi{J.
Giapintzakis \etal, \journal Phys. Rev. B
&50&15967(94).}\refsend\refmark{,\giap}

{\bf \section{Doping studies}}

It is generally believed that Ni and Zn are the only substitutes for Cu
which reside on the CuO$_2$ planes, although Al also resides
on both planes and chains.  Naively, one would expect Zn to
behave as a non-magnetic impurity, whereas Ni would behave as a magnetic
impurity.  Early measurements\Ref\ishearly{K. Ishida \etal, \journal
Physica C &179&29(91).} as well as the more recent
measurements\refmark{\ishidanmr} of the nuclear spin-lattice relaxation
time $T_1$ and the $^{63}$Cu Knight shift in Zn-doped \Yba\ indicated
that the low temperature Knight shift increases with Zn content, and that
the relaxation rate becomes Korringa-like at low temperatures.  Both of
these results are consistent with the notion that there is a non-zero
density of states at the Fermi surface in the doped crystal, as would be
expected for an unconventional singlet state.  The Ni doping does not
produce the same results: the Knight shift is essentially unchanged,
and the relaxation time saturates at low temperature, rather than
obeying the Korringa behaviour found in the Zn-doped samples.  Thus, it
seems that Ni induces localised magnetic moments.  The origin of the
difference between these two behaviours is not well-understood.
Nevertheless, these results appear inconsistent with an s-wave pairing
state.

The effects of Ni doping on \Yba\ films have been explored by infra-red
measurements\rlap.\refmark{\sumner} At the high concentrations
studied, namely 4 at. \%, the elastic scattering rate is observed to
increase in the normal state to such an extent that if the same
scattering is assumed to be present in the superconducting state, a
large absorption onset would have been visible in the infra-red
reflectance.  This was not observed, suggesting that there is indeed no
gap function minimum in the \Yba\ films studied.  Similar results were
later confirmed for the Zn-doped films\rlap.\refmark{\kimohio}\

The changes in \Tc\ with doping have been extensively studied
by many groups. For a review see, for example, Ref. \REF\westerholt{
K. Westerholt and B. vom Hedt, \journal J. Low Temp. Phys.
&95&123(94).} \westerholt. What is immediately striking about the data
is that \Tc\ decreases {\it linearly} with dopant concentration
in almost all cases, especially in the concentration $\rightarrow 0$
limit.
Obviously one must be cautious in interpreting this result
since there is no generally accepted theory of \Tc\ itself
in the pure material. Nevertheless it would be hard to understand
this behaviour in an isotropic s-wave picture, since 
as is well known, elastic (s-wave) non-magnetic impurity scattering 
has no effect on \Tc\ for s-wave superconductors\rlap,\Ref\rickayzen{
See e.g. G. Rickayzen, {\sl Green's Functions and Condensed Matter}
(Academic Press, London, 1980), sec 8.5 and references therein.} 
a result which is usually known as ``Anderson's theorem''.
In s-wave superconductors \Tc\ decreases linearly with
magnetic impurities only.  The linear decrease in \Tc\ with
impurity concentration 
is thus not surprising for the magnetic
impurities (Ni and Fe) but was unexpected for a non-magnetic
impurity such as Zn.  However, this result is not hard to
understand in qualitative terms.  ``Anderson's
theorem'' relies upon a cancellation that occurs only if the gap
is exactly isotropic. In an anisotropic superconductor (whether
anisotropic s-wave or d-wave) this cancellation does not occur, and in 
general \Tc\ will decrease linearly with impurity concentration.
The results are thus qualitatively
consistent with either d-wave pairing, or anisotropic s-wave, and
they cannot be said to provide any stronger constraints on the
pairing state than to eliminate isotropic s-wave.

{\bf \section{Irradiation studies}}

Irradiation studies are possibly a more clear cut test of the
sensitivity of the superconductivity to disorder, since 
unlike doping studies there is no danger of substantially altering
the electronic structure at the same time as introducing disorder.
Basov \etal\Ref\basovexp{D.N. Basov 
\etal, \journal Phys. Rev. B &49&12165(94).}
measured the evolution of the infra-red optical conductivity
in a \Yba\ crystal both before and after irradiation with low-energy
He ions. 
The irradiation suppressed \Tc\ from 93.5K to 80K. Basov \etal\
observed that  after irradiation the optical conductivity was finite at
all frequencies, and even developed a Drude-like low frequency peak.
Recently, detailed calculations of the far-infra-red conductivity
were performed for a \dx2y2\ superconductor with both elastic and
inelastic scattering\rlap.\REFS\carbotteii{J.P. Carbotte,
 C. Jiang, D.N. Basov and 
T. Timusk, \journal Phys. Rev. B &51&11798(95).}\REFSCON\graf{
M.J. Graf, M. Palumbo, D. Rainer and J.A. Sauls, \journal
Phys. Rev. B &52&10588(95).}\REFSCON\quinlan{
S.M. Quinlan, P.J. Hirschfeld and D.J. Scalapino, preprint.}\refsend\
These calculations show that the qualitative evolution of
the spectra with increasing disorder, including the Drude-like peak,
could be understood naturally in a d-wave picture. In contrast
the optical conductivity of disordered isotropic s-wave  
is qualitatively different\rlap.\Ref\akis{ R. Akis, 
J.P. Carbotte and T. Timusk, \journal Phys. Rev. B &43&12804(91).}\
It is particularly noteworth that, at least within the Eliashberg
formalism of the calculations, there is a reasonable {\it quantitative}
agreement between the decrease in \Tc\
under irradiation, and the observed spectral shape\rlap.\refmark{\carbotteii}\
In other words, there is a consistent choice of
parameters such that the same defect scattering rate
leads to both the observed  \Tc\ decrease and the spectral shape.
Carbotte \etal\ emphasize that the calculated spectral shape
would be essentially the same for any gap function $\Delta({\bf k})$
with a node, including extended s-wave states, but 
would be inconsistent with states in which $\Delta({\bf k})$
does not change sign (such as $| k_x^2 - k_y^2 | $).
Thus, we conclude that the d-wave
pairing state accounts qualitatively (and possibly even quantitatively)
for the experimentally observed
features of the far infra-red conductivity.  Perhaps the main question
about the theoretical calculations for this comparison is the use of the
Eliashberg theory: it is by no means obvious that this is applicable,
given that there is presently no understanding of the microscopic
mechanism, and hence no justification to assume Migdal's theorem
applies.  Nevertheless we believe that the comparison is probably
qualitatively correct, in the sense that it is hard to see how a
more detailed theory could restore agreement between the s-wave
calculations and the experimental observations.

A very elegant technique to explore the effects of impurities on the
cuprates is to use electron irradiation to displace atoms from the
crystal lattice.  These studies\refmark{\giap,\giapi} were performed
on high quality, detwinned single crystals, and the electron energy was
chosen so as to selectively remove the planar oxygens O(2,3); a careful
study was undertaken to demonstrate that this had in fact occurred.  The
resulting defects do not contribute a Curie-Weiss tail to the normal
state susceptibility, neither do they affect the plasma frequency nor
the carrier concentration.  Thus, the resulting samples contained
non-magnetic defects in the CuO$_2$ planes, without any change in
carrier concentration or plasma frequency, at least to the extent
that could be determined experimentally.  The a-axis resistivity,
measured as a function of irradiation dose, indicates clearly that
for 4.1\% displacement of planar oxygens, the superconducting transition
temperature is zero, or at least less than 12K, the lowest
temperature attained in the experiment.  If we follow the
authors, and assume that \Tc\ actually
is zero, or would have become zero at the appropriate extrapolated
irradiation dose, then it appears that these
data are
in sharp contrast to the
expectation based upon a s-wave state and non-magnetic scattering, viz.
an originally anisotropic s-wave
state would become isotropised by the impurity scattering, leading to a
non-zero value of \Tc\ above some critical concentration of impurities.
We conclude two things: first, the pairing state must have nodes, but
could be either d-wave or an extended s-wave state (with 8 nodes).  The
latter state is ruled out by (\eg) the photoemission data on gap
anisotropy.  Secondly, the d-wave state may have a small admixture of $s$,
but not so much that the gap function symmetry reverts to s-wave before
the superconductivity is destroyed.  Thus, this experiment places an
upper bound on any non-zero value for the gap function integrated over
the Fermi surface.

The results of electron irradiation experiments agree qualitatively with
the earlier study of Sun \etal\Ref\sunpr{A.G. Sun \etal, \journal Phys.
Rev. B &50&3266(94).} who also found that \Tc\ gets driven to zero in
both Pr-doped and ion-beam damaged \Yba\ single crystals.  These authors
did not attempt to determine if the resulting impurity scattering was
magnetic or not.  In both experiments, taken in isolation from the body
of other data described in this Chapter, it would be conceivable that
the pairing state is s-wave, and that there is a small magnetic
component to the scattering: a quantitative comparison of the
\Tc\ dependence and the experimentally implied upper bounds on magnetic
scattering would be needed to explore this possibility.  A potentially
serious issue for the \dx2y2\ scenario, which is still not resolved, 
is the large quantitative
discrepancy between the observed \Tc\ in the experiments of Sun
\etal\ and calculations based on Eliashberg theory combined with the
observed residual resistance at
\Tc\rlap.\refmark{\sunpr,}\Ref\radtkee{R.J. Radtke \etal, \journal Phys.
Rev. B &48&653(93).}

There are three logically distinct points at which any of the above
arguments could fail.
Firstly, the applicability of Abrikosov-Gor'kov theory to the cuprates
is certainly not obvious {\it a priori}. 
It should be emphasised that the Abrikosov-Gor'kov theory is in
essence a variational calculation, which compares the free energy of one
possible kind of pairing state, namely that in which pairs form in
plane-wave states, despite the fact that these are no longer eigenstates
of the single-particle Hamiltonian in the presence of impurities, with
the normal-state free energy.  While it is clear that some obvious
alternatives, e.g. pairing in the basis of single-particle
eigenstates, do worse than the Abrikosov-Gor'kov ansatz,it is not
entirely obvious that there is no possible alternative which might do
better in the rather unusual environment of the high temperature
superconductors, and this must remain a caveat with respect to
conclusions drawn on this basis.

Secondly, as mentioned earlier,
Eliashberg theory itself is dubious in the cuprates
since it is unlikely that Migdal's theorem is valid if the
pairing mechanism is either purely electronic (e.g. spin fluctuation
exchange) or involves the lattice (e.g. bipolaron theories).

Thirdly, even assuming that the above two points can be dealt with,
it is far from clear that all the conclusions of
Abrikosov-Gor'kov theory are then strictly valid in a disordered
two-dimensional system.
In particular Wenger and Nersesyan\Ref\wenger{F. Wenger and
A.A. Nersesyan, {\it Strong vertex corrections from weak disorder
in two-dimensional d-wave superconductors}, preprint.}\ have shown that the
vertex corrections are large for two-dimensional  d-wave superconductors,
and hence Abrikosov-Gor'kov theory may fail for the quasiparticle states
near the gap-node.  In particular they also found\Ref\tsvelik{
A.A. Nersesyan \etal, \journal Phys. Rev. Lett.
&72&2628(94);
A.A. Nersesyan \etal, \journal Nucl. Phys. 
B &438&561(95).}
 a density
of quasiparticle states near the d-wave gap node of the form 
$\rho(\omega) \sim | \omega | ^\alpha$ as $\omega\rightarrow 0$ with $\alpha < 1$
which is qualitatively different from the d-wave Abrikosov-Gor'kov
result $\rho(\omega) \sim {\rm const}$. On the other hand, numerical
calculations\Ref\xiang{T. Xiang and J.M. Wheatley, \journal
Phys. Rev. B &51&11721(95).}\
of the density of states in a disordered two-dimensional d-wave
superconductor were qualitatively 
similar to the Abrikosov-Gor'kov predictions (cf. Ref. \xiang, Fig. 8).

{\bf\chapter{Conclusions}}

No single pairing state, whether a pure
irreducible representation  or a mixture, 
is consistent with all of the experimental results, taken 
at face value, 
which we have described above. Nevertheless we feel that
a clear picture is definitely emerging from the data taken as
whole. In fact there is a remarkable degree of consistency
between quite different experiments performed on the whole range of
the cuprate superconductors (except the Nd compounds).
The temperature dependence of the penetration depth clearly
indicates that the gap function  has nodes on the Fermi surface,
and furthermore, that $\Delta({\bf k})$ changes sign there.
On the other hand the angle resolved photoemission experiments 
show that $| \Delta({\bf k})|$ is small  or zero
at (or very close to) the \dx2y2\ nodal position. 
Taken together these two results alone imply that the
pairing state is predominantly \dx2y2\ with possibly
small admixtures of $s^+$, $s^-$, or $d_{xy}$. Any such admixtures
could be real or imaginary, but in either case would appear to be no
more than $10\%$ of the \dx2y2\ gap function.  A wide range of
other experimental observations, such as: the electronic Raman scattering,
the infra-red conductivity, 
and the neutron scattering, all can be understood naturally 
in the context of such a gap function.  There appear to be no spectroscopic
experiments which cannot be understood within this framework.
Also the extreme sensitivity
of  d-wave pairing to impurities is consistent
with such a gap function, since the impurity dependence of such a
state is indeed at least qualitatively consistent with what is observed.
Some of these spectroscopic measurements, for example the
photoemission, could also be consistent with a gap which did not
change sign, such as $ | k_x^2 - k_y^2 | $, but such a
state would not agree with either the impurity dependence of 
penetration depth of infra-red conductivity, or the Raman
scattering. The spectroscopic evidence, taken as a whole, 
is thus quite clearly in favour of a predominantly \dx2y2\
gap function.

As to the Josephson experiments taken at face value, we have seen
in Section 7 that if we use the thermodynamic argument of Section 4
to exclude the possibility of more than one irreducible representation, then
there is no single assignment of pairing states which will account for
all the data.  However, the preponderance of the evidence clearly points
to a \dx2y2\ pairing state.  If we relax the thermodynamic constraint so
as to allow a mixed pairing state, the data can all be accommodated by a real
mixture of $s^+$ and \dx2y2, with possibly also an admixture of $s^-$
and $d_{xy}$.  It should be emphasised that this possibility does {\it not\/}
arise from the orthorhombic distortion of YBCO; nevertheless, it would
clearly be valuable to repeat the type-I and and type-II experiments
on the purely tetragonal Tl compound.

In conclusion, the only type of state which is compatible with all the
experiments we have reviewed is, then, a mixed state with an appreciable
\dx2y2\ component.  The most plausible candidate is a real mixture of 
\dx2y2\ and $s^+$, with the former being the dominant component.
This case would require there to be a second transition, for which
there is presently no evidence in single crystal samples.  Given this,
the pairing state which is compatible with by far the highest proportion
of experiments to date on YBCO and similar compounds (but
not the Nd compound) is the pure \dx2y2\
state.

\noindent 

{\bf \ACK} 
We wish to thank the following individuals who have generously contributed
their thoughts, insights and findings to us 
during the writing of this review article
and at other times:
P. Anderson,
A. Barone,
D. Bonn,
P. Chaudhari,
J. Clarke,
G. Crabtree,
R. Dynes,
V. Emery,
D. Ginsberg,
L. Greene,
B. Gyorffy,
P. Hirschfeld, 
J. Kirtley,
R. Klemm,
I. Kosztin,
K. Levin,
K. Moler,
M. Norman,
M. Onellion,
D. Pines,
R. Radtke,
J. Sauls,
Z. Shen,
D. Van Harlingen,
J. Wilson,
D. Wollman
S. Yip.
We thank Don Ginsberg especially for inviting us to write this Chapter,
and for his incomparable patience.


We gratefully acknowledge the following support: National Science
Foundation grants NSF-DMR-93-14938 (NG) and NSF-DMR-91-20000 through the Science and Technology Center for Superconductivity (AJL) and the Office of Naval Research grant
ONR-N00014-95-1-0398 (JA).

\refout 
\vfill\eject
\figout

\end